\documentclass[12pt,preprint]{aastex}

\newif\ifcolorfigure
\colorfiguretrue

\usepackage{lineno}
\linenumbers

\usepackage{amsmath}
\usepackage{graphicx}
\usepackage{xspace}
\usepackage{url}

\usepackage{hyperref}

\newcommand{\mev}{\text{MeV}\xspace}
\newcommand{\gev}{\text{GeV}\xspace}
\newcommand{\tev}{\text{TeV}\xspace}
\newcommand{\s}{\text{s}\xspace}

\newcommand{\cm}{\text{cm}\xspace}
\newcommand{\km}{\text{km}\xspace}
\newcommand{\phflux}{\ensuremath{\text{ph}\,\text{cm}^{-2}\,\text{s}^{-1}}\xspace}
\newcommand{\tsext}{{\ensuremath{\text{TS}_{\text{ext}}}}\xspace}
\newcommand{\tsinc}{\ensuremath{\text{TS}_{\text{2pts}}}\xspace}

\newcommand{\likelihood}{\ensuremath{\mathcal{L}}\xspace}

\newcommand{\chandra}{\text{{\em Chandra}}\xspace}
\newcommand{\swiftxrt}{\text{{\em Swift}/XRT}\xspace}
\newcommand{\rosat}{\text{{\em ROSAT}}\xspace}
\newcommand{\suzaku}{\text{{\em Suzaku}}\xspace}
\newcommand{\asca}{\text{{\em ASCA}}\xspace}

\newcommand{\xmmnewton}{\text{{\em XMM-Newton}}\xspace}
\newcommand{\fermi}{\textit{Fermi}\xspace}

\newcommand{\rsixeight}{{\ensuremath{\text{r}_{68}}}\xspace}

\newcommand{\ts}{\text{TS}\xspace}
\newcommand{\aic}{\text{AIC}\xspace}
\newcommand{\glon}{\text{GLON}\xspace}
\newcommand{\glat}{\text{GLAT}\xspace}
\newcommand{\altdiff}{\text{alt,diff}\xspace}
\newcommand{\altpsf}{\text{alt,psf}\xspace}
\newcommand{\sys}{\text{sys}\xspace}
\newcommand{\stat}{\text{stat}\xspace}

\newcommand{\gtlike}{\ensuremath{\mathtt{gtlike}}\xspace}
\newcommand{\pointlike}{\ensuremath{\mathtt{pointlike}}\xspace}
\newcommand{\gtobssim}{\ensuremath{\mathtt{gtobssim}}\xspace}
\newcommand{\minuit}{\ensuremath{\mathtt{minuit}}\xspace}

\newcommand{\degree}{\ensuremath{^\circ}\xspace}

\usepackage{color}

\begin{document}

\title{Search for Spatially Extended \fermi-LAT Sources Using Two Years of Data}
\shorttitle{Search for Extended LAT Sources}


\author{
J.~Lande\altaffilmark{1,2}, 
M.~Ackermann\altaffilmark{3,4}, 
A.~Allafort\altaffilmark{1}, 
J.~Ballet\altaffilmark{5}, 
K.~Bechtol\altaffilmark{1}, 
T.~H.~Burnett\altaffilmark{6}, 
J.~Cohen-Tanugi\altaffilmark{7}, 
A.~Drlica-Wagner\altaffilmark{1}, 
S.~Funk\altaffilmark{1,8}, 
F.~Giordano\altaffilmark{9,10}, 
M.-H.~Grondin\altaffilmark{11,12}, 
M.~Kerr\altaffilmark{1}, 
M.~Lemoine-Goumard\altaffilmark{13,14}
}
\altaffiltext{1}{W. W. Hansen Experimental Physics Laboratory, Kavli Institute for Particle Astrophysics and Cosmology, Department of Physics and SLAC National Accelerator Laboratory, Stanford University, Stanford, CA 94305, USA}
\altaffiltext{2}{email: joshualande@gmail.com}
\altaffiltext{3}{Deutsches Elektronen Synchrotron DESY, D-15738 Zeuthen, Germany}
\altaffiltext{4}{email: markus.ackermann@desy.de}
\altaffiltext{5}{Laboratoire AIM, CEA-IRFU/CNRS/Universit\'e Paris Diderot, Service d'Astrophysique, CEA Saclay, 91191 Gif sur Yvette, France}
\altaffiltext{6}{Department of Physics, University of Washington, Seattle, WA 98195-1560, USA}
\altaffiltext{7}{Laboratoire Univers et Particules de Montpellier, Universit\'e Montpellier 2, CNRS/IN2P3, Montpellier, France}
\altaffiltext{8}{email: funk@slac.stanford.edu}
\altaffiltext{9}{Dipartimento di Fisica ``M. Merlin" dell'Universit\`a e del Politecnico di Bari, I-70126 Bari, Italy}
\altaffiltext{10}{Istituto Nazionale di Fisica Nucleare, Sezione di Bari, 70126 Bari, Italy}
\altaffiltext{11}{Max-Planck-Institut f\"ur Kernphysik, D-69029 Heidelberg, Germany}
\altaffiltext{12}{Landessternwarte, Universit\"at Heidelberg, K\"onigstuhl, D 69117 Heidelberg, Germany}
\altaffiltext{13}{Universit\'e Bordeaux 1, CNRS/IN2p3, Centre d'\'Etudes Nucl\'eaires de Bordeaux Gradignan, 33175 Gradignan, France}
\altaffiltext{14}{Funded by contract ERC-StG-259391 from the European Community}

\begin{abstract}
Spatial extension is an important characteristic for correctly
associating $\gamma$-ray-emitting sources with their counterparts at other wavelengths
and for obtaining an unbiased model of their spectra.  We present a new
method for quantifying the spatial extension of sources detected by the Large
Area Telescope (LAT), the primary science instrument on the {\em \fermi
Gamma-ray Space Telescope} (\fermi).  We perform a series of Monte Carlo
simulations to validate this tool and calculate the LAT threshold for
detecting the spatial extension of sources.  
We then test all sources
in the second \fermi-LAT catalog (2FGL) for extension. 
We report the detection of seven new spatially extended sources.
\end{abstract}

\section{Introduction}

A number of astrophysical source classes including supernova remnants
(SNRs), pulsar wind nebulae (PWNe), molecular clouds, normal galaxies,
and galaxy clusters are expected to be spatially resolvable 
by the Large Area Telescope (LAT), the primary instrument on the {\em \fermi
Gamma-ray Space Telescope} (\fermi).
Additionally, dark matter satellites are also hypothesized to
be spatially extended. See \cite{atwood_fermi} for pre-launch predictions.
The LAT has
detected seven SNRs which are significantly extended
at \gev energies: 
W51C, W30, 
IC~443, W28, W44, RX\,J1713.7$-$3946,
and the Cygnus Loop
\citep{w51c,w30_lat,ic443,w28,w44,rx_j1713_lat,cygnus_loop_lat}. In addition, 
three extended PWN have been detected by the LAT: MSH\,15$-$52, Vela~X, and
HESS\,J1825$-$137 \citep{msh1552,velax,fermi_hess_j1825}. Two
nearby galaxies, the Large and Small Magellanic Clouds, 
and the lobes of one
radio galaxy, Centaurus A, were spatially resolved at \gev energies
\citep{lmc,smc,cen_a_lat}.  A number of additional sources detected
at \gev energies are positionally coincident with sources that exhibit
large enough extension at other wavelengths to be spatially
resolvable by the LAT at \gev energies.
In particular, there are 59 \gev sources in the second Fermi Source Catalog (2FGL)
that might be associated with extended SNRs \citep[2FGL,][]{second_cat}.
Previous analyses of extended LAT sources were performed as dedicated
studies of individual sources so we expect that a systematic scan of
all LAT-detected sources could uncover additional spatially extended
sources. 

The current generation of air Cherenkov detectors have
made it apparent that many sources can be spatially resolved
at even higher energies.
Most
prominent was a survey of the Galactic plane using the High Energy
Stereoscopic System (H.E.S.S) which reported 14 spatially extended
sources with extensions varying from $\sim0\fdg1$ to $\sim0\fdg25$
\citep{hess_plane_survey}.  Within our Galaxy very few
sources detected at \tev energies, most notably the $\gamma$-ray binaries
LS\,5039 \citep{HESSLS5039}, LS I+61$-$303 \citep{MAGICLSI, VERITASLSI},
HESS\,J0632+057 \citep{HESS0632}, and the Crab nebula \citep{crab_weekes},
have no detectable extension.  High-energy $\gamma$-rays from
\tev sources are produced by the decay of $\pi^0$s produced by hadronic
interactions with interstellar matter and by relativistic electrons
due to Inverse Compton (IC) scattering and bremsstrahlung radiation.  
It is plausible that the \gev and
\tev emission from these sources originates from the same population of
high-energy particles and so at least some of these sources should be
detectable at \gev energies.  Studying 
these \tev sources at \gev energies would help to
determine the emission mechanisms producing these high energy photons.

The LAT is a pair conversion telescope that has been surveying the
$\gamma$-ray sky since 2008 August.  The LAT has broad energy coverage
(20 \mev to $>300$ \gev), wide field of view ($\sim 2.4$ sr), and large
effective area ($\sim 8000\ \cm^2$ at $>1$ \gev) 
Additional information about the performance of the LAT can be found in
\cite{atwood_LAT_mission}.

Using 2 years of all-sky survey data, the LAT Collaboration published
2FGL \citep[2FGL,][]{second_cat}.
The possible counterparts of many of these sources can be spatially resolved
when observed at other frequencies. But detecting the spatial extension
of these sources at \gev energies is difficult because the size of the
point-spread function (PSF) of the LAT is comparable to the typical size
of many of these sources.

The capability to spatially resolve \gev $\gamma$-ray
sources is important for several reasons.  
Finding a coherent source extension across different energy bands can
help to associate a LAT source to an otherwise confused counterpart.
Furthermore, $\gamma$-ray emission from dark matter
annihilation has been predicted to be detectable by the LAT.
Some of the dark matter substructure in our Galaxy 
could be spatially resolvable by the LAT \citep{pre_luanch_dark_matter_fermi}.  
Characterization of spatial extension could help to identify this substructure.
Also,
due to the strong energy dependence of the LAT PSF, the spatial and
spectral characterization of a source cannot be
decoupled. An inaccurate
spatial model will bias the spectral model of the source and vice versa. Specifically,
modeling a spatially extended source as point-like will systematically
soften measured spectra. Furthermore, correctly
modeling source extension is important for 
understanding an entire region of the sky. For example,
an imperfect model of the spatially extended LMC introduced
significant residuals in the surrounding region \citep{first_cat,second_cat}.
Such residuals can bias the significance and measured spectra of
neighboring sources in the densely populated Galactic plane.

 For these reasons, in Section~\ref{analysis_methods_section}
we present a new systematic method for analyzing spatially extended
LAT sources.  
In Section~\ref{validate_ts}, we demonstrate
that this method can be used to test the statistical significance of the
extension of a LAT source and we assess the expected level of bias 
introduced by assuming an incorrect spatial model.
In Section~\ref{extension_sensitivity},
we calculate the LAT detection threshold to resolve the extension
of a source.
In Section~\ref{dual_localization_method}, we
study the ability of the LAT to 
distinguish between a single extended source and unresolved closely-spaced point-like sources
In Section~\ref{test_2lac_sources}, we further demonstrate that our
detection method does not misidentify point-like sources as being
extended by testing the extension of active Galactic nuclei (AGN)
believed to be unresolvable.  In Section~\ref{validate_known},
we systematically reanalyze the twelve extended sources included
in the 2FGL catalog and in Section~\ref{systematic_errors_on_extension}
we describe a way to estimate systematic errors on the measured extension of a source.
In Section~\ref{extended_source_search_method}, we
describe a search for new spatially extended LAT sources. Finally,
in Section~\ref{new_ext_srcs_section} we present the detection of the
extension of nine spatially extended sources that were reported in the 2FGL catalog
but treated as point-like in the analysis.  Two of these sources have been previously analyzed in
dedicated publications.

\section{Analysis Methods}
\label{analysis_methods_section}

Morphological studies of sources using the LAT are challenging
because of the strongly energy-dependent PSF that is comparable in
size to the extension of many sources expected to be detected at
\gev energies.  Additional complications arise for sources along
the Galactic plane due to systematic uncertainties in the model for
Galactic diffuse emission.  

For energies below $\sim$300~\mev, the angular resolution is limited by
multiple scattering in the silicon strip tracking section
of the detector and is several degrees at 100 \mev.  The PSF improves
with energy approaching a 68\% containment radius of $\sim0\fdg2$ at
the highest energies (when averaged over the acceptance of the LAT)
and is limited by the ratio of the strip pitch to the height of the tracker
\citep{atwood_LAT_mission,on_orbit_calibration,lat_on_orbit_psf}.\footnote{More
information about the performance of the LAT can be found at the \fermi
Science Support Center (FSSC, \url{http://fermi.gsfc.nasa.gov}).} However,
since most high energy astrophysical sources have spectra that decrease
rapidly with increasing energy, there are typically fewer higher
energy photons with improved angular resolution. Therefore sophisticated
analysis techniques are required to maximize the sensitivity of the LAT
to extended sources.

\subsection{Modeling Extended Sources in the \pointlike Package}

A new maximum-likelihood analysis tool has been developed to address the
unique requirements for studying spatially extended sources with the LAT.
It works by maximizing the Poisson 
likelihood to detect the observed distributions of $\gamma$-rays (referred to as counts)
given a parametrized spatial and spectral model of the sky.  
The data are binned spatially, using a HEALPix pixellization, and spectrally 
\citep{healpix} and the likelihood is maximized over all bins in
a region.
The extension of a source can be modeled by a geometric shape
(e.g. a disk or a two-dimensional Gaussian) and the position, extension,
and spectrum of the source can be simultaneously fit.

This type of analysis is unwieldy using the standard LAT likelihood
analysis tool \gtlike\footnote{\gtlike is distributed publicly by the
FSSC.} because it can only fit the spectral parameters of the model
unless a more sophisticated iterative procedure is used.  
We note that \gtlike has been used in the
past in several studies of source extension in the LAT Collaboration
\citep{lmc,smc,w28,w51c}.  In these studies, a 
set of \gtlike maximum likelihood fits at fixed extensions was used
to build a profile of the likelihood as a function of extension.
The \gtlike likelihood profile approach has been shown to correctly
reproduce the extension of simulated extended sources assuming that the
true position is known \citep{francesco_2011}.  But it is not optimal
because the position, extension, and spectrum of the source must be
simultaneously fit to find the best fit parameters and to maximize the
statistical significance of the detection.  Furthermore, because the \gtlike
approach is computationally intensive, no large-scale Monte Carlo
simulations have been run to calculate its false detection rate.

The approach presented here is based on a second maximum likelihood
fitting package developed in the LAT Collaboration called \pointlike
\citep{first_cat,matthew_kerr_thesis}.  The choice to base the
spatial extension fitting on \pointlike rather than \gtlike was made
due to considerations of computing time.  The \pointlike algorithm was
optimized for speed to handle larger numbers of sources efficiently,
which is important for our catalog scan and for being able
to perform large-scale Monte Carlo simulations to validate the analysis.
Details on the \pointlike package can be
found in \cite{matthew_kerr_thesis}.  We extended the code to allow a
simultaneous fit of the source extension together with the position and
the spectral parameters.

\subsection{Extension Fitting}
\label{extension_fitting}

In \pointlike, one can fit the position and extension 
of a source under the assumption that the source model
can be factorized:
$M(x,y,E)=S(x,y)\times X(E)$, where $S(x,y)$ is the spatial distribution
and $X(E)$ is the spectral distribution.  To fit an extended source,
\pointlike convolves the extended source shape with the PSF (as a function
of energy) and uses the \minuit library \citep{minuit_documentation}
to maximize the likelihood by simultaneously varying the position,
extension, and spectrum of the source.  As will be described in
Section~\ref{monte_carlo_validation}, simultaneously fitting the
position, extension, and spectrum is important to maximize
the statistical significance of the detection of the extension of a source.
To avoid projection effects, the longitude and latitude 
of the source are
not directly fit but instead the displacement of the source 
in a reference frame centered on the source.

The significance of the extension of a source can be calculated from the
likelihood-ratio test. The likelihood ratio defines the
test statistic (TS) by comparing the likelihood of a simpler hypothesis to
a more complicated one:
\begin{equation}
  \ts=2\log(\likelihood(H_1)/\likelihood(H_0)),
\end{equation}
where $H_1$ is the more complicated hypothesis and $H_0$ the simpler one.
For the case of the extension test, we compare the likelihood
when assuming the source has either a point-like or spatially extended spatial model:
\begin{equation}
  \tsext=2\log(\likelihood_\text{ext}/\likelihood_\text{ps}).
\end{equation}
\pointlike calculates \tsext by fitting a source first with a spatially
extended model and then as a point-like source.  The interpretation
of \tsext in terms of a statistical significance is discussed in
Section~\ref{monte_carlo_validation}.

For extended sources with an assumed radially-symmetric shape,
we optimized the calculation by performing one
of the integrals analytically.
The expected photon 
distribution can be written as
\begin{equation}
  \text{PDF}(\vec r) = \int  \text{PSF}(|\vec r - \vec r'|)I_\text{src}(\vec r') r' dr' d\phi'
\end{equation}
where $\vec r$ represents the position in the sky and
$I_\text{src}(\vec r)$ is the spatial distribution of the
source.
The PSF of the LAT is currently parameterized 
in the Pass~7\_V6 (P7\_V6) Source Instrument
Response Function \citep[IRFs,][]{lat_on_orbit_psf} by a King function \citep{king_function}:
\begin{equation}
  \text{PSF}(r) = 
  \frac{1}{2\pi\sigma^2}
  \left(1-\frac{1}{\gamma}\right)
  \left(1+\frac{u}{\gamma}\right)^{-\gamma},
\end{equation}
where $u=(r/\sigma)^2/2$ and $\sigma$ and $\gamma$ are free parameters
\citep{matthew_kerr_thesis}.  For radially-symmetric extended sources,
the angular part of the integral can be evaluated analytically
\begin{align}
  \text{PDF}(u) & = \int_0^\infty r' dr'
  I_\text{src}(v) 
  \int_0^{2\pi} d\phi' 
  \text{PSF}(\sqrt{2\sigma^2(u+v-2\sqrt{uv}\cos(\phi-\phi'))})
  \\
  & = \int_0^\infty dv
  I_\text{src}(v) 
  \left(\frac{\gamma-1}{\gamma}\right)
  \left( \frac{\gamma}{\gamma + u + v}\right)^\gamma 
  \times {}_2F_1 \left(\gamma/2,\frac{1+\gamma}{2},1,\frac{4uv}{(\gamma+u+v)^2}\right),
\end{align}
where $v=(r'/\sigma)^2/2$ and ${}_2F_1$ is the Gaussian hypergeometric
function.  This convolution formula reduces the expected photon
distribution to a single numerical integral.

There will always be a small numerical discrepancy between the expected
photon distribution derived from a true point-like source and a very small
extended source due to numerical error in the convolution.  In most
situations, this error is insignificant.  But in particular for
very bright sources, this numerical error has the potential to bias the
TS for the extension test. Therefore, when calculating
\tsext, we compare the likelihood fitting the source with an extended
spatial model to the likelihood when the extension is
fixed to a very small value ($10^{-10}$ degrees in radius 
for a uniform disk model).

We estimate the error on the extension of a source by fixing
the position of the source and varying the extension until the
log of the likelihood has decreased by 1/2, corresponding to a $1\sigma$ error
\citep{Statistical_methods_book}.  Figure~\ref{four_plots_ic443}
demonstrates this method by showing the change in the log of the
likelihood when 
varying the modeled extension of the SNR IC~443.  The localization
error is calculated by fixing the extension and spectrum of the source
to their best fit values and then
fitting the log of the likelihood to
a 2D Gaussian as a function
of position. This localization error algorithm is further described in
\cite{second_cat}.

\subsection{\gtlike Analysis Validation}
\label{gtlike_crosscheck}

\pointlike is important for analyses of LAT data that require many iterations
such as source localization and extension fitting.  On the other hand,
because \gtlike makes fewer approximations in calculating the likelihood
we expect the spectral parameters found with \gtlike to be slightly more
accurate.  Furthermore, because \gtlike is the 
standard likelihood analysis package for LAT data, 
it has been more extensively validated for spectral analysis.
For those reasons, in the following analysis we used \pointlike to
determine the position and extension of a source and subsequently derived
the spectrum using \gtlike. Both \gtlike and \pointlike can be used to
estimate the statistical significance of the extension of a source and we
required that both methods agree for a source to be considered extended.
There was good agreement between the two methods.  Unless explicitly
mentioned, all \ts, \tsext, and spectral parameters were calculated using
\gtlike with the best-fit positions and extension found by \pointlike.

\subsection{Comparing Source Sizes}

\label{compare_source_size}

\begin{figure}
    \ifcolorfigure
      \plotone{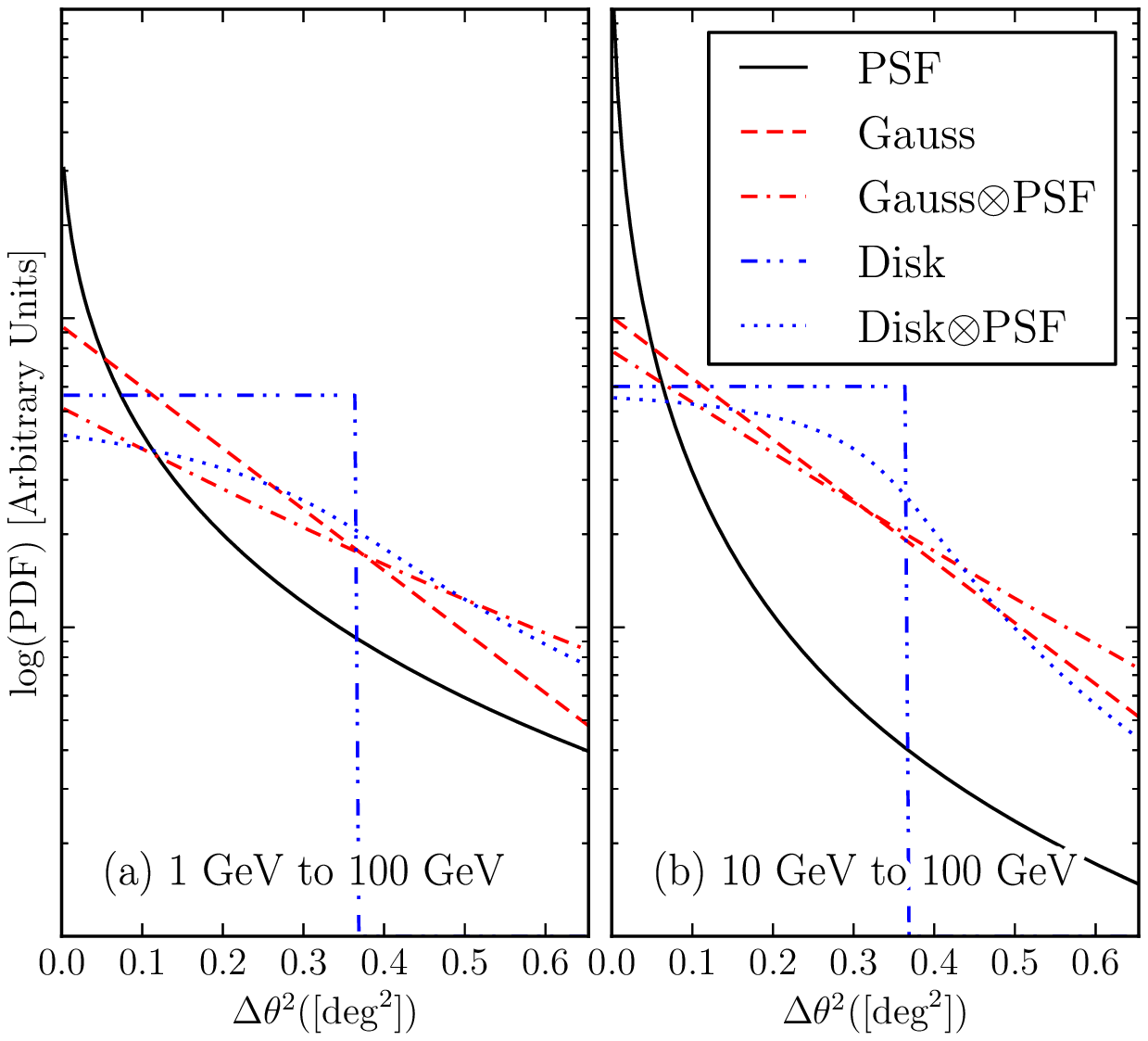}
    \else
      \plotone{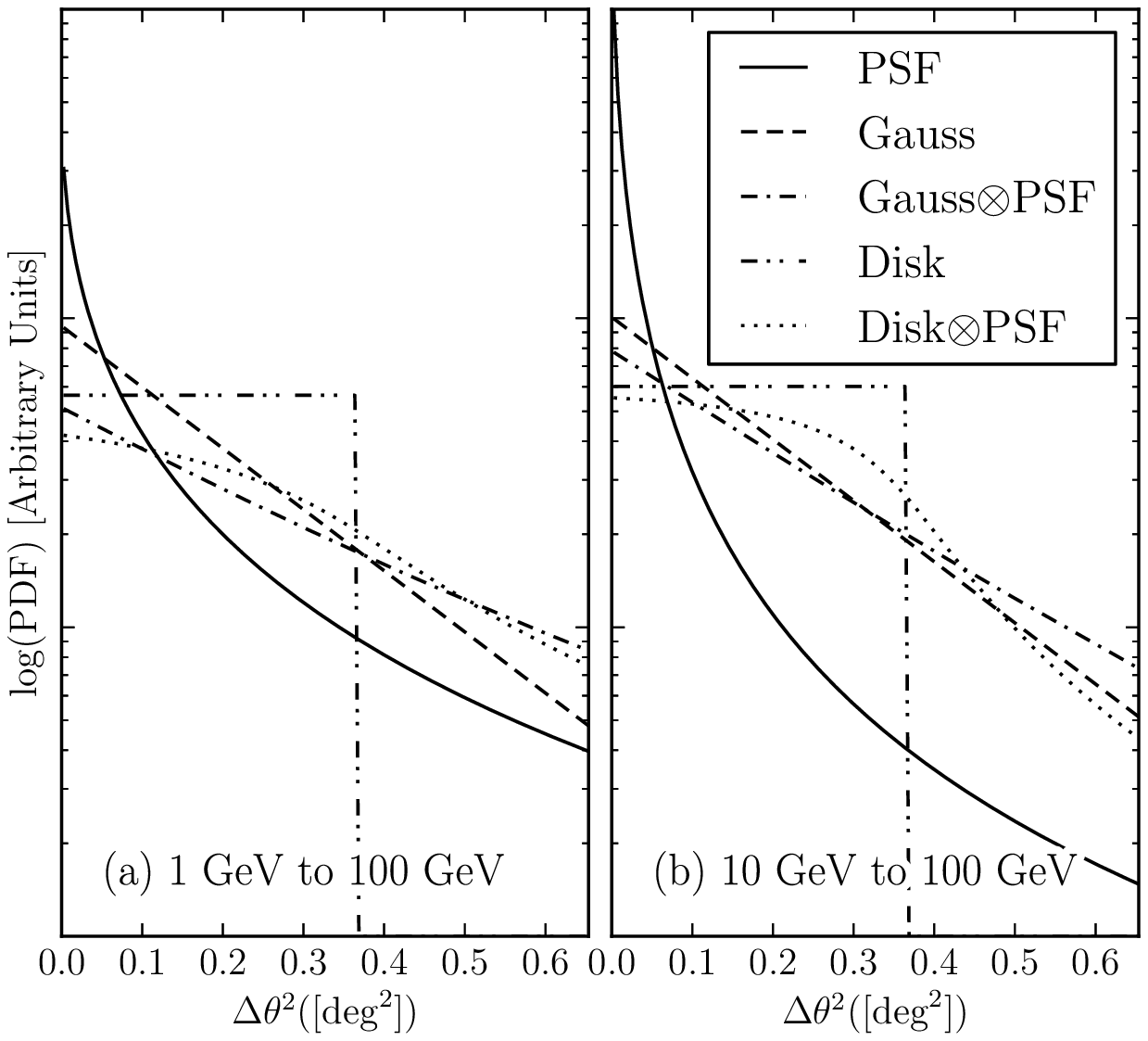}
    \fi
    \caption{
    A comparison of a 2D Gaussian and uniform disk spatial model
    of extended sources before and after convolving with the PSF for two
    energy ranges.  The solid black line is the PSF that would be observed
    for a power-law source of spectral index 2. The dashed line
    and the dash-dotted lines are 
    the brightness profile of a Gaussian with $\rsixeight=0\fdg5$
    and the convolution of this profile with the LAT PSF respectively
    (colored red in the online version).
    The dash-dot-dotted and the dot-dotted lines are the brightness profile
    of a uniform disk with $\rsixeight=0\fdg5$ and the convolution
    of this profile with the LAT PSF respectively (colored blue in the online version).
    }\label{compare_disk_gauss}
  \end{figure}

We considered two models for the
surface brightness profile for extended sources: a 2D Gaussian model
\begin{equation}\label{gauss_pdf}
  I_\text{Gaussian}(x,y)=\tfrac{1}{2\pi\sigma^2}\exp\left(-(x^2+y^2)/2\sigma^2\right)
\end{equation}
or a uniform disk model
\begin{equation}\label{disk_pdf}
  I_\text{disk}(x,y)=
  \begin{cases}
    \frac{1}{\pi\sigma^2} & x^2+y^2\le\sigma^2 \\
    0                      & x^2+y^2>\sigma^2.
  \end{cases}
\end{equation}
Although these shapes are significantly different,
Figure~\ref{compare_disk_gauss} shows that, after convolution with the
LAT PSF, their PDFs are similar for a source that has a 0\fdg5 radius
typical of LAT-detected extended sources.  To allow a valid comparison
between the Gaussian and the uniform disk models,
we define the source size as the radius containing 68\% of the
intensity ($\rsixeight$). 
By direct integration, we find
\begin{align}
\rsixeight_\text{,Gaussian}=&1.51\sigma, \\
\rsixeight_\text{,disk}=&0.82\sigma, 
\end{align}
where $\sigma$ is defined
in Equation~\ref{gauss_pdf} and Equation~\ref{disk_pdf} respectively.
For the example above, $\rsixeight=0\fdg5$ so $\sigma_\text{disk}=0.61\degree$
and $\sigma_\text{Gaussian}=0.33\degree$.

For sources that are comparable in size to the PSF,
the differences in the PDF for
different spatial models are lost in the noise and the LAT is not sensitive
to the detailed spatial structure of these sources.  
In section \ref{bias_wrong_spatial_model}, we perform a dedicated Monte Carlo simulation
that shows there is little bias due to incorrectly modeling the spatial structure
of an extended source.
Therefore, in our search for extended sources we use only a radially-symmetric uniform
disk spatial model. Unless otherwise noted,
we quote the radius to the edge ($\sigma$) as the size of the source.

\section{Validation of the \ts Distribution}
\label{validate_ts}

\subsection{Point-like Source Simulations Over a Uniform Background}
\label{monte_carlo_validation}

We tested the theoretical distribution for \tsext
to evaluate the false detection probability for measuring source extension.
To do so, we tested simulated point-like sources for extension. 
\cite{mattox_egret}
discuss that the \ts distribution for a likelihood-ratio test
on the existence of a source at a given position is
\begin{equation}\label{ts_ext_distribution}
  P(\ts)=\onehalf(\chi^2_1(\ts)+\delta(\ts)),
\end{equation}
where $P(\ts)$ is the probability density to get a particular value of TS,
$\chi^2_1$ is the chi-squared distribution with one degree of freedom, and
$\delta$ is the Dirac delta function.
The particular form of Equation \ref{ts_ext_distribution} is due to the
null hypothesis (source flux $\Phi=0$) residing on the edge of parameter
space and the model hypothesis adding a single degree of freedom (the source flux).
It leads to the often quoted result $\sqrt{TS}=\sigma$, where 
$\sigma$ here refers to the significance of the detection. It is plausible
to expect a similar distribution for the TS in the test for
source extension since the same conditions apply (with the source flux
$\Phi$ replaced by the source radius $r$ and $r<0$ being unphysical).
To verify Equation~\ref{ts_ext_distribution}, we evaluated the
empirical distribution function of \tsext computed from simulated sources.

We simulated point-like sources with various spectral forms using
the LAT on-orbit simulation tool
\gtobssim\footnote{\gtobssim is distributed publicly by the FSSC.} and fit the sources
with \pointlike using both point-like
and extended source hypotheses.  These point-like sources were simulated with a power-law
spectral model with integrated fluxes above 100 \mev ranging from $3\times10^{-9}$ 
to $1\times10^{-5}$ \phflux and spectral
indices ranging from 1.5 to 3.  These values
were picked to represent typical parameters of LAT-detected
sources. The point-like sources were simulated on top of an isotropic
background with a power-law spectral model with
integrated flux above 100 \mev of $1.5\times10^{-5}$ \phflux sr$^{-1}$
and spectral index 2.1.
This was
taken to be the same as the isotropic spectrum measured by EGRET
\citep{sreekumar_isotropic}.  This spectrum is comparable
to the high-latitude background intensity seen by the LAT.
The Monte Carlo simulation was performed
over a one-year observation period using a representative 
spacecraft orbit and livetime.
The reconstruction was performed
using the P7\_V6 Source class event selection and IRFs \citep{lat_on_orbit_psf}. For each 
significantly detected point-like source ($\ts\ge25$), we used \pointlike
to fit the source as an extended source and calculate \tsext.
This entire procedure was performed twice, once fitting in the 1 \gev
to 100 \gev energy range and once fitting in the 10 \gev to 100 \gev
energy range.

\begin{figure}
    \ifcolorfigure
    \plotone{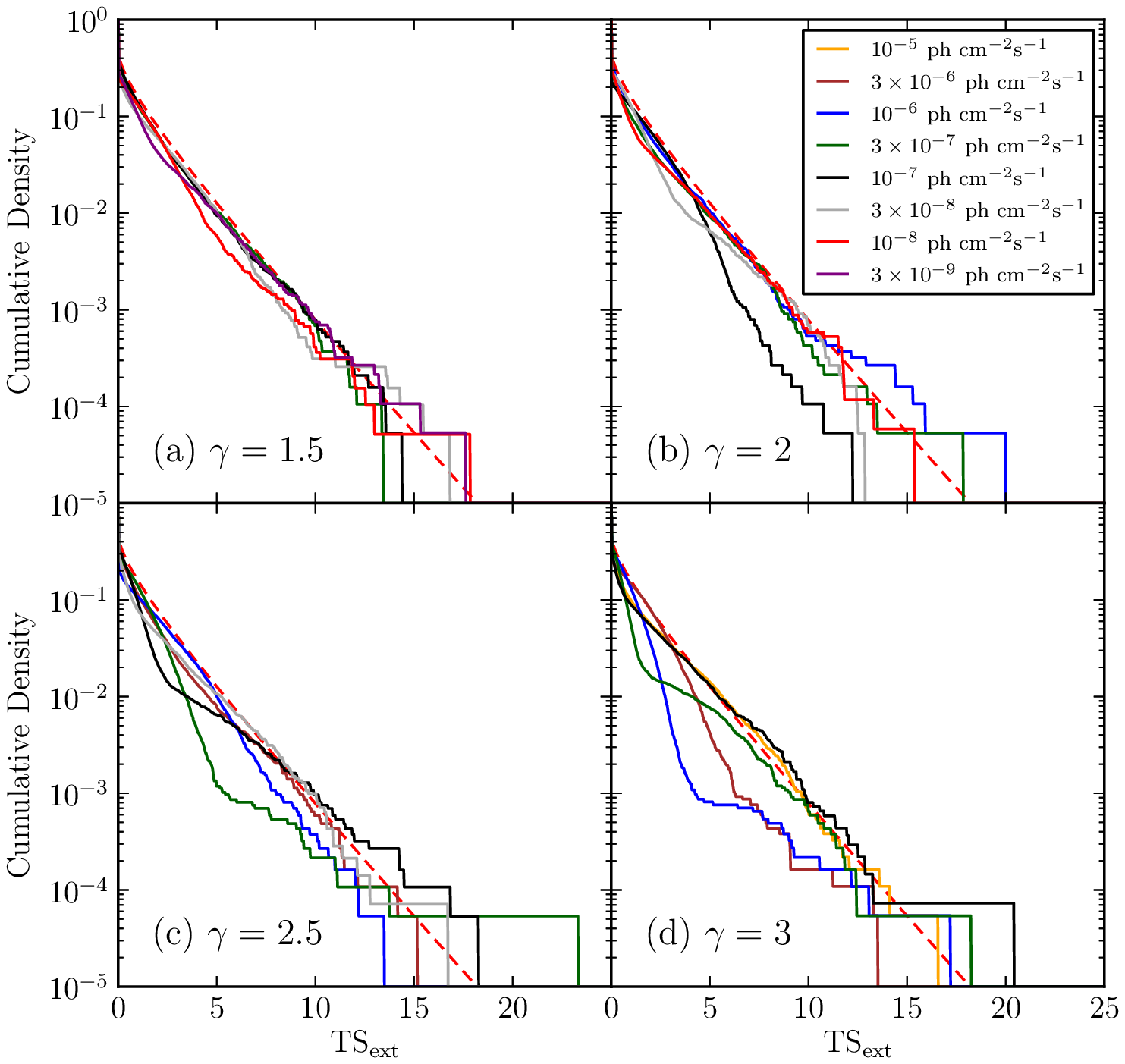}
    \else
    \plotone{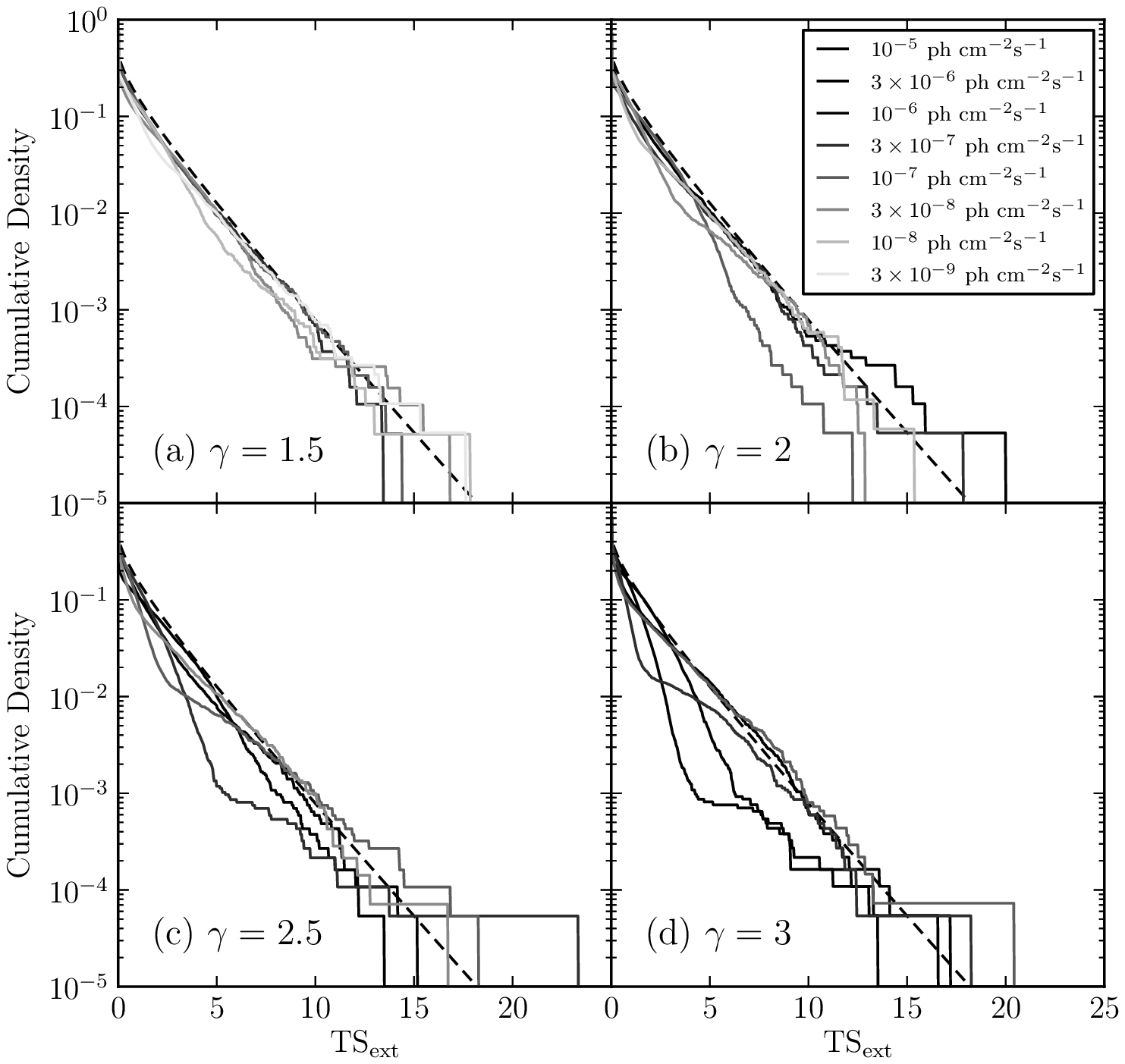}
    \fi
    \caption{
    Cumulative distribution of the TS for the extension test when fitting
    simulated point-like sources in the 1 GeV to 100 GeV energy range.
    The four plots
    represent simulated sources of different spectral indices and
    the different lines (colored in the online version) 
    represent point-like sources with different 100 \mev
    to 100 \gev integral fluxes.  The dashed line (colored red)
    is the cumulative
    density function of Equation~\ref{ts_ext_distribution}.
    }\label{ts_ext_mc_1000}
  \end{figure}

\begin{figure}
    \ifcolorfigure
    \plotone{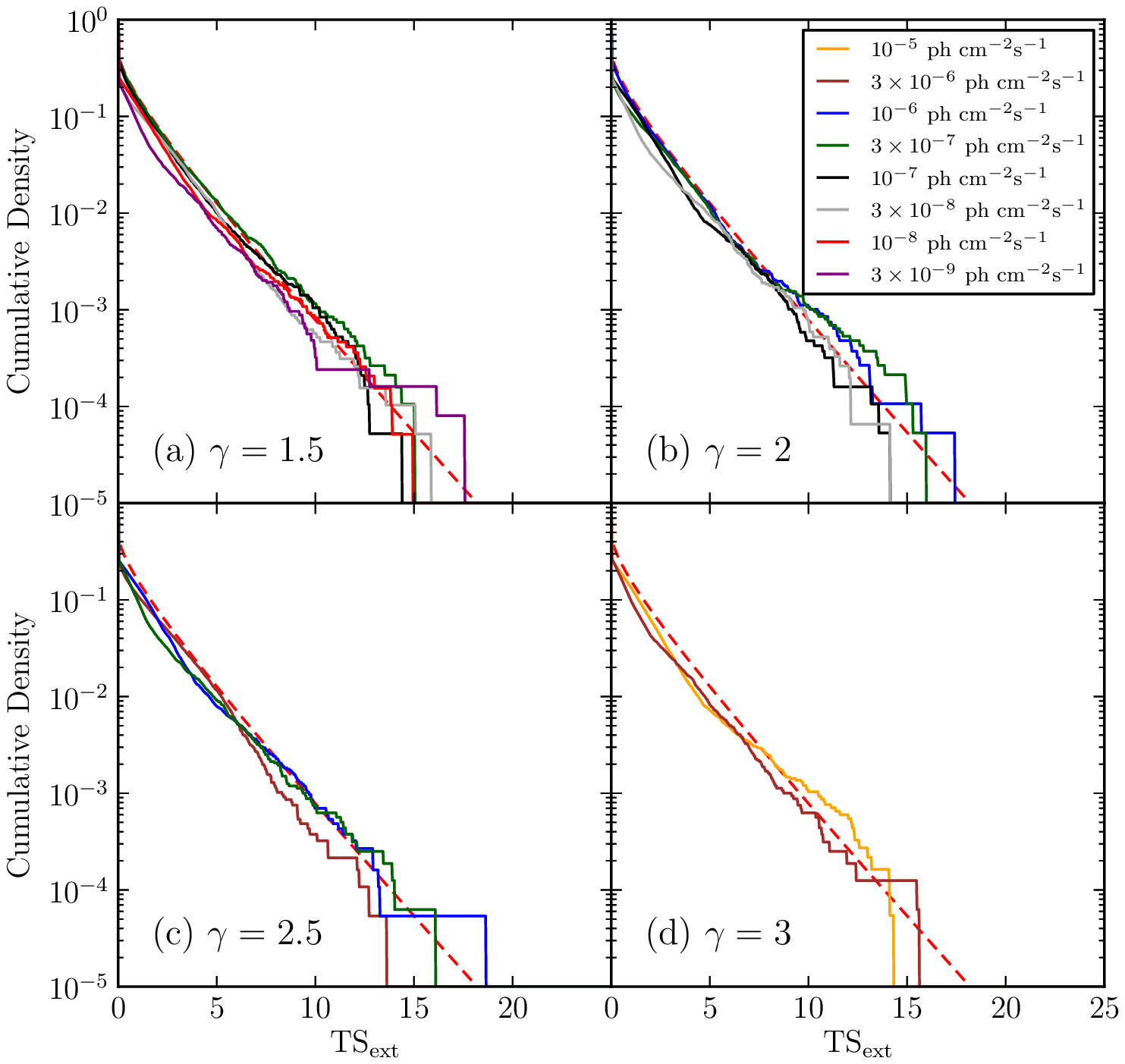}
    \else
    \plotone{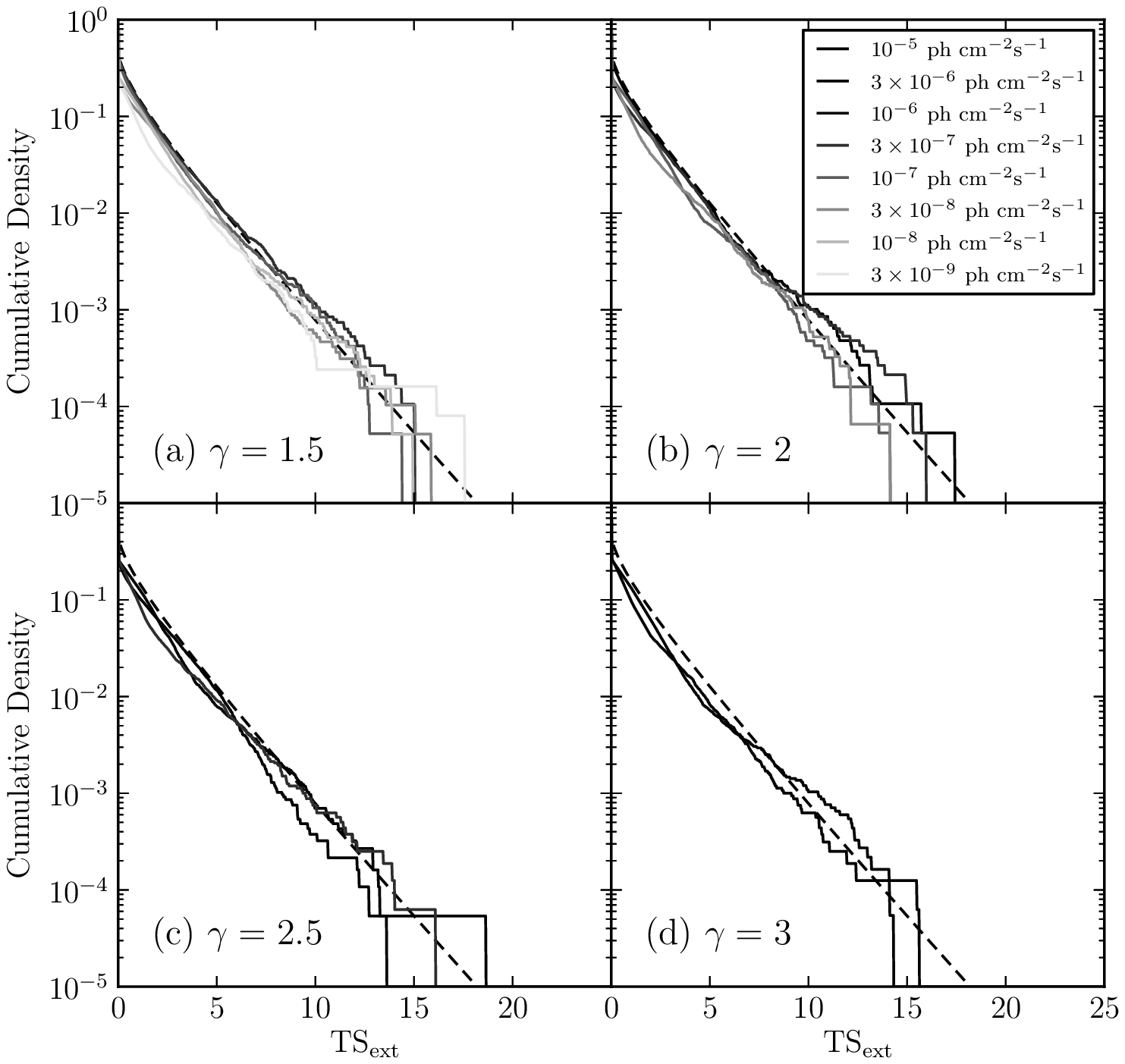}
    \fi
    \caption{
    The same plot as Figure~\ref{ts_ext_mc_1000} but fitting in the 10 \gev to 100 \gev energy 
    range.
    }\label{ts_ext_mc_10000}
  \end{figure}

For each set of spectral parameters, $\sim20,000$ statistically independent
simulations were performed. For lower-flux spectral models, many of the
simulations left the source insignificant ($\ts<25$)
and were discarded.  Table~\ref{ts_ext_num_sims}
shows the different spectral models used in our study as well as the
number of simulations and the average point-like source
significance.  The cumulative density of \tsext is plotted in
Figures~\ref{ts_ext_mc_1000} and \ref{ts_ext_mc_10000} 
and compared to the $\chi^2_1/2$ distribution of
Equation~\ref{ts_ext_distribution}.

Our study shows broad agreement between simulations and
Equation~\ref{ts_ext_distribution}. To the extent that there is
a discrepancy, the simulations tended to produce smaller than expected
values of \tsext which would make the formal significance conservative.
Considering the distribution in Figures~\ref{ts_ext_mc_1000} and
\ref{ts_ext_mc_10000}, the choice of a threshold \tsext set to 16
(corresponding to a formal $4\sigma$ significance) is reasonable.

\subsection{Point-like Source Simulations Over a Structured Background}
\label{validation_over_plane}

We performed a second set of simulations to show that the theoretical distribution
for \tsext is still preserved when the point-like sources are present over
a highly-structured diffuse background.
Our simulation setup was the same as above except that the sources were
simulated on top of and analyzed assuming the presence of the standard
Galactic diffuse and isotropic background models used in 2FGL.  In our
simulations, we selected our sources to have random positions on the sky
such that they were within 5\degree of the Galactic plane. This probes the 
brightest and most strongly contrasting areas of the Galactic background.

To limit the number of tests, we selected only one flux
level for each of the four spectral indices and we performed
this test only in the 1 \gev to 100 \gev energy range. 
As described below, the fluxes were selected so that $\ts\sim50$. We do not
expect to be able to spatially resolve sources 
at lower fluxes than these, and the results for much brighter sources
are less likely to be affected by the 
structured background.

Because the Galactic diffuse emission is highly structured with
strong gradients, the point-source
sensitivity can vary significantly across the Galactic plane.
To account for this, we scaled the flux (for a given spectral index)
so that the source always has approximately the same signal-to-noise ratio:
\begin{equation}
  \label{scale_flux_by_background}
  F(\vec{x}) = F(\text{GC}) \times \left(
  \frac{B(\vec{x})}{B(\text{GC})}\right)^{1/2}.
\end{equation}
Here, $\vec{x}$ is the position of the simulated source, $F$ is the integral
flux of the source from 100 \mev to 100 \gev, $F(\text{GC})$
is the same quantity if the source was at the Galactic center, $B$
is the integral of the Galactic diffuse and isotropic emission
from 1 \gev to 100 \gev at the position of the source, and $B(\text{GC})$ is the same quantity
if the source was at the Galactic center.  For the four spectral models,
Table~\ref{ts_ext_num_sims} lists $F(\text{GC})$ and the average value of \ts.

\begin{figure}
    \ifcolorfigure
    \plotone{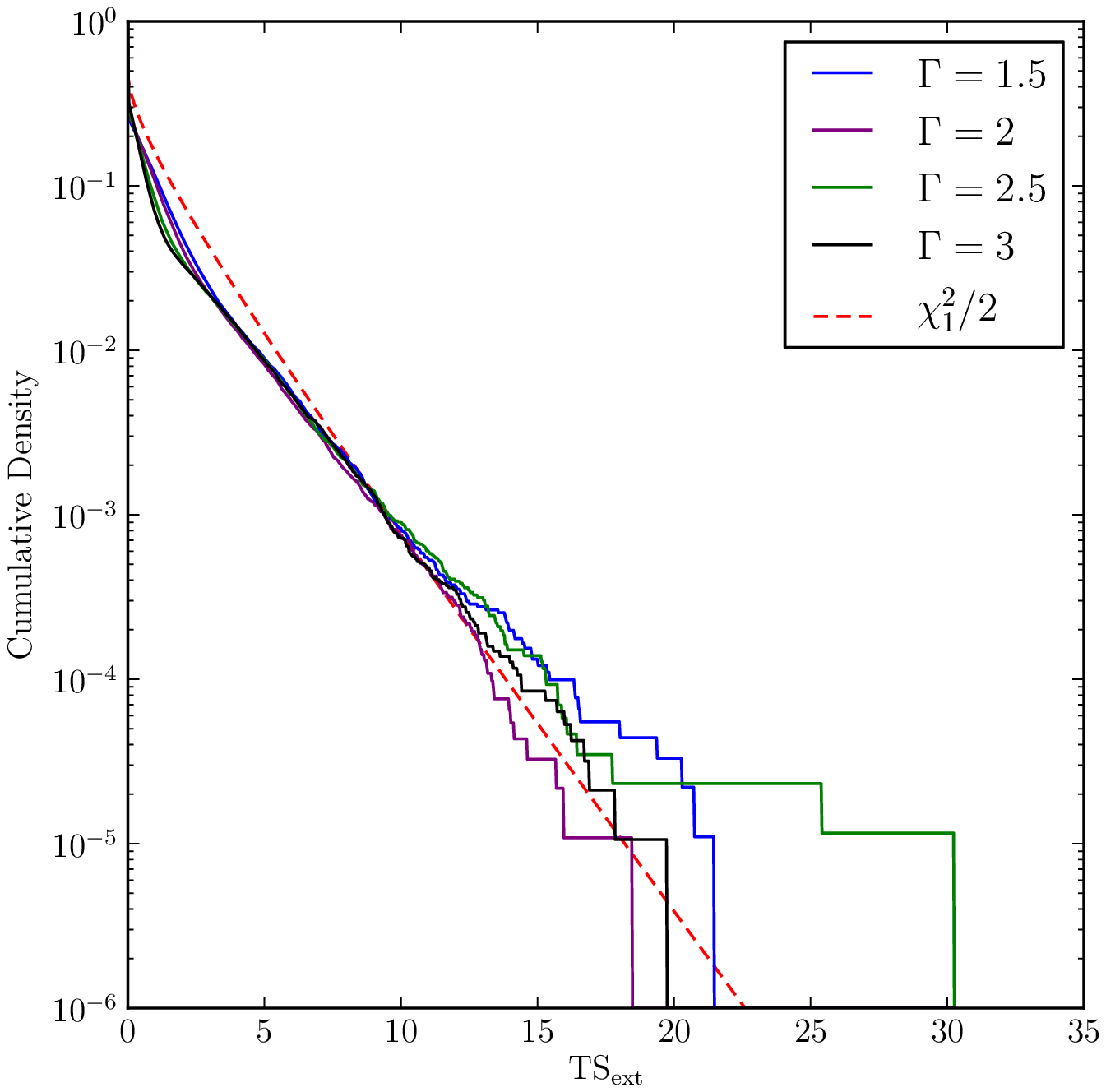}
    \else
    \plotone{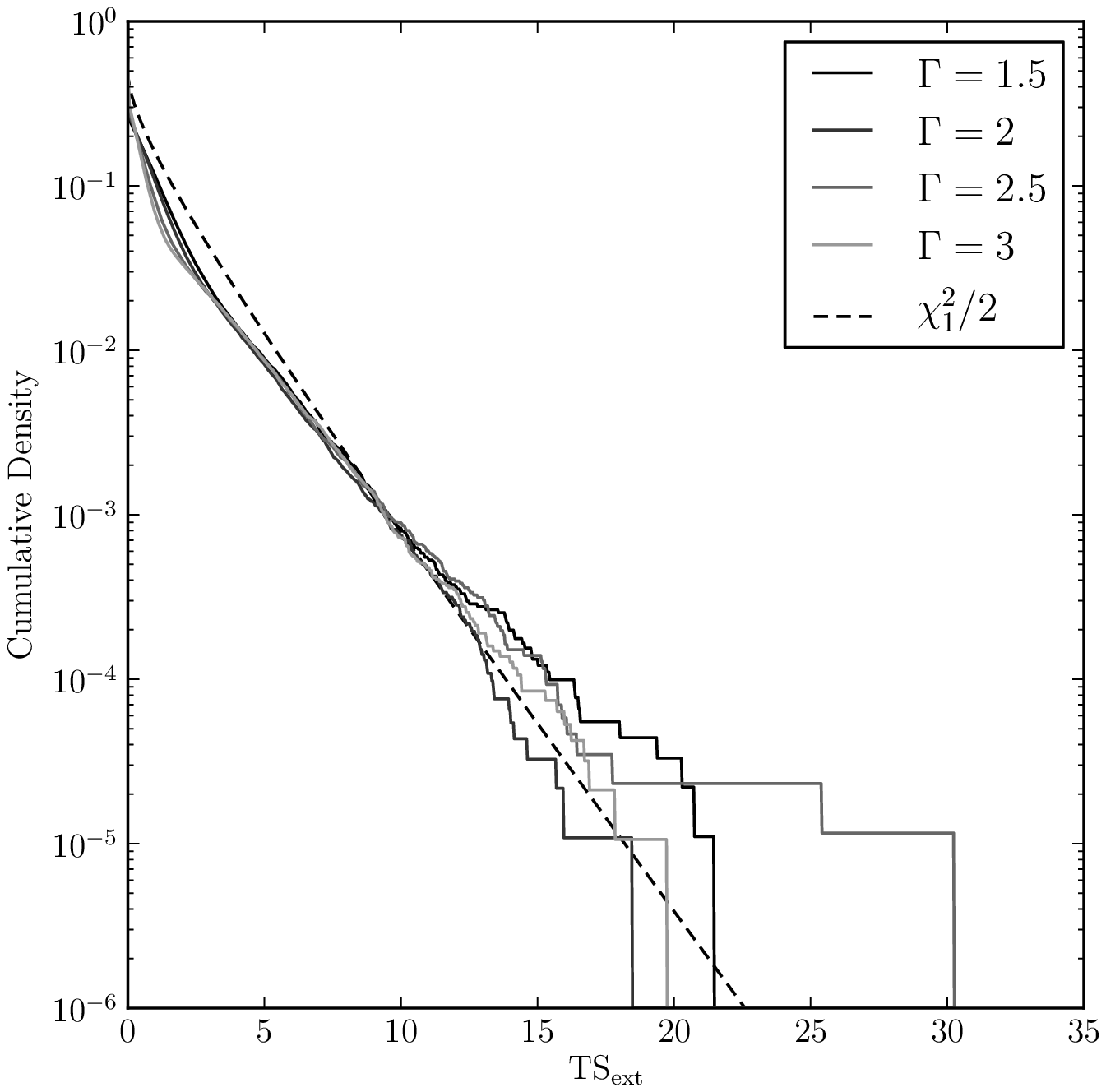}
    \fi
    \caption{
    Cumulative distribution of \tsext
    for sources simulated on top of the Galactic diffuse and isotropic background.
    }\label{tsext_plane_plot}
  \end{figure}

For each spectrum, we performed $\sim90,000$ simulations.
Figure~\ref{tsext_plane_plot} shows the cumulative density
of \tsext for each spectrum. For small values of \tsext,
there is good agreement between the simulations and
theory.  For the highest values of \tsext, there is possibly
a small discrepancy, but the discrepancy is not statistically significant.
Therefore, we are confident we can use \tsext as a robust measure of
statistical significance when testing LAT-detected sources for extension.

\subsection{Extended Source Simulations Over a Structured Background}
\label{bias_wrong_spatial_model}

We also performed a Monte Carlo study to show that incorrectly modeling the
spatial extension of an extended source does not substantially bias
the spectral fit of the source, although it does alter the value of the \ts.
To assess this, we simulated the spatially extended ring-type SNR W44.
We selected W44 because it is the most significant extended source detected by the LAT 
that has a non-radially symmetric photon distribution \citep{w44}. 

W44 was simulated with a power-law spectral model with an integral flux
of $7.12\times10^{-8}$ \phflux in the energy range from 1 \gev to 100
\gev and a spectral index of 2.66 (see Section~\ref{validate_known}).

W44 was simulated with the elliptical ring spatial model described in
\cite{w44}. For reference, the ellipse has a semi-major axis of 0\fdg3,
a semi-minor axis of 0\fdg19, a position angle of $147\degree$
measured East of celestial North, and the ring's inner radius is 75\% of
the outer radius.

We used a simulation setup similar to that described in
Section~\ref{validation_over_plane}, but the simulations
were over the 2-year interval of the 2FGL catalog.
In the simulations, 
we did not include the finite energy resolution of the LAT
to isolate any effects due to changing the assumed spatial model. 
The fitting code we use also ignores this energy dispersion and the
potential bias introduced by this will be discussed in an upcoming paper
by the LAT collaboration \citep{lat_on_orbit_psf}.
In total, we performed 985 independent simulations.

The simulated sources were fit using a point-like spatial model,
a radially-symmetric Gaussian spatial model, a uniform disk spatial model, 
an elliptical disk spatial model, and finally with an elliptical
ring spatial model.
We obtained the best fit spatial parameters using \pointlike and, 
with these parameters, obtained the best fit spectral parameters 
using \gtlike. 

\begin{figure}
    \ifcolorfigure
    \plotone{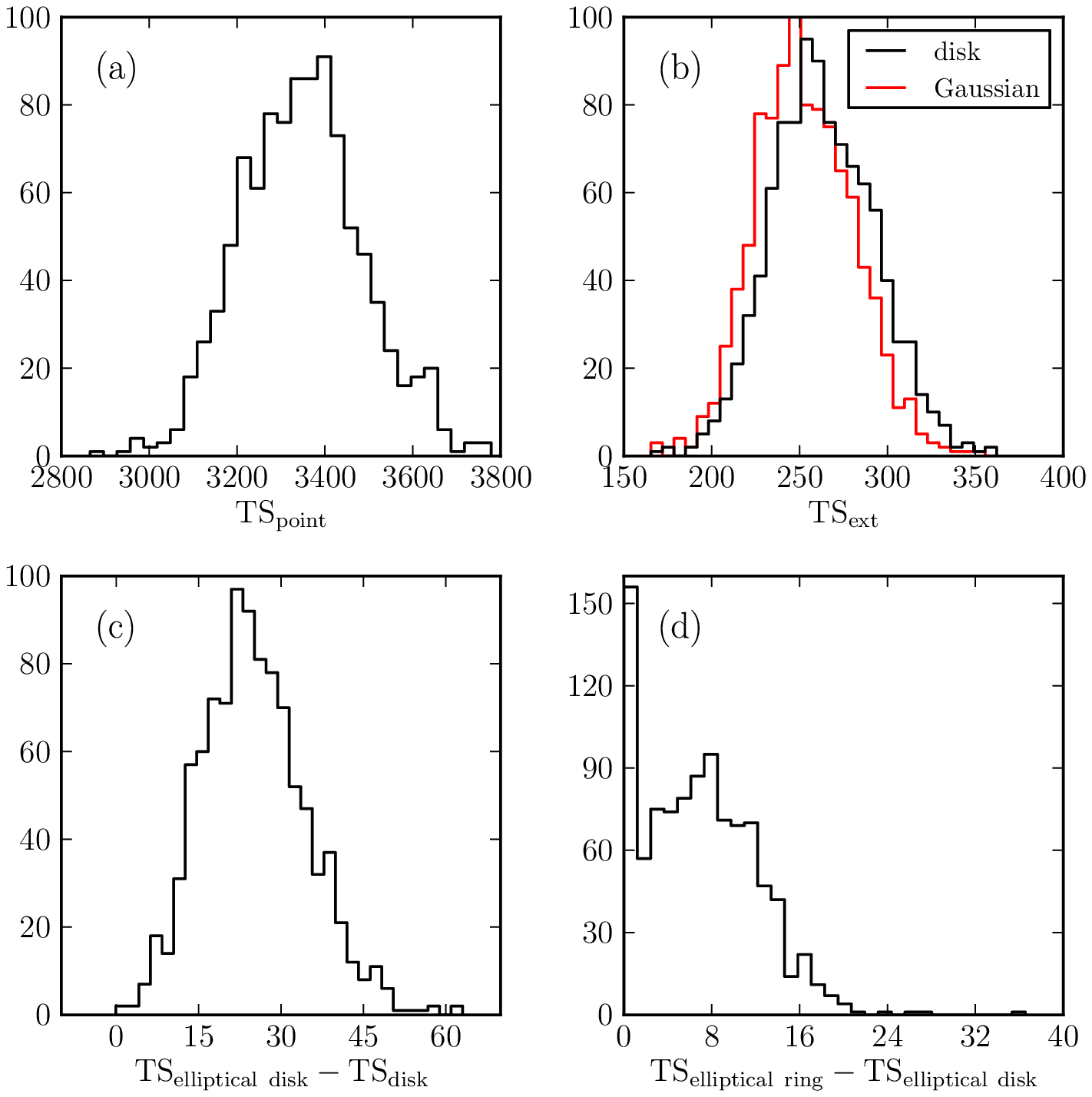}
    \else
    \plotone{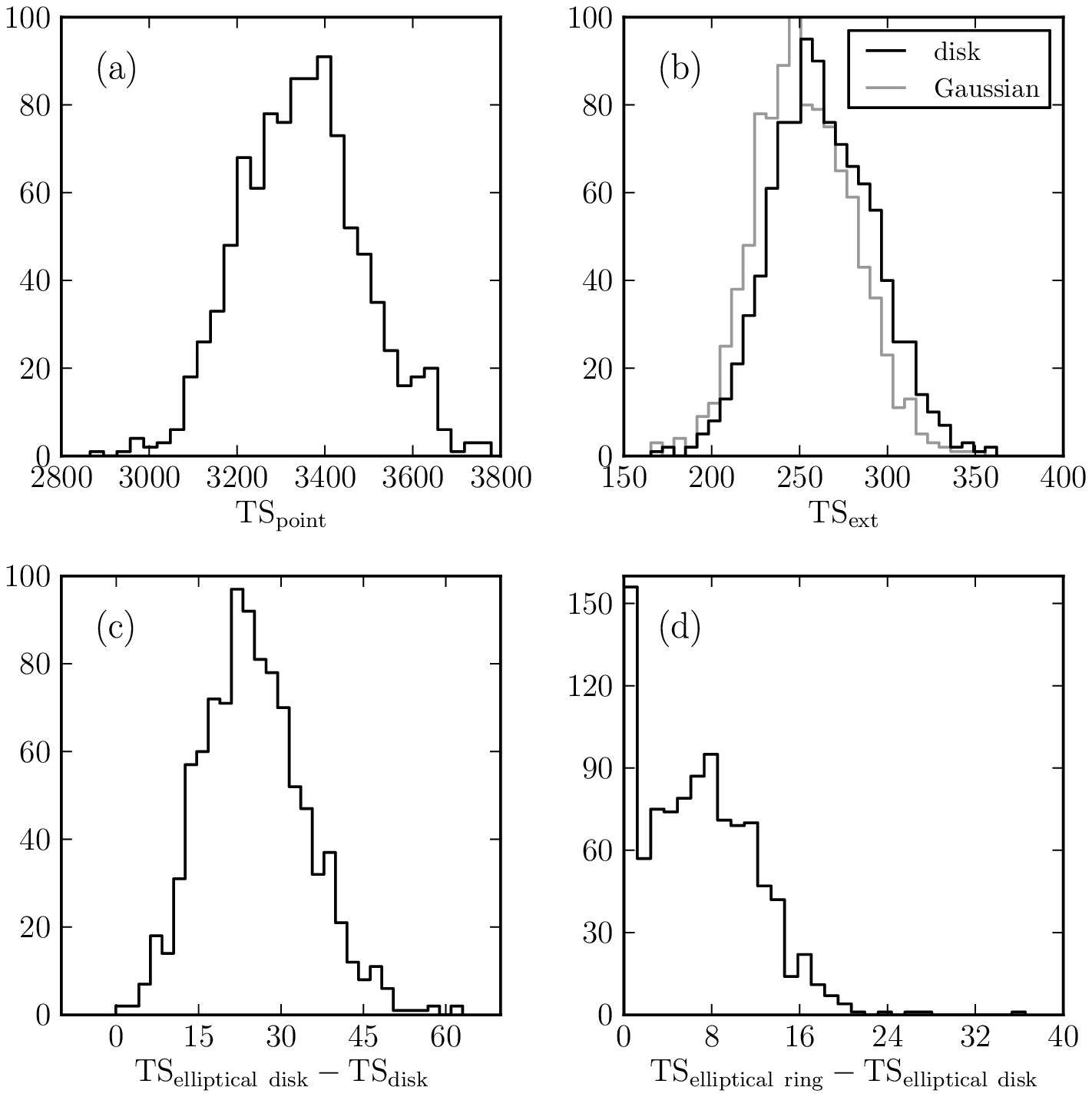}
    \fi
    \caption{
    The distribution of \ts values when fitting 985 statistically
    independent simulations of W44. (a) is the distribution of \ts values
    when fitting W44 as a point-like source and (b) is the
    distribution of \tsext when fitting the source with a uniform disk or a 
    radially-symmetric Gaussian
    spatial model. (c) is the distribution of the change in TS when
    fitting the source with an elliptical disk spatial model compared to
    fitting it with a radially-symmetric disk spatial model and (d) 
    when fitting the source with an elliptical ring spatial model compared
    to an elliptical disk spatial model.
    }\label{ts_comparison_w44sim}
\end{figure}

Figure~\ref{ts_comparison_w44sim}a 
shows that
the significance of W44 in the simulations is very large ($\ts\sim3500$) 
for a model with a point-like source hypothesis.
Figure~\ref{ts_comparison_w44sim}b shows that 
the significance of the spatial extension is also large
($\tsext\sim250$).  
On average \tsext is somewhat larger when fitting
the sources with more accurate spatial models.  This shows that
assuming an incorrect spatial model will cause the source's
significance to be underestimated.  Figure~\ref{ts_comparison_w44sim}c
shows that the sources were fit better when assuming an elliptical
disk spatial model compared to a uniform disk spatial model
($\ts_\text{elliptical\ disk}-\ts_\text{disk}\sim30$).  Finally,
Figure~\ref{ts_comparison_w44sim}d shows that the sources were
fit somewhat better assuming an elliptical ring spatial model
compared to an elliptical disk spatial model ($\ts_\text{elliptical\
ring}-\ts_\text{elliptical\ disk}\sim9$). This shows that the LAT has
some additional power to resolve substructure in bright extended sources.

\begin{figure}
    \ifcolorfigure
    \plotone{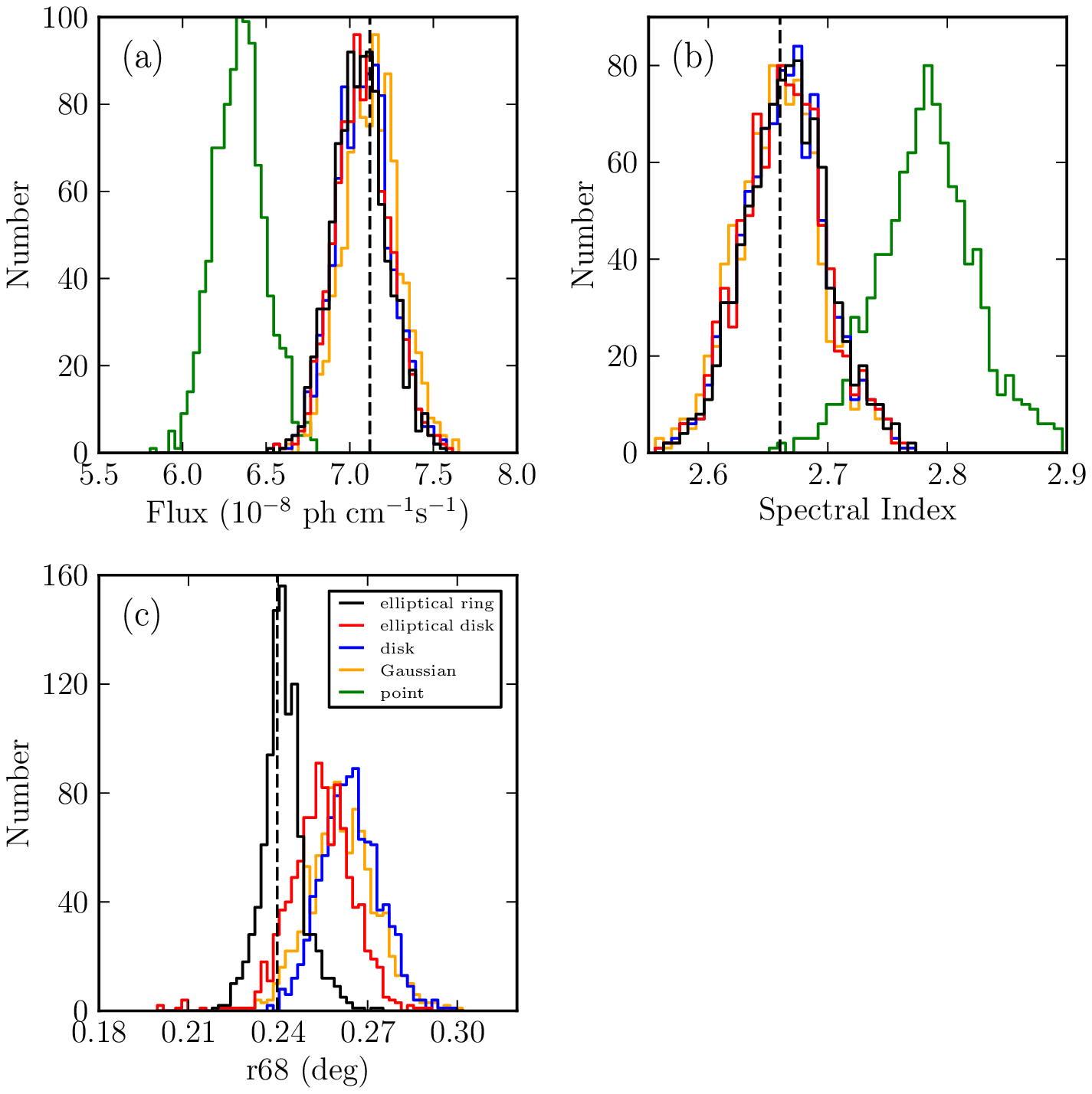}
    \else
    \plotone{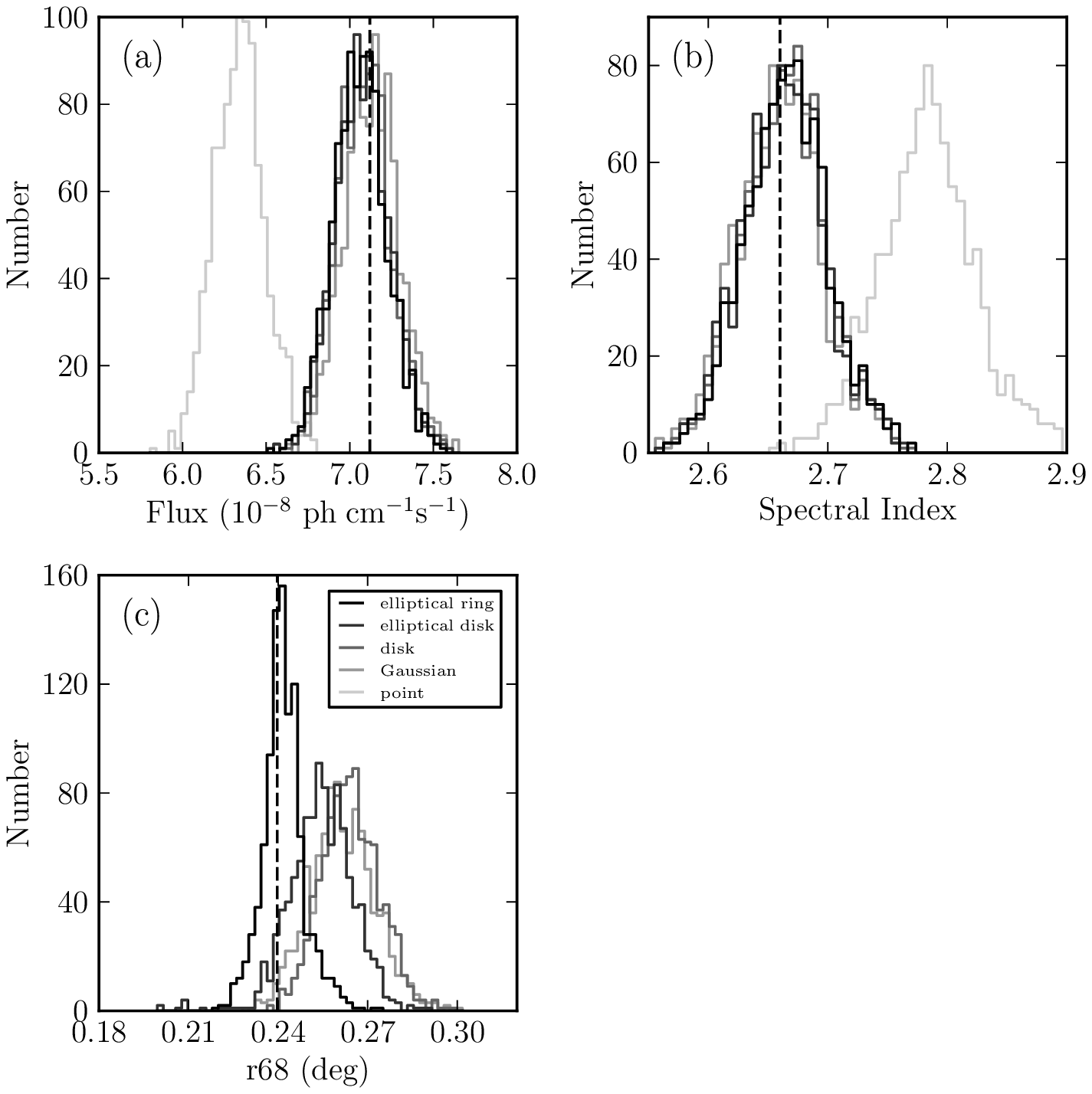}
    \fi
    \caption{
    The distribution of fit parameters
    for the Monte Carlo simulations of W44.
    The plots show the distribution of 
    best fit (a) flux (b) spectral index and (c) 68\% containment
    radius. The dashed vertical lines represent
    the simulated values of the parameters.
    }\label{bias_w44sim}
\end{figure}

Figure~\ref{bias_w44sim}a and Figure~\ref{bias_w44sim}b clearly show that
no significant bias is introduced by modeling the source as extended
but with an inaccurate spatial model, while a point-like source modeling
results in a $\sim10\%$ and $\sim0.125$ bias in the fit flux and index,
respectively.  
Furthermore, Figure~\ref{bias_w44sim}c shows that the \rsixeight estimate of
the extension size is very mildly biased ($\sim10\%$) toward higher values
when inaccurate spatial models are used, and thus represents a reasonable
measurement of the true 68\% containment radius for the source.
For the elliptical spatial models, \rsixeight is computed by numeric integration.

\section{Extended Source Detection Threshold}
\label{extension_sensitivity}

We calculated the LAT flux 
threshold to detect spatial extent. We define the detection threshold as the flux at
which the value of $\tsext$ averaged over many statistical realizations is
$\langle\tsext\rangle=16$ 
(corresponding to a formal $4\sigma$ significance)
for a source of a given extension.

We used a simulation setup similar to that described in
Section~\ref{monte_carlo_validation}, but instead of point-like sources
we simulated extended sources with radially-symmetric uniform disk spatial 
models. Additionally, we simulated our sources over the two-year
time range included in the 2FGL catalog.  For each extension and spectral index,
we selected a flux range which bracketed $\tsext=16$ and performed an
extension test for $>100$ independent realizations of ten fluxes in
the range.  We calculated $\langle\tsext\rangle=16$ by fitting a line
to the flux and $\tsext$ values in the narrow range.

\begin{figure}
    \ifcolorfigure
    \plotone{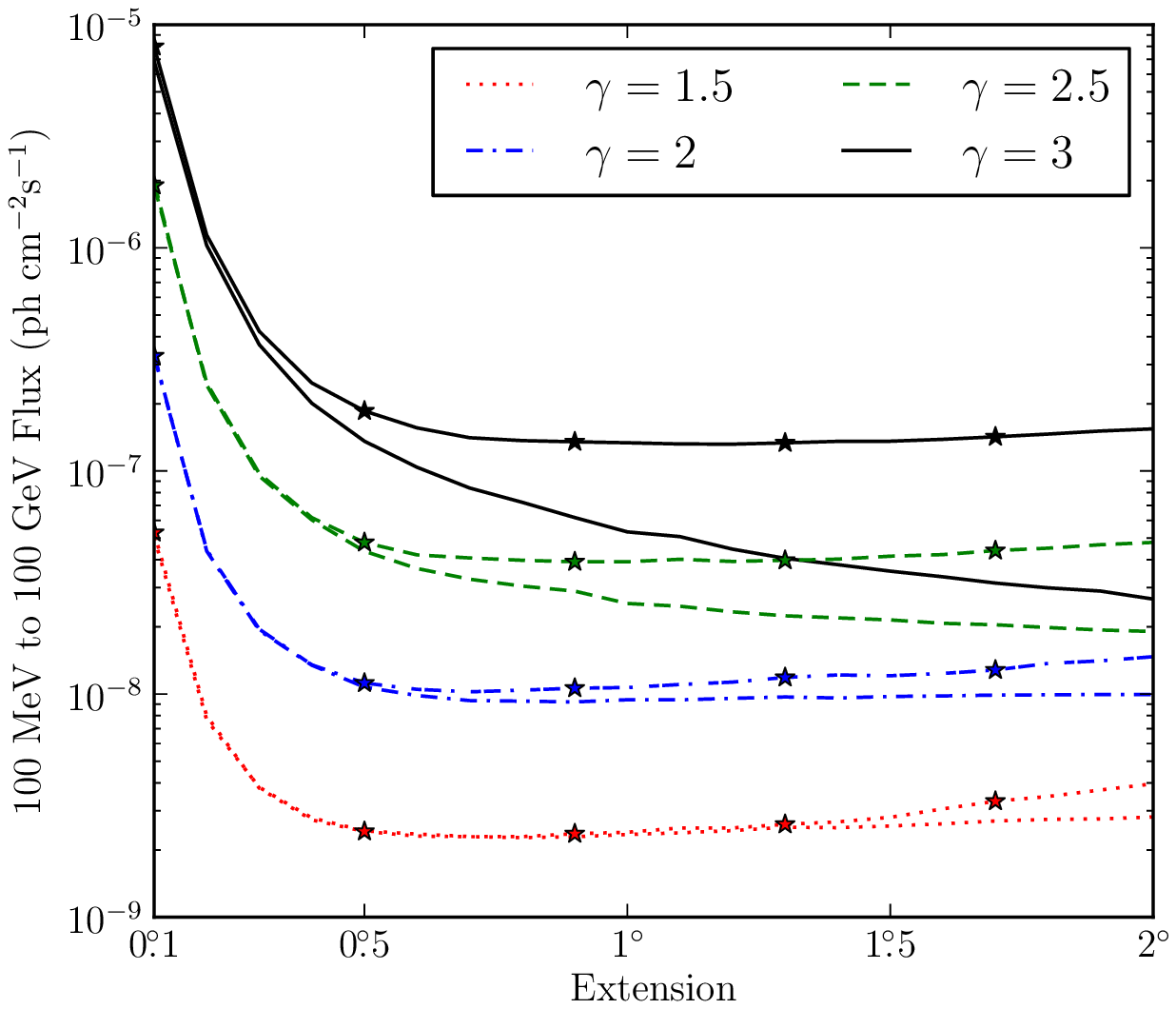}
    \else
    \plotone{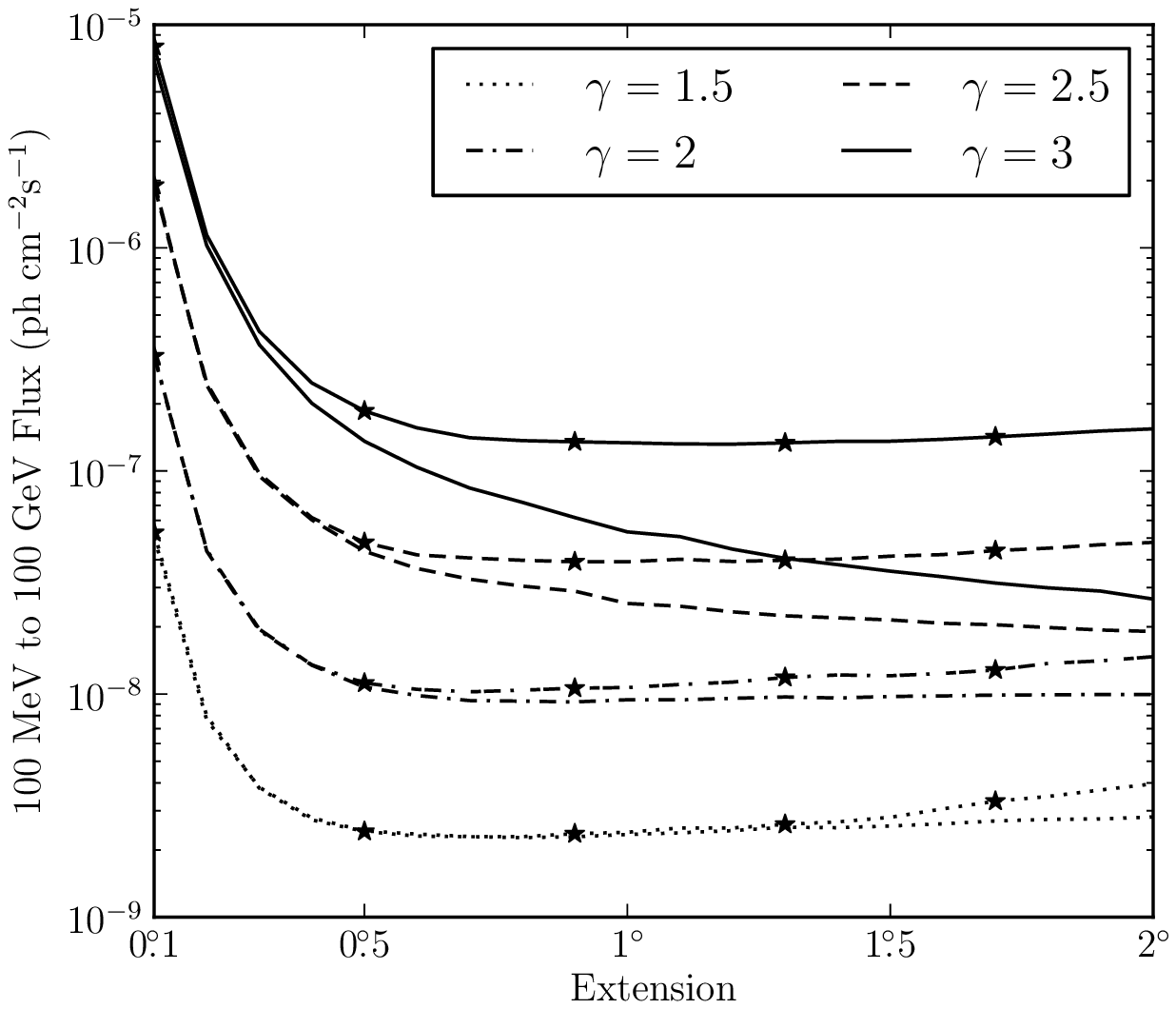}
    \fi
    \caption{
    The detection threshold to resolve an
    extended source with a uniform disk model for a two-year exposure.  All sources have an
    assumed power-law spectrum and the different line styles (colors in
    the electronic version) correspond to different simulated spectral
    indices.  The lines with no markers correspond to the detection
    threshold using photons with energies between 100 \mev and 100 \gev,
    while the lines with star-shaped markers correspond to the threshold
    using photons with energies between 1 \gev and 100 \gev.
    }\label{index_sensitivity}
  \end{figure}

Figure~\ref{index_sensitivity} shows the threshold for sources of four
spectral indices from 1.5 to 3 and extensions varying from $\sigma=0\fdg1$
to $2\fdg0$.  
The threshold is high for small extensions when the
source is 
small compared to the size of the PSF. 
It drops quickly with increasing source size and reaches
a minimum around 0\fdg5. 
The threshold increases
for large extended sources because the source becomes
increasingly diluted by the background.
Figure~\ref{index_sensitivity} shows
the threshold using photons with energies between 100 \mev and 100 \gev
and also using only photons with energies between 1 \gev and 100 \gev.
Except for very large or very soft
sources, the threshold is
not substantially improved by including photons with energies between 100 \mev and
1 \gev.  This is also demonstrated in Figure~\ref{four_plots_ic443}
which shows \tsext for the SNR IC~443 computed independently in twelve
energy bins between 100 \mev and 100 \gev. For IC~443, which has a
spectral index $\sim2.4$ and an extension $\sim0\fdg35$, 
almost the entire 
increase in likelihood from optimizing the source extent in the model
comes
from energies above 1 \gev.  Furthermore, other systematic errors
become increasingly large at low energy. For our extension search
(Section~\ref{extended_source_search_method}),
we therefore used only photons with energies above 1 \gev.

\begin{figure}
    \ifcolorfigure
    \plotone{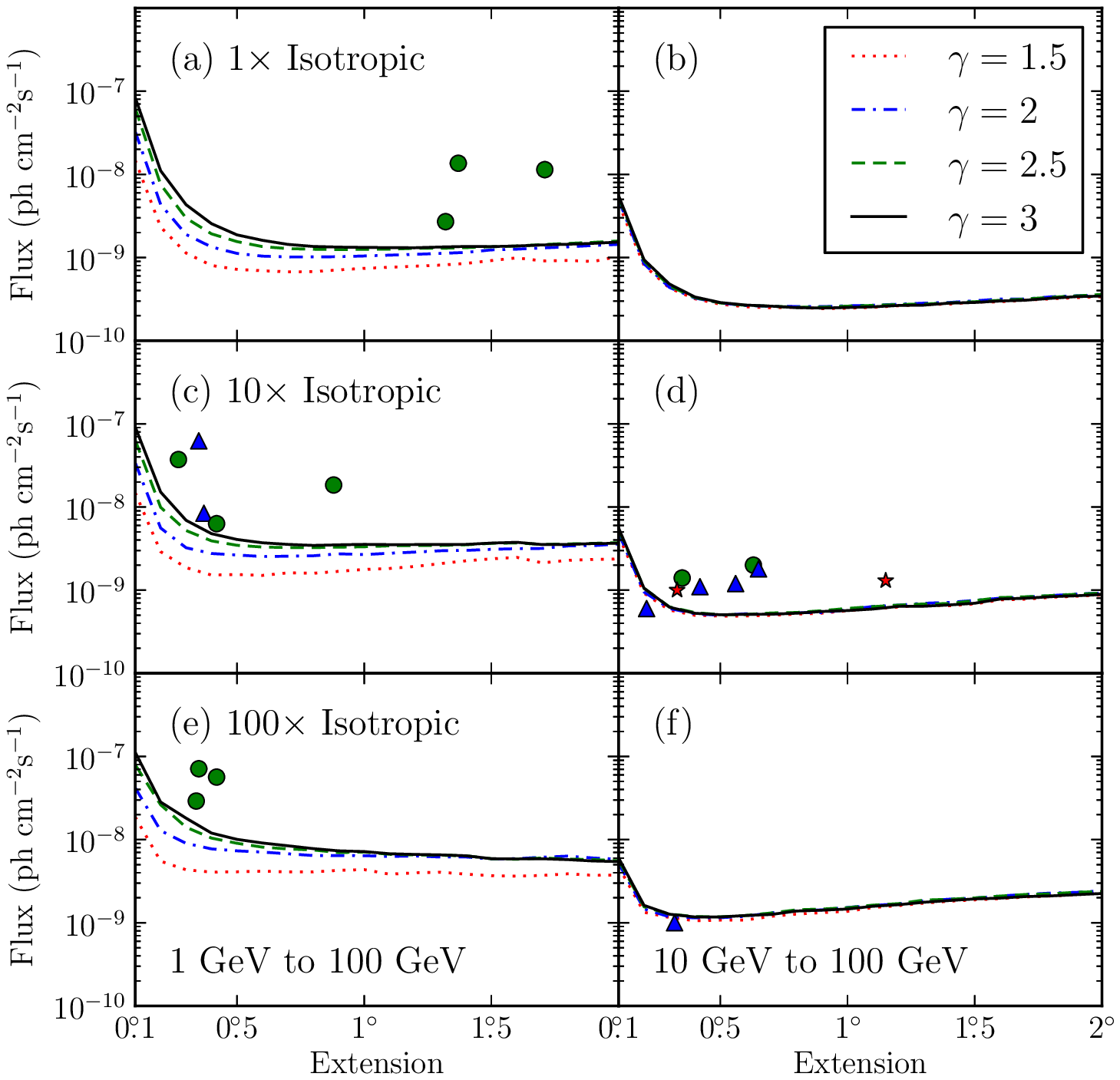}
    \else
    \plotone{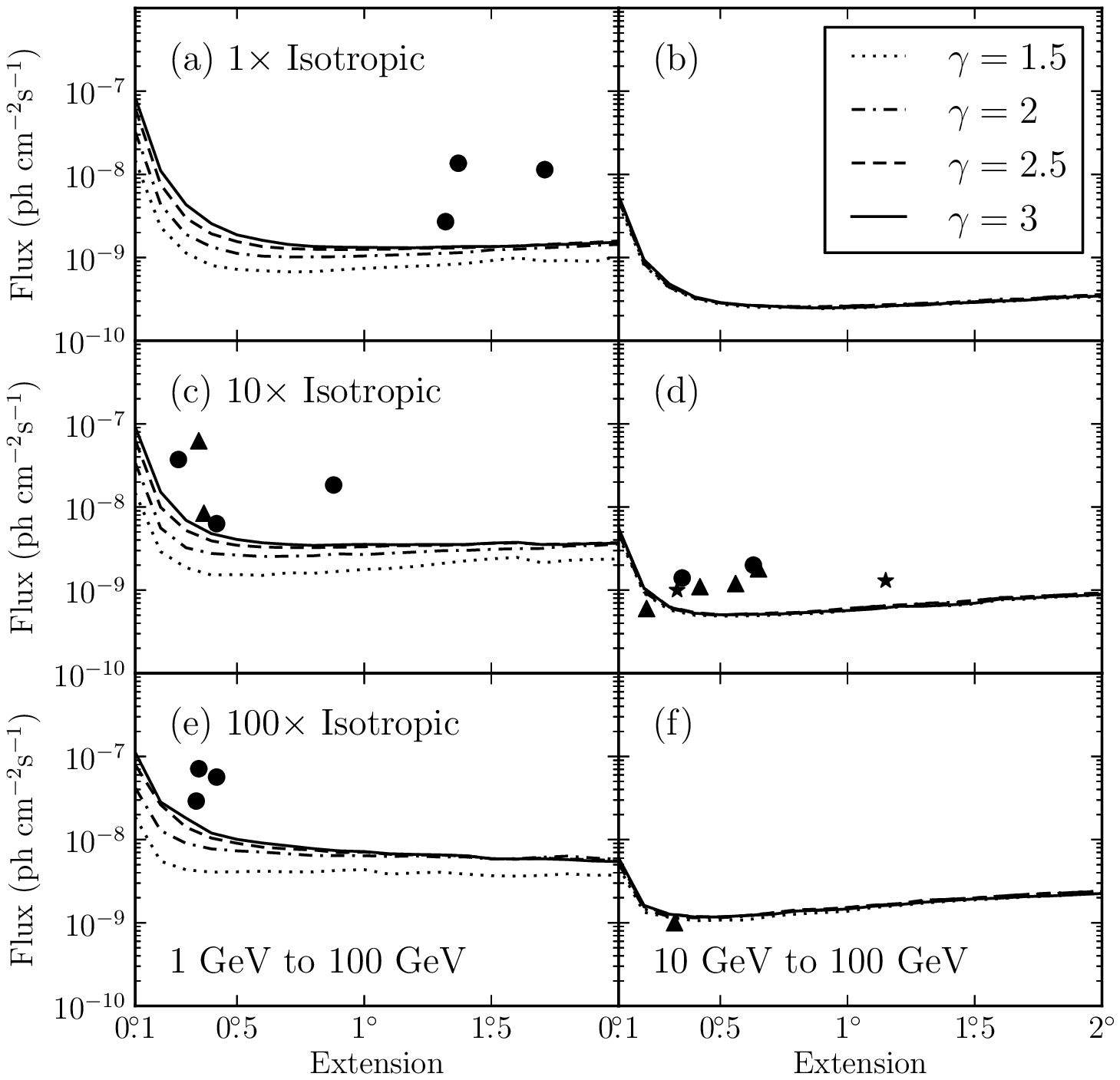}
    \fi
    \caption{The LAT detection threshold for four spectral indices
    and three backgrounds ($1\times$, $10\times$, and $100\times$ the
    Sreekumar-like isotropic background) for a two-year exposure. The
    left-hand plots are the detection threshold when using 
    photons with energies between
    1 \gev and 100 \gev
    and the right-hand plots are the detection threshold when using
    photons with energies between
    10 \gev and 100 \gev.  The flux is integrated only in the
    selected energy range.  Overlaid on this plot are the LAT-detected
    extended sources placed by the magnitude of the nearby Galactic
    diffuse emission and the energy range they were analyzed with.
    The star-shaped markers (colored red in the electronic version)
    are sources with a spectral index closer to 1.5, the triangular
    markers (colored blue) an index closer to 2, and the circular markers
    (colored green) an index closer to 2.5.  The triangular marker in plot
    (d) below the sensitivity line is MSH\,15$-$52.
    }\label{all_sensitivity} 
  \end{figure}

Figure~\ref{all_sensitivity} shows the flux threshold as a function of
source extension for different background levels ($1\times$, $10\times$,
and $100\times$ the nominal background), different spectral indices,
and two different energy ranges (1 \gev to 100 \gev and 10 \gev to
100 \gev).  The detection threshold is higher for sources in regions of
higher background.  When studying sources only at energies above 1 \gev,
the LAT detection threshold (defined as the 1 \gev to 100 \gev flux at
which $\langle\tsext\rangle=16$) depends less strongly on the 
spectral index of the source. 
The index dependence of the detection threshold is
even weaker when considering only photons with energies above 10 \gev
because the PSF changes little from 10 \gev to 100 \gev.
Overlaid on Figure~\ref{all_sensitivity} are the LAT-detected extended
sources that will be discussed in Sections~\ref{validate_known} and
\ref{new_ext_srcs_section}.  The extension thresholds are tabulated in
Table~\ref{all_sensitivity_table}.

Finally, Figure~\ref{time_sensitivity} shows the projected
detection threshold of the LAT to extension with a 10 year
exposure against 10 times
the isotropic background measured by EGRET. This background is
representative of the background near the Galactic plane.  For small
extended sources, the threshold improves by a factor larger
than the square root of the relative exposures because the LAT is signal-limited
at high energies where the present analysis is most sensitive. For large
extended sources, the relevant background is over a larger spatial range
and so the improvement is closer to a factor corresponding
to the square root of the relative exposures that is caused by Poisson fluctuations in the background.

\section{Testing Against Source Confusion}
\label{dual_localization_method}

It is impossible to discriminate using only LAT data between a
spatially extended source and multiple point-like sources separated by
angular distances comparable to or smaller than the size of the LAT PSF. 
To assess the plausibility of source confusion for sources with
$\tsext\ge16$, we developed an algorithm to test if a region contains
two point-like sources.  The algorithm works by simultaneously fitting
in \pointlike the positions and spectra of the two point-like sources.
To help with convergence, it begins by dividing the source into two
spatially coincident point-like sources and then fitting the sum and
difference of the positions of the two sources without any limitations
on the fit parameters.

After simultaneously fitting the two positions and two spectra,
we define \tsinc as twice the increase in the log of the likelihood
fitting the region with two point-like sources compared to fitting the
region with one point-like source:
\begin{equation}
  \tsinc=2\log(\likelihood_\text{2pts}/\likelihood_\text{ps}).
\end{equation} 
For the following analysis of LAT data, \tsinc was computed
by fitting the spectra of the two point-like sources in \gtlike using the best fit positions
of the sources found by \pointlike.

\tsinc cannot be quantitatively compared to \tsext using a simple
likelihood-ratio test to evaluate which model is significantly better
because the models are not nested \citep{statistics_with_care}.
Even though the comparison of \tsext with \tsinc is not a calibrated
test, $\tsext>\tsinc$ indicates that the likelihood for the extended
source hypothesis is higher than for two point-like sources and we only
consider a source to be extended if $\tsext>\tsinc$.

We considered using 
the Bayesian information criterion \citep[BIC,][]{BIC_statistical_test} as
an alternative Bayesian formulation for this test, but it is difficult to apply
to LAT data because it contains a term including the number of data points. 
For studying $\gamma$-ray sources in LAT data, we analyze relatively large
regions of the sky to better define the contributions from diffuse
backgrounds and nearby point sources. This is important for accurately
evaluating source locations and fluxes but the fraction of data directly
relevant to the evaluation of the parameters for the source of interest
is relatively small.

As an alternative, we considered the Akaike information criterion test \citep[\aic,][]{AIC_statistical_test}.
The \aic is defined as $\aic=2k-2\log\likelihood$, where $k$ is the number of parameters in the model. 
In this formulation, the best hypothesis is considered to be the one that minimizes the \aic.
The first term penalizes models with additional parameters. 

The two point-like sources hypothesis has three more parameters than
the single extended source hypothesis (two more spatial parameters and
two more spectral parameters compared to one extension parameter), so the
comparison $\aic_\text{ext} < \aic_\text{2pts}$  is formally equivalent to
$\tsext + 6 > \tsinc$.  Our criterion for accepting extension ($\tsext > \tsinc$) 
is thus equivalent to requesting that the AIC-based empirical
support for the two point-like sources model be ``considerably less''
than for the extended source model, following the classification by
\cite{aic_stats_book}.

We assessed the power of the $\tsext>\tsinc$ test with a Monte Carlo
study.  We simulated one spatially extended source and fit it as both
an extended source and as two point-like sources using \pointlike.
We then simulated two point-like sources and fit them with the same two
hypotheses. By comparing the distribution of \tsinc and \tsext computed by
\pointlike for the two cases, we evaluated how effective the $\tsext>\tsinc$
test is at rejecting cases of source confusion as well as how
likely it is to incorrectly reject that an extended source is spatially
extended.  All sources were simulated using the same time range as in
Section~\ref{extension_sensitivity} against a background 10 times the
isotropic background measured by EGRET, representative of the background
near the Galactic plane.

\begin{figure}
    \ifcolorfigure
    \plotone{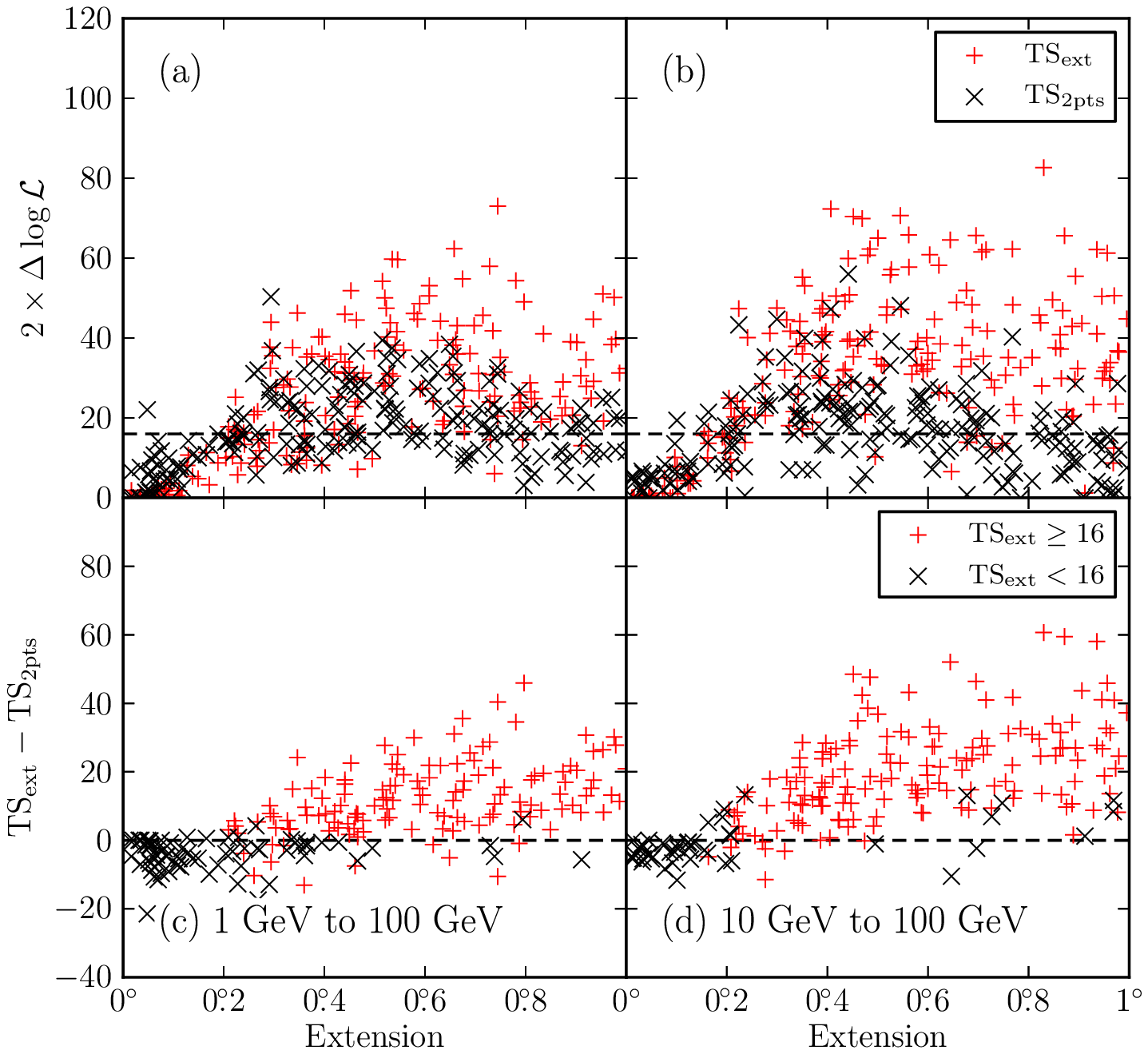}
    \else
    \plotone{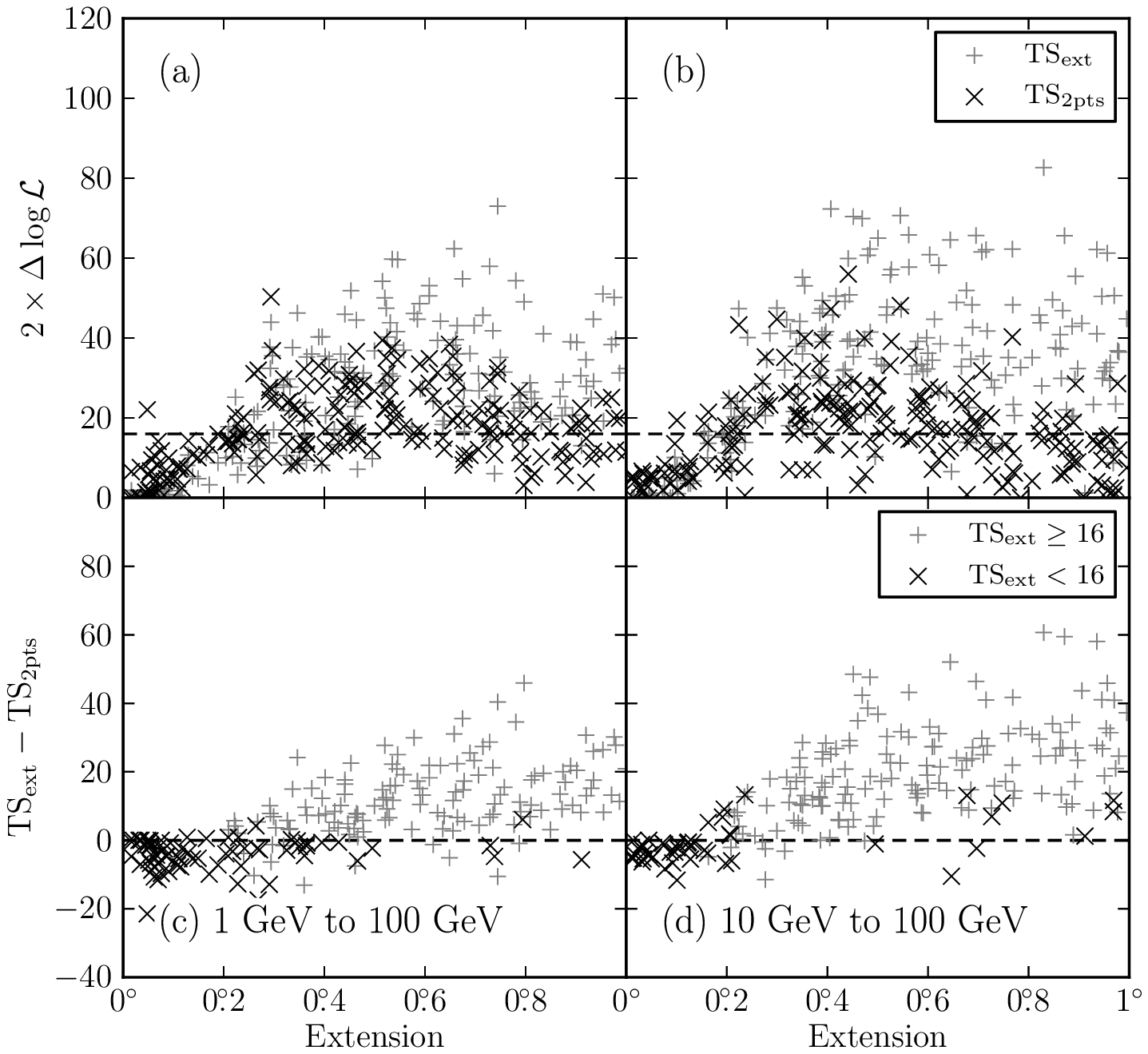}
    \fi
    \caption{
    (a) and (b) are the distribution of \tsext and of \tsinc when
    fitting simulated spatially extended sources of varying sizes as
    both an extended source and as two point-like sources.  (c) and
    (d) are the distribution of $\tsext-\tsinc$ for the same simulated sources.
    (a) and (c) represent sources fit in the 1 \gev to 100 \gev energy
    range and (b) and (d) represent sources fit in the 10 \gev to
    100 \gev energy range.  In (c) and (d), the plus-shaped markers
    (colored red in the electronic version) are fits where $\tsext\ge16$.
    }\label{confusion_extended_plot}
  \end{figure}

We did this study first in the energy range from 1 \gev to 100 \gev by
simulating extended sources of flux $4\times10^{-9}$ \phflux integrated
from 1 \gev to 100 \gev and a power-law spectral model with
spectral index 2.  This spectrum was picked to be representative of the
new extended sources that were discovered in the following analysis
when looking in the 1 \gev to 100 \gev energy range
(see Section~\ref{new_ext_srcs_section}).
We simulated these sources using uniform disk spatial models
with extensions varying
up to $1\degree$.  
Figure~\ref{confusion_extended_plot}a shows the distribution
of \tsext and \tsinc and 
Figure~\ref{confusion_extended_plot}c shows 
the distribution of $\tsext-\tsinc$ as a
function of the simulated extension of the source
for 200 statistically independent simulations.

\begin{figure}
    \ifcolorfigure
    \plotone{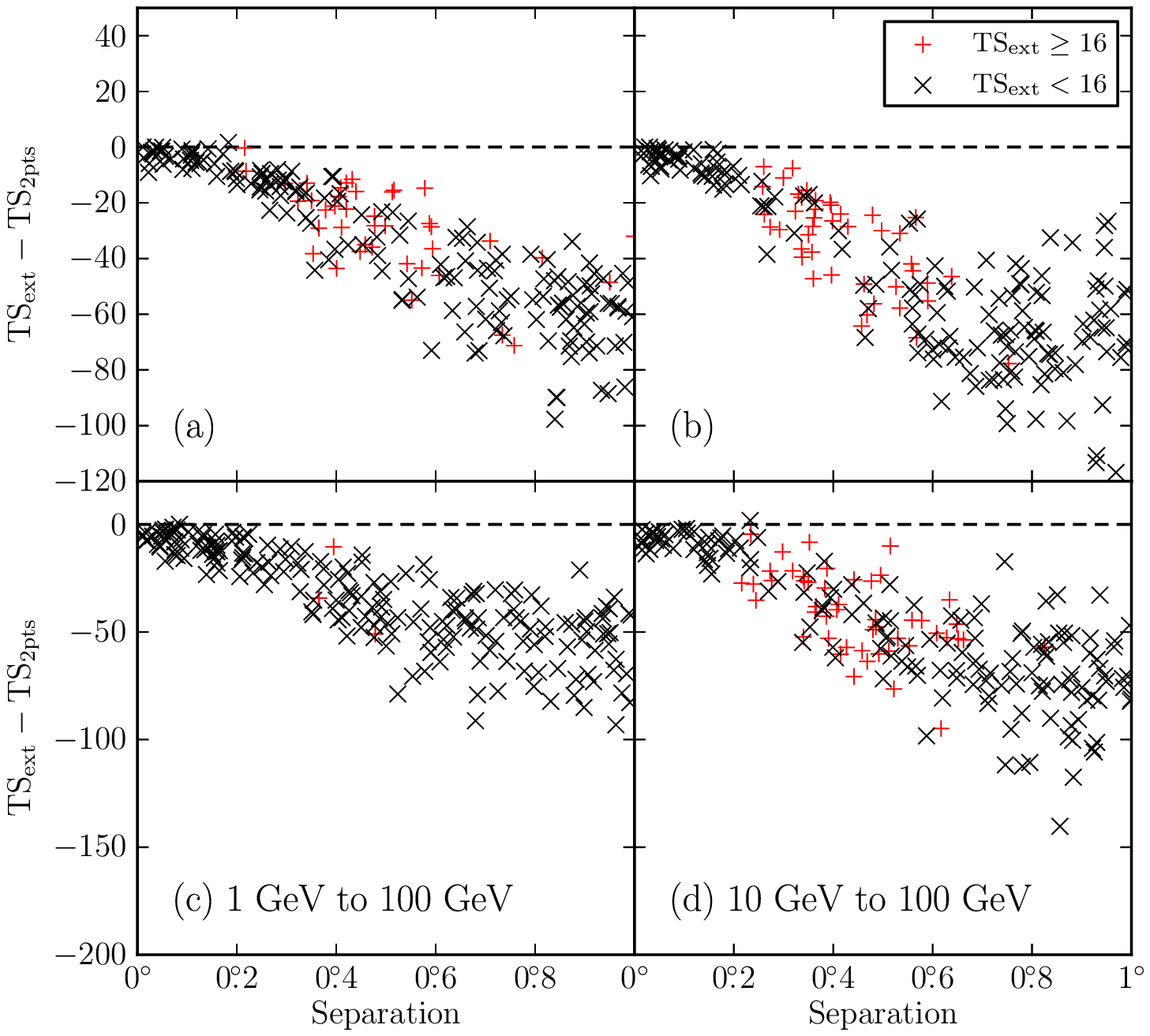}
    \else
    \plotone{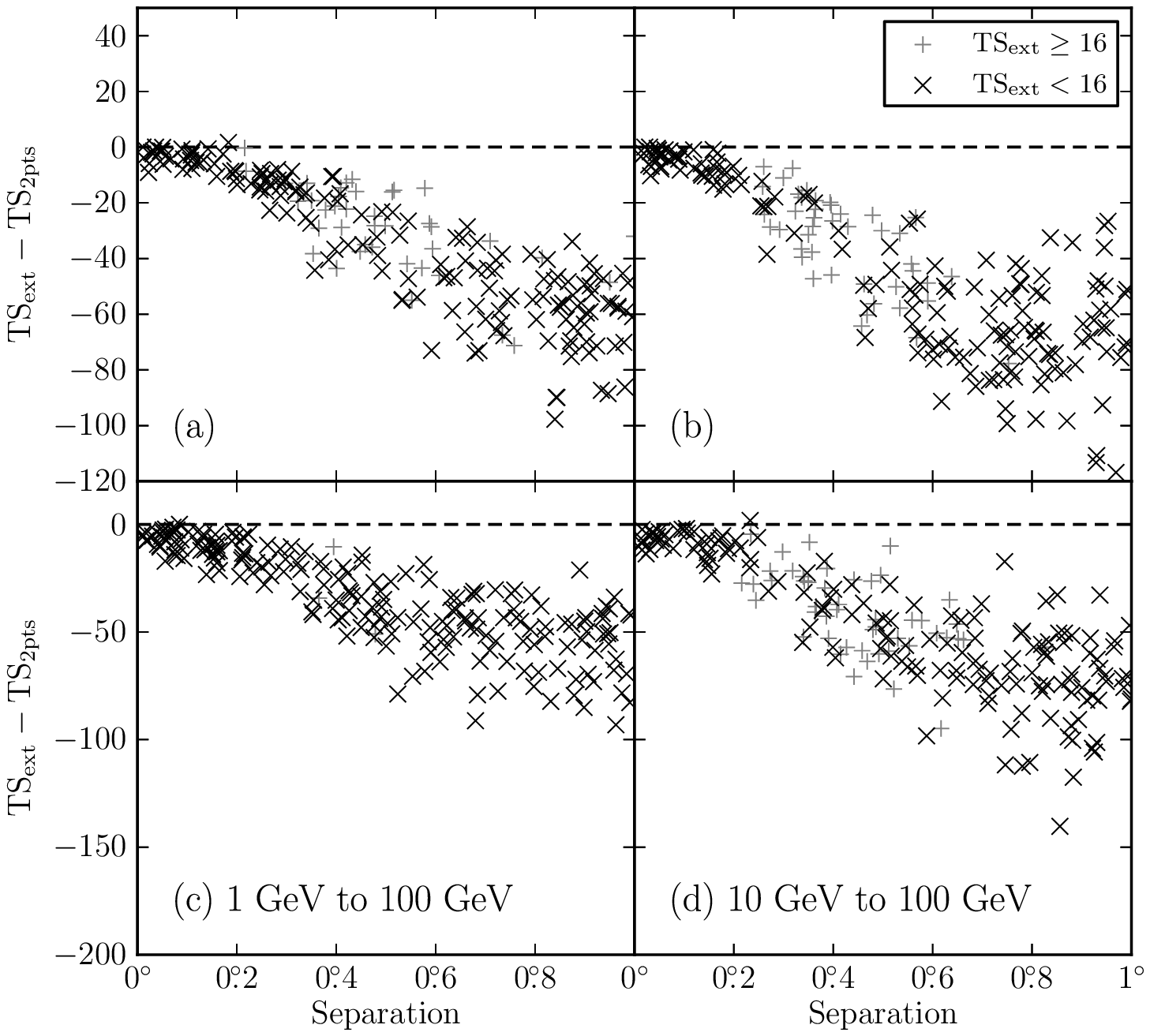}
    \fi
    \caption{
    The distribution of $\tsext-\tsinc$ when fitting two simulated
    point-like sources of varying separations as
    both an extended source and as two point-like sources.  (a),
    and (b) represent simulations of two point-like sources with the
    same spectral index and (c) and (d) represent simulations of two
    point-like sources with different spectral indices.  (a) and (c)
    fit the simulated sources in the 1 \gev to 100 \gev energy range
    and (b) and (d) fit in the 10 \gev to 100 \gev energy range.
    The plus-shaped markers (colored red in the electronic version)
    are fits where $\tsext\ge16$.
    }\label{confusion_2pts_plot}
  \end{figure}

Figure~\ref{confusion_2pts_plot}a shows the same plot but when fitting
two simulated point-like sources each with half of the flux of
the spatially extended source and with the same spectral index as the
extended source.  Finally, Figure~\ref{confusion_2pts_plot}c shows the same
plot with each point-like source having the same flux but different
spectral indices.  One point-like source had a spectral index of 1.5
and the other an index of 2.5.
These indices are representative of the range of indices of LAT-detected
sources.

The same four plots are shown in Figure~\ref{confusion_extended_plot}b, \ref{confusion_extended_plot}d, 
\ref{confusion_2pts_plot}b, and \ref{confusion_2pts_plot}d but this
time when analyzing a source of flux $10^{-9}$ \phflux (integrated from
10 \gev to 100 \gev) only in the 10 \gev to 100 \gev energy range.
This flux is typical of the new extended sources discovered
using only photons with energies between 10 \gev and 100 \gev (see
Section~\ref{new_ext_srcs_section}).

Several interesting conclusions can be made from this study.  As one would
expect, $\tsext-\tsinc$ is mostly positive when fitting the simulated
extended sources.  In the 1 \gev to 100 \gev analysis, only 11 of the
200 simulated extended sources had $\tsext>16$ but were incorrectly
rejected due to \tsinc being greater than \tsext.  In the 10 \gev to 100 \gev
analysis, only 7 of the 200 sources were incorrectly rejected. From this,
we conclude that this test is unlikely to incorrectly reject truly
spatially extended sources.

On the other hand, 
it is often
the case that $\tsext>16$ when testing the 
two simulated point-like sources
for extension.  This is especially the case
when the two sources had the same spectral index. Forty out of 200 sources
in the 1 \gev to 100 \gev energy range and 43 out of 200 sources in the
10 \gev to 100 \gev energy range had $\tsext>16$.  But in these cases,
we always found the single extended source fit to be worse than
the two point-like source fit.  From this, we conclude that the $\tsext>\tsinc$
test is powerful at discarding cases in which the true emission comes
from two point-like sources.

The other interesting feature in Figure~\ref{confusion_extended_plot}a
and \ref{confusion_extended_plot}b is that for simulated extended
sources with typical sizes ($\sigma\sim0\fdg5$), one can often obtain
almost as large an increase in likelihood fitting the source as two
point-like sources ($\tsinc\sim\tsext$).  This is because although the
two point-like sources represent an incorrect spatial model, the second
source has four additional degrees of freedom (two spatial and two
spectral parameters) and can therefore easily model much of the extended
source and statistical fluctuations in the data.  This effect is most
pronounced when using photons with energies between 1 \gev and 100 \gev
where the PSF is broader.

From this Monte Carlo study, we can see the limits of an analysis with
LAT data of spatially extended sources.  Section~\ref{monte_carlo_validation}
showed that we have a statistical test that finds when a LAT source is
not well described by the PSF.  But this test does not uniquely prove
that the emission originates from spatially extended emission instead
of from multiple unresolved sources.  Demanding that $\tsext>\tsinc$
is a powerful second test to avoid cases of simple confusion of two
point-like sources. But it could always be the case that an extended
source is actually the superposition of multiple point-like or
extended sources that could be resolved with deeper observations of the
region.  There is nothing about this conclusion unique to analyzing LAT data,
but the broad PSF of the LAT and the density of sources expected to be
\gev emitters in the Galactic plane makes this issue more significant
for analyses of LAT data.  When possible, multiwavelength information should be
used to help select the best model of the sky.

\section{Test of 2LAC Sources}
\label{test_2lac_sources}

For all following analyses of LAT data, we used the same two-year dataset
that was used in the 2FGL catalog spanning from 2008 August 4 to 2010 August 1. We
applied the same acceptance cuts and we used the same P7\_V6 Source class
event selection and IRFs \citep{lat_on_orbit_psf}.  
When analyzing sources in \pointlike, we used a circular $10\degree$ region of
interest (ROI) centered on our source and eight energy bins per
logarithmic decade in energy.
When refitting the region in \gtlike using the best fit spatial and
spectral models from \pointlike, we used the `binned likelihood' mode of
\gtlike on a $14\degree\times14\degree$ ROI with a pixel size of 0\fdg03.

Unless explicitly
mentioned, we used the same background model as 2FGL to represent the
Galactic diffuse, isotropic, and Earth limb emission.  To compensate for
possible residuals in the diffuse emission model, the Galactic emission
was scaled by a power-law and the normalization
of the isotropic component
was left free.  
Unless explicitly mentioned,
we used all 2FGL sources within $15\degree$ of our source as our list
of background sources and we refit the spectral parameters of all sources
within $2\degree$ of the source.

To validate our method, we tested LAT sources associated with AGN for
extension.  \gev emission from AGN is believed to originate from collimated jets.  Therefore AGN are
not expected to be spatially resolvable by the LAT and provide a good
calibration source to demonstrate that our extension detection method
does not misidentify point-like sources as being extended.  We note that
megaparsec-scale $\gamma$-ray halos around AGNs have been hypothesized
to be resolvable by the LAT \citep{pair_halo_paper}. However, no such
halo has been discovered in the LAT data so far \citep{neronov_agn_halo}.

Following 2FGL, the LAT Collaboration published the Second LAT AGN
Catalog (2LAC), a list of high latitude ($|b|>10\degree$) sources
that had a high probability association with AGN \citep{second_agn_cat}.
2LAC associated 1016 2FGL sources with AGN.  To avoid systematic problems
with AGN classification, we selected only the 885 AGN which made it into
the clean AGN sub-sample defined in the 2LAC paper.  An AGN association is considered clean only
if it has a high probability of association $P\ge 80\%$, if it is the
only AGN associated with the 2FGL source, and if 
no analysis flags have been set for the source
in the 2FGL catalog. These last two conditions are important for our
analysis. Source confusion may look like a spatially extended source
and flagged 2FGL sources may correlate with unmodeled structure in
the diffuse emission.

\begin{figure}
    \ifcolorfigure
      \plotone{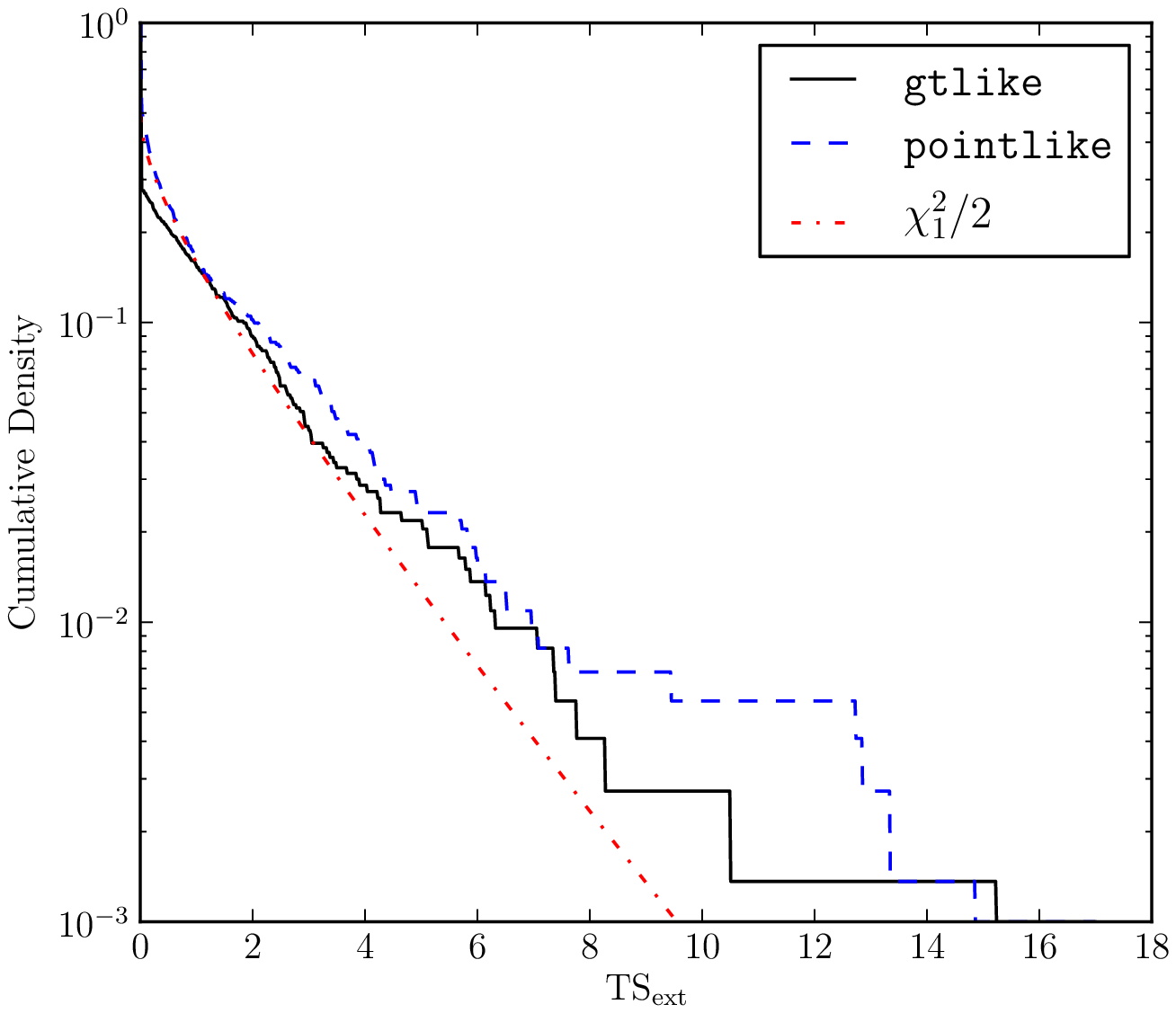}
    \else
      \plotone{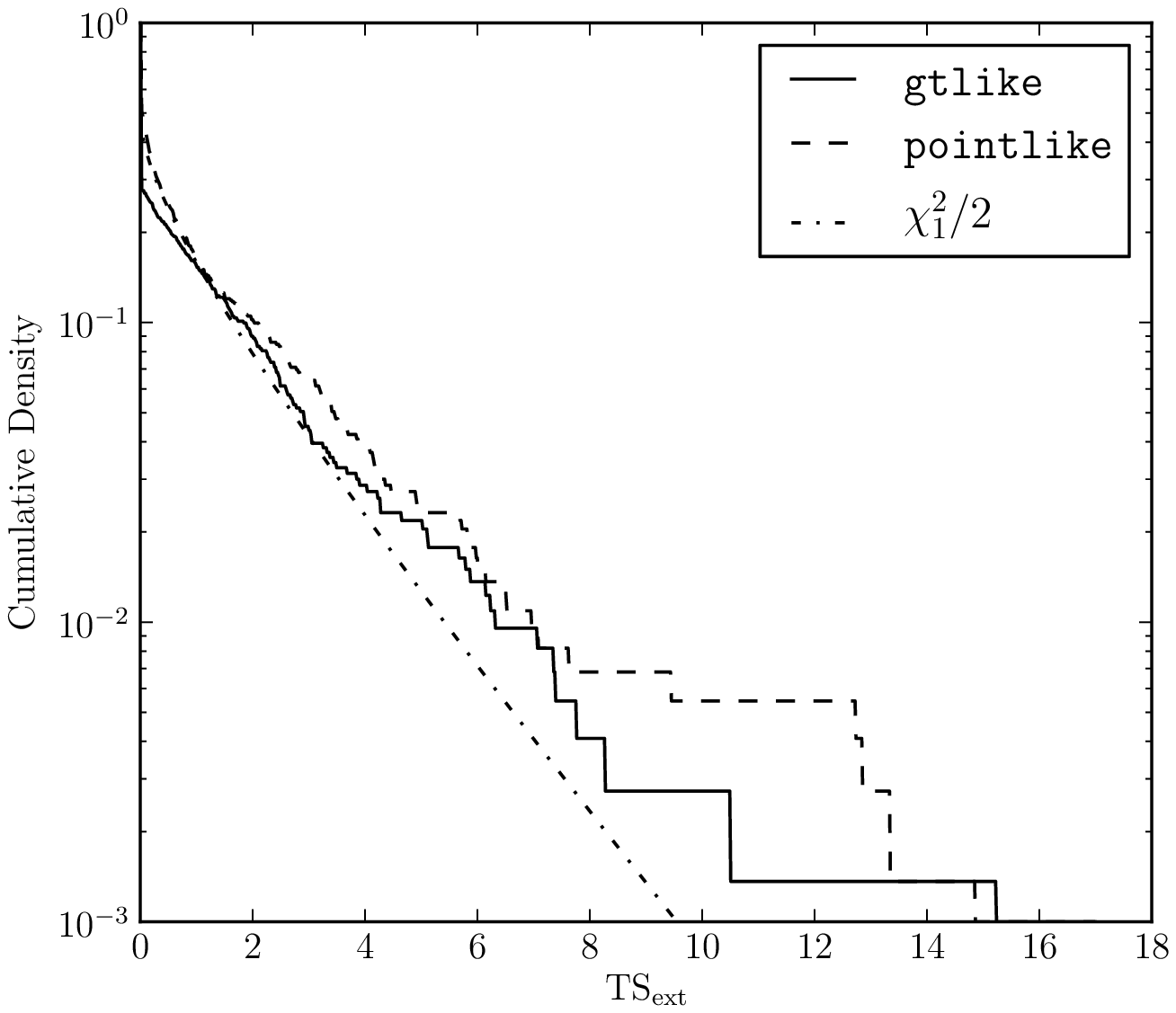}
    \fi
    \caption{The cumulative density of \tsext for the 733 clean
    AGN in 2LAC that were significant above 1 \gev calculated with \pointlike (dashed line
    colored blue in the electronic version)
    and with \gtlike (solid line colored black).  AGN are too far
    and too small to be resolved by the LAT. Therefore, the cumulative
    density of $\tsext$ is expected to follow a $\chi^2_1/2$ distribution
    (Equation~\ref{ts_ext_distribution}, the dash-dotted line colored
    red).
    }\label{agn_ts_ext}
  \end{figure}

Of the 885 clean AGN, we selected the 733 of these 2FGL sources which
were significantly detected above 1 \gev and fit each of them for
extension.  The cumulative density of \tsext for these AGN is compared
to the $\chi^2_1/2$ distribution of Equation~\ref{ts_ext_distribution}
in Figure~\ref{agn_ts_ext}.  The \tsext distribution
for the AGN shows reasonable agreement with the theoretical
distribution and no AGN was found to be significantly extended
($\tsext>16$).  The observed discrepancy from the theoretical
distribution is likely due to small systematics
in our model of the LAT PSF and the Galactic diffuse emission (see
Section~\ref{systematic_errors_on_extension}).  
The discrepancy could
also in a few cases be due to confusion with a nearby undetected source.
We note that the Monte Carlo study
of section~\ref{monte_carlo_validation} effectively used perfect
IRFs and a perfect model of the sky.
The overall agreement with the expected distribution demonstrates that
we can use \tsext as a measure of the statistical significance of the
detection of the extension of a source.

We note that the LAT PSF used in this study was determined
empirically by fitting the distributions of gamma rays around bright AGN (see
Section~\ref{systematic_errors_on_extension}). Finding that the AGN we
test are not extended is not surprising.  This validation analysis is
not suitable to reject any hypotheses about the existence of megaparsec-scale
halos around AGN.

\section{Analysis of Extended Sources Identified in the 2FGL Catalog}
\label{validate_known}

As further validation of our method for studying
spatially extended sources, we reanalyzed the twelve spatially extended
sources which were included in the 2FGL catalog \citep{second_cat}.  
Even though these sources had all been the subjects of dedicated
analyses and separate publications, and had been fit with a variety
of spatial models,
it is valuable to show that
these sources are significantly extended using our systematic 
method assuming radially-symmetric uniform disk spatial models.  On the other hand, for some of
these sources a uniform disk spatial model does not well describe the
observed extended emission and so the dedicated 
publications by the LAT collaboration provide better models of these sources.


Six extended SNRs were
included in the 2FGL catalog: W51C, IC~443, W28, W44, the Cygnus Loop,
and W30
\citep{w51c,ic443,w28,w44,cygnus_loop_lat,w30_lat}.
Using photons
with energies between
1 \gev and 100 \gev, our analysis significantly detected
that these six SNRs are spatially extended.


Two nearby satellite galaxies of the Milky Way the Large Magellanic Cloud (LMC)
and the Small Magellanic
Cloud (SMC) were included in the 2FGL catalog as spatially extended sources \citep{lmc,smc}.  
Their extensions were significantly
detected using photons with energies between
1 \gev and 100 \gev. Our
fit extensions are comparable to the published result, but we note that
the previous LAT Collaboration publication on the LMC used a more complicated two 2D Gaussian surface
brightness profile when fitting it \citep{lmc}.

Three PWNe, MSH\,15$-$52, Vela X, and HESS\,J1825$-$137, were fit as
extended sources in the 2FGL analysis \citep{msh1552,velax,fermi_hess_j1825}.  
In the present analysis, HESS\,J1825$-$137
was significantly detected using photons with energies between 10
\gev and 100 \gev.  To avoid confusion with the nearby bright pulsar
PSR\,J1509$-$5850, MSH\,15$-$52 had to be analyzed at high energies.
Using photons with energies above 10 \gev, we fit the extension of
MSH\,15$-$52 to be consistent with the published size but with \tsext=6.6.

Our analysis was unable to resolve Vela X which would have required first
removing the pulsed photons from the Vela pulsar which was beyond the
scope of this paper.  Our analysis also failed to detect a significant
extension for the Centaurus A Lobes because
the shape of the source is significantly different from a uniform
radially-symmetric disk\citep{cen_a_lat}.

Our analysis of these sources is summarized in
Table~\ref{known_extended_sources}.  This table includes the best fit
positions and extensions of these sources when fitting them 
with a radially-symmetric uniform disk model. It also
includes the best fit spectral parameters for each source.  The positions
and extensions of Vela X and the Centaurus A Lobes were taken from
\cite{velax,cen_a_lat} and are included in this table for completeness.

\section{Systematic Errors on Extension}
\label{systematic_errors_on_extension}


We developed two criteria for estimating systematic errors
on the extensions of the sources.
First, we estimated a systematic error due to
uncertainty in our knowledge of the LAT PSF.  
Before launch, the LAT PSF
was determined by detector simulations which were verified in accelerator
beam tests \citep{atwood_LAT_mission}. However, in-flight data revealed
a discrepancy above 3 \gev in the PSF compared to the angular
distribution of photons from bright AGN \citep{lat_on_orbit_psf}.
Subsequently, the PSF was fit empirically to bright AGN and this
empirical parameterization is used in the P7\_V6 IRFs.  To account for
the uncertainty in our knowledge of the PSF, we refit our extended source
candidates using the pre-flight Monte Carlo representation of the PSF
and consider the difference in extension found using the two PSFs as a
systematic error on the extension of a source.  The same approach was used
in \cite{ic443}.  We believe that our parameterization
of the PSF from bright AGN is substantially better than the Monte Carlo
representation of the PSF so this systematic error is conservative.


We estimated a second systematic error on the extension of a source
due to uncertainty in our model of the Galactic diffuse emission by
using an alternative 
approach to modeling the diffuse emission
which takes as input templates
calculated by
GALPROP\footnote{GALPROP is a software package for calculating the
Galactic $\gamma$-ray emission based on a model of cosmic-ray propagation
in the Galaxy and maps of the distributions of the 
components of the interstellar medium \citep{galprop1998,galprop2011}. 
See also \url{http://galprop.stanford.edu/} for details.} 
but then fits each template locally in
the surrounding region.
The particular GALPROP model that was used as input is described in
the analysis of the isotropic diffuse
emission with LAT data \citep{isotropic_lat}.  
The intensities of various components
of the Galactic diffuse emission were fitted individually using a
spatial distribution predicted by
the model.  
We considered separate contributions from
cosmic-ray interactions with the
molecular hydrogen, the atomic and ionized hydrogen, residual gas traced
by dust \citep{isabelle_dark_gass}, and the interstellar radiation
field. We further split the contributions from interactions with molecular
and atomic hydrogen to the Galactic diffuse emission according to the
distance from the Galactic center in which they are produced. Hence, we
replaced the standard diffuse emission model by 18 individually fitted
templates to describe individual components of the diffuse emission.
A similar crosscheck was used in an analysis of RX\,J1713.7$-$3946 
by the LAT Collaboration \citep{rx_j1713_lat}.

It is not expected that this diffuse model is superior to the standard
LAT model obtained through an all-sky fit.  However, adding degrees of
freedom to the background model can remove likely spurious sources that
correlate with features in the Galactic diffuse emission.  Therefore,
this tests systematics that may be due to imperfect modeling of the
diffuse emission in the region. 
Nevertheless, this alternative approach to modeling the diffuse emission
does not test all systematics related to the diffuse emission model. In
particular, because the alternative approach uses the same underlying gas
maps, it is unable to be used to assess systematics due to insufficient
resolution of the underlying maps. Structure in the diffuse emission that
is not correlated with these maps will also not be assessed by this test.

We do not expect the systematic error due to uncertainties in the PSF
to be correlated with the systematic error due to uncertainty in the
Galactic diffuse emission. Therefore, the total systematic error on the
extension of a source was obtained by adding the two errors in quadrature.

There is another systematic error on the size of a source due to issues
modeling nearby sources in crowded regions of the sky. It is beyond the
scope of this paper to address this systematic error. Therefore, 
for sources in crowded regions the systematic
errors quoted in this paper 
may not represent the full set of systematic errors associated with this analysis.

\section{Extended Source Search Method}
\label{extended_source_search_method}

Having demonstrated that we understand the statistical
issues associated with analyzing spatially extended sources
(Section~\ref{monte_carlo_validation} and~\ref{test_2lac_sources}) and
that our method can correctly analyze the extended sources included in
2FGL (Section~\ref{validate_known}), we applied this method to search for
new spatially extended \gev sources.
The data and general analysis setting is as described in Section~\ref{test_2lac_sources}.

Ideally, we would apply a completely blind and uniform search that
tests the extension of each 2FGL source in the presence of all other
2FGL sources to find a complete list of all spatially extended sources.
As our test of AGN in Section~\ref{test_2lac_sources} showed, at high
Galactic latitude where the source density is not as large and the
diffuse emission is less structured, this method works well.

But this is infeasible in the Galactic plane where we 
are most likely to discover new spatially extended sources.  In the Galactic plane,
this analysis is challenged by our imperfect model of the diffuse
emission and by an imperfect model of nearby sources.  The Monte Carlo
study in Section~\ref{dual_localization_method}
showed that the overall likelihood would greatly increase by fitting
a spatially extended source as two point-like sources so we expect
that spatially extended sources would be modeled in the 2FGL catalog as
multiple point-like sources. Furthermore, the positions of other nearby sources
in the region close to an extended source could be biased by not correctly
modeling the extension of the source.  The 2FGL catalog contains a
list of sources significant at energies above 100 \mev whereas we are
most sensitive to spatial extension at higher energies.
We therefore expect that at higher energies our analysis would be complicated
by 2FGL sources no longer significant and by 2FGL
sources whose positions were biased by diffuse emission at lower energies.

To account for these issues, we first produced a large list of possibly
extended sources employing very liberal search criteria and then
refined the analysis of the promising candidates on a case by case basis.
Our strategy was to test all point-like 2FGL sources for extension assuming
they had a uniform radially-symmetric disk spatial model
and a power-law spectral model.  Although not all extended sources are
expected to have a shape very similar to a uniform disk, Section~\ref{compare_source_size} showed that for many spatially
extended sources the wide PSF of the LAT and limited statistics makes
this a reasonable approximation.  On the other hand, choosing this
spatial model biases us against finding extended sources that are not
well described by a uniform disk model such as shell-type SNRs.

Before testing for extension, we automatically removed from the background
model all other 2FGL sources within 0\fdg5 of the source.  This distance
is somewhat arbitrary, but was picked in hopes of finding extended
sources with sizes on the order of $\sim1\degree$ or smaller. On the
other hand, by removing these nearby background sources we expect to
also incorrectly add to our list of extended source candidates
point-like sources that
are confused with nearby sources.  To screen out obvious cases of source
confusion, we performed the dual localization procedure described in
Section~\ref{dual_localization_method} to compare the extended source
hypothesis to the hypothesis of two independent point-like sources.

As was shown in Section~\ref{extension_sensitivity}, little sensitivity is gained by
using photons with energies below 1 \gev. In addition,
the broad PSF at low energy makes the analysis more susceptible to systematic
errors arising from source confusion due to nearby soft point-like sources
and by uncertainties in our modeling of the Galactic diffuse emission. 
For these reasons,
we performed our search using only photons with energies between 1 \gev
and 100 \gev.

We also performed a second search for extended sources using only
photons with energies between 10 \gev and 100 \gev.  Although this
approach tests the same sources, it is complementary because the Galactic
diffuse emission is even less dominant above 10 \gev and because source
confusion is less of an issue.  A similar procedure was used to detect
the spatial extensions of MSH\,15$-$52 and
HESS\,J1825$-$137 with the LAT \citep{msh1552,fermi_hess_j1825}.

When we applied this test to the 1861 point-like sources in the 2FGL catalog, our
search found 117 extended source candidates in the 1 \gev to 100 \gev
energy range and 11 extended source candidates in the 10 \gev to 100
\gev energy range. Most of the extended sources found above 10 \gev were
also found above 1 \gev and in many cases multiple nearby point-like
sources were found to be extended even though they fit the same emission region.
For example, the sources 2FGL\,J1630.2$-$4752, 2FGL\,J1632.4$-$4753c 2FGL\,J1634.4$-$4743c,
and 2FGL\,J1636.3$-$4740c were all found to be spatially extended in the
10 \gev to 100 \gev energy range even though they all fit to similar
positions and sizes.  For these situations, we manually discarded all
but one of the 2FGL sources.

\begin{figure}
    \ifcolorfigure
    \plotone{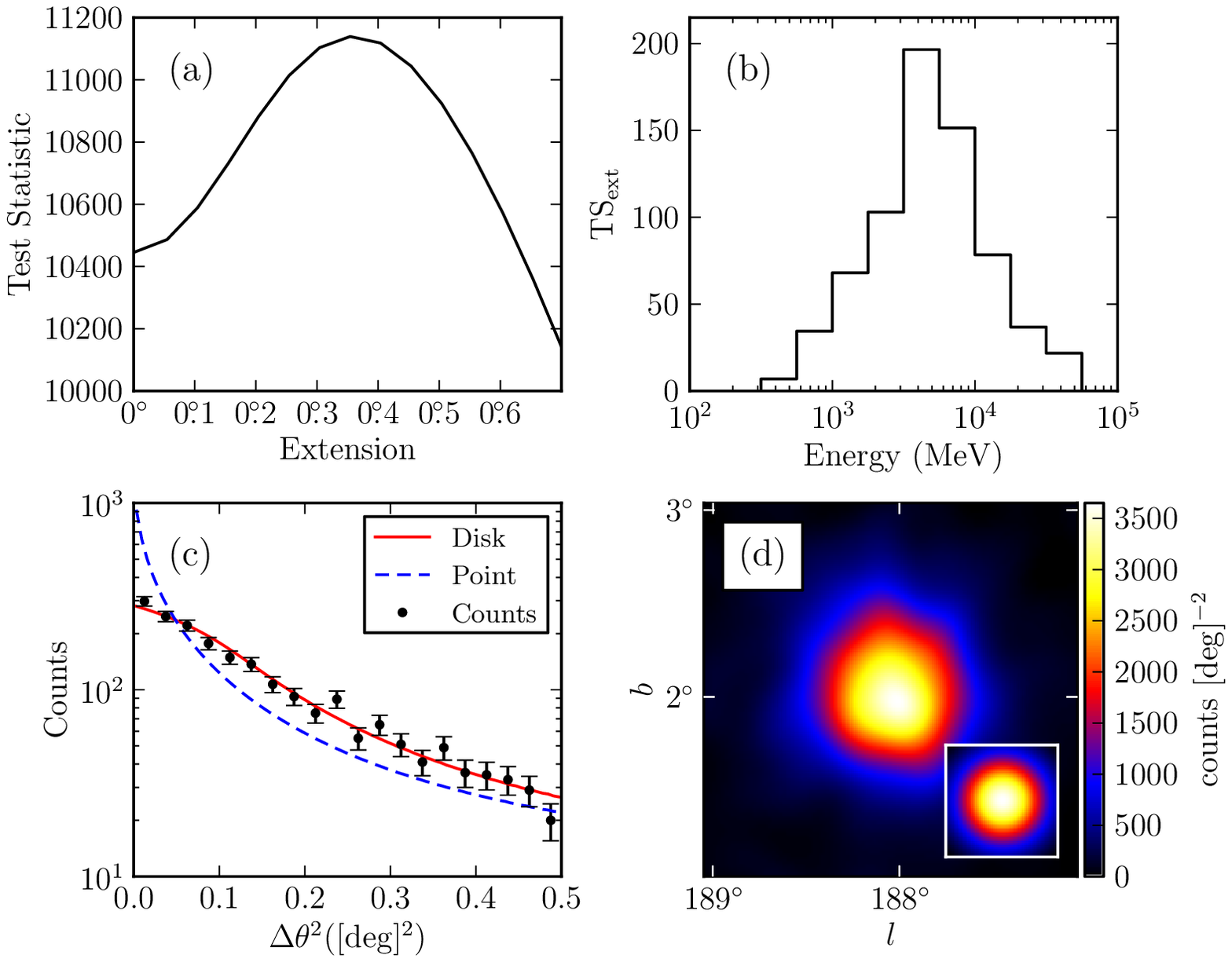}
    \else
    \plotone{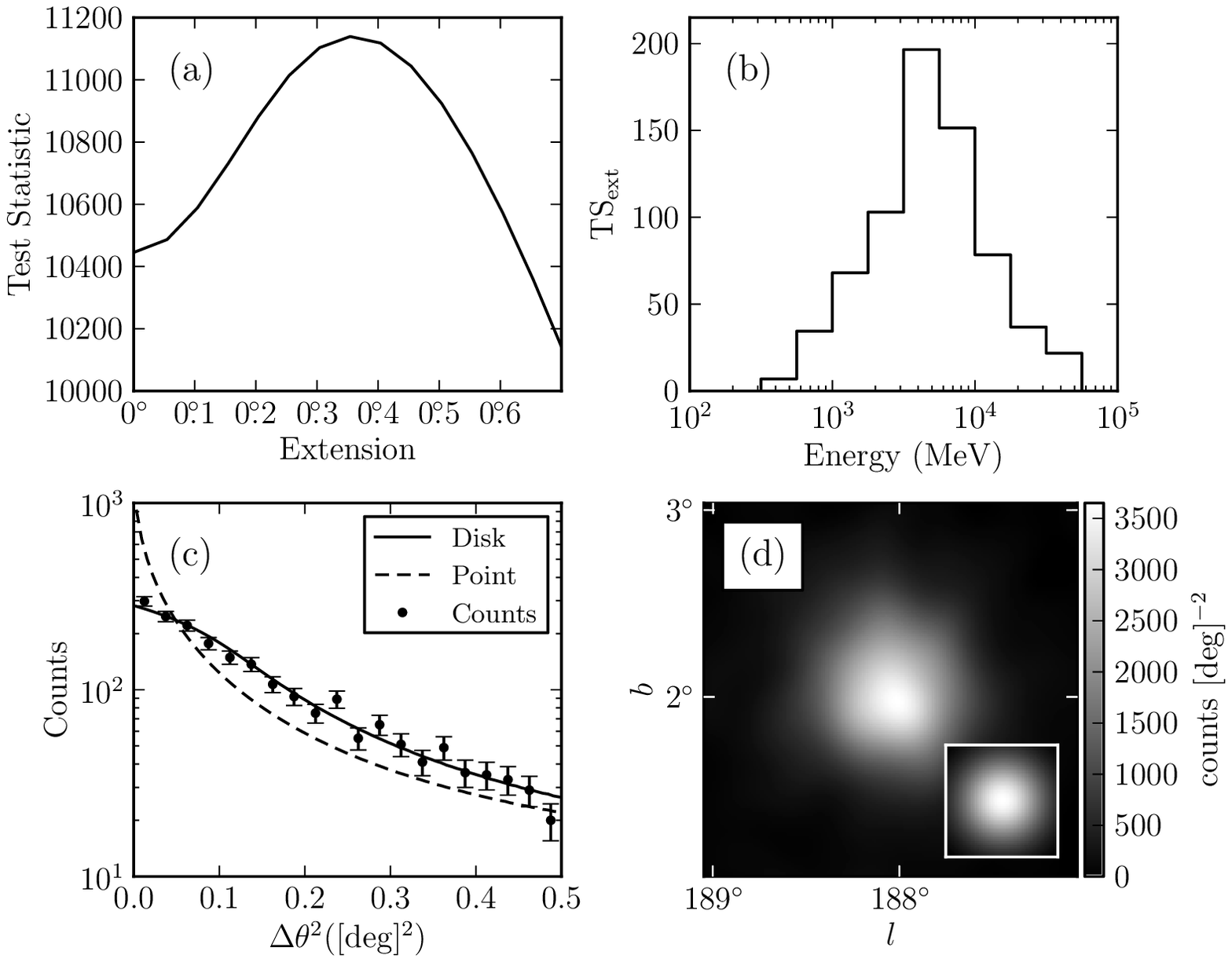}
    \fi
    \caption{
    Counts maps and TS profiles for the SNR IC~443. (a) \ts
    vs. extension of the source. (b) \tsext for individual energy
    bands. (c) observed radial profile of counts in comparison to the
    expected profiles for a spatially extended source (solid and colored
    red in the online version) and for a point-like source (dashed and colored
    blue in the online version).  (d) smoothed counts map after subtraction
    of the diffuse emission compared to the smoothed
    LAT PSF (inset). Both were smoothed by a 0\fdg1 2D Gaussian kernel.
    Plots (a),
    (c), and (d) use only 
    photons with energies between
    1 \gev and 100 \gev.  Plots (c) and (d) include
    only photons which converted in the front part of the tracker and
    have an improved angular resolution \citep{atwood_fermi}.
    }
    \label{four_plots_ic443}
\end{figure}

\begin{figure}
  \ifcolorfigure
    \plotone{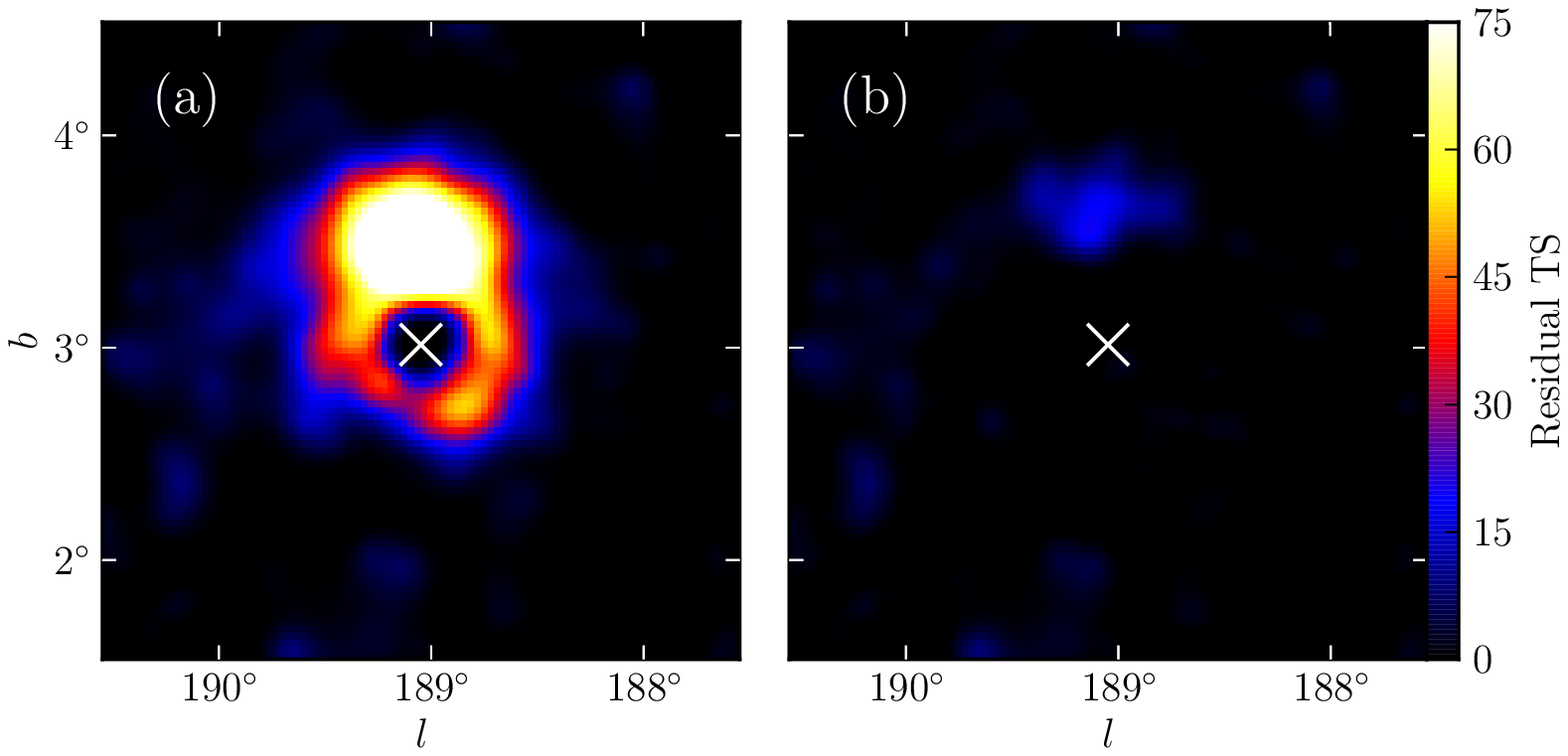}
    \else
    \plotone{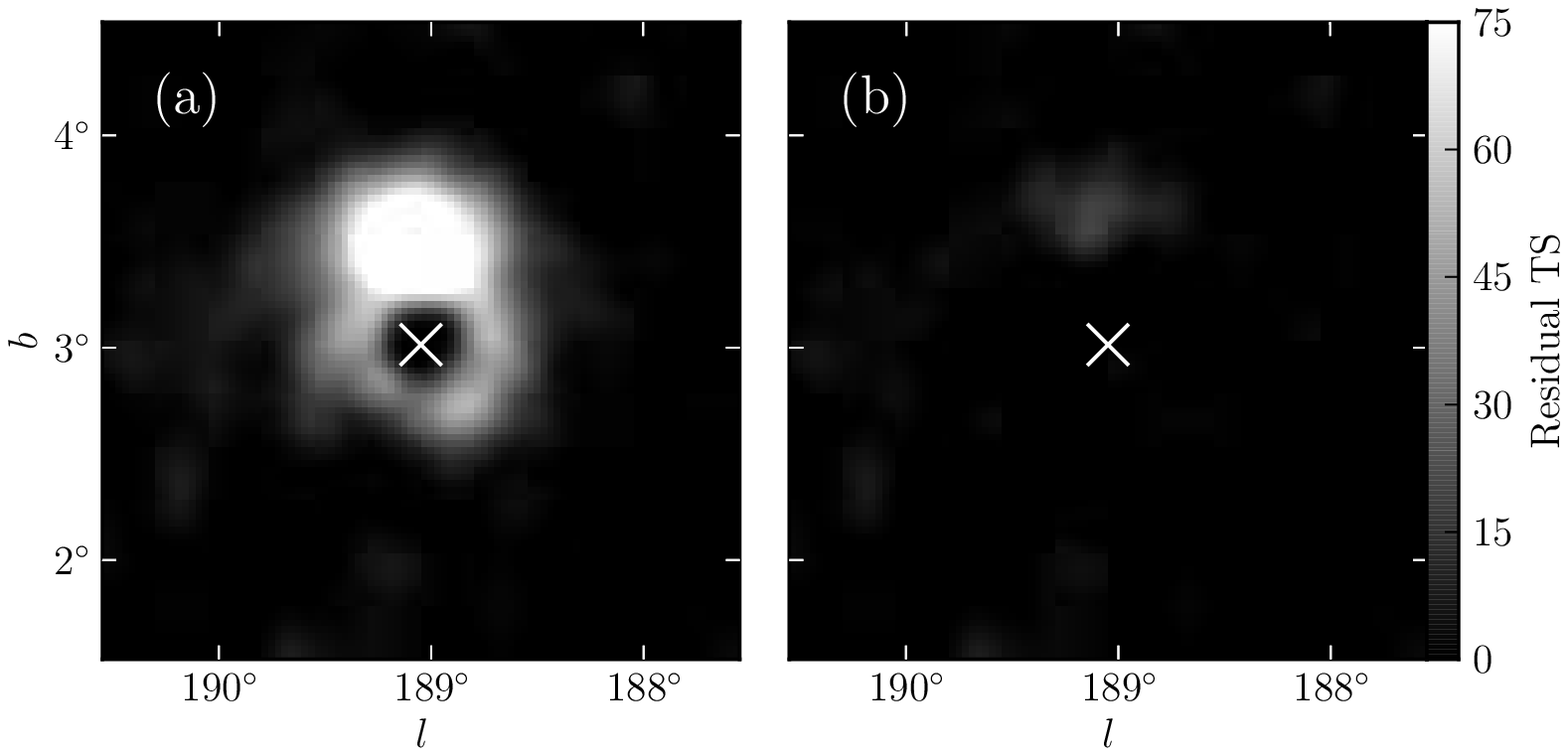}
    \fi

  \caption{
  A TS map generated for the region around the SNR 
  IC~443 using 
  photons with energies between
  1 \gev and 100 \gev.  (a) TS map after
  subtracting IC~443 modeled as a point-like source. (b) same as (a), but
  IC~443 modeled as an extended source. The cross represents the best
  fit position of IC~443.
  }
  \label{res_tsmaps}
\end{figure}

Similarly, many of these sources were confused with nearby
point-like sources or influenced by large-scale residuals in the
diffuse emission.  To help determine which of these fits found
truly extended sources and when the extension was
influenced by source confusion and diffuse emission, we generated a
series of diagnostic plots.  For each candidate, we generated a map
of the residual TS by adding a new point-like source of spectral index 2 into the
region at each position and finding the increase in likelihood when fitting
its flux. Figure~\ref{res_tsmaps} shows this map around the
most significantly extended source IC~443 when it is modeled both as a
point-like source and as an extended source.  The residual TS
map indicates
that the spatially extended model for IC~443 is a significantly better
description of the observed photons and that there is no $\ts>25$
residual in the region after modeling the source as being spatially extended.
We also generated plots of the sum of all counts within a given distance of
the source and compared them to the model predictions assuming the emission
originated from a point-like source.  An example radial integral plot
is shown for the extended source IC~443 in Figure~\ref{four_plots_ic443}.
For each source, we also made diffuse-emission-subtracted smoothed counts
maps (shown for IC~443 in Figure~\ref{four_plots_ic443}).

\begin{figure}
    \ifcolorfigure
    \plotone{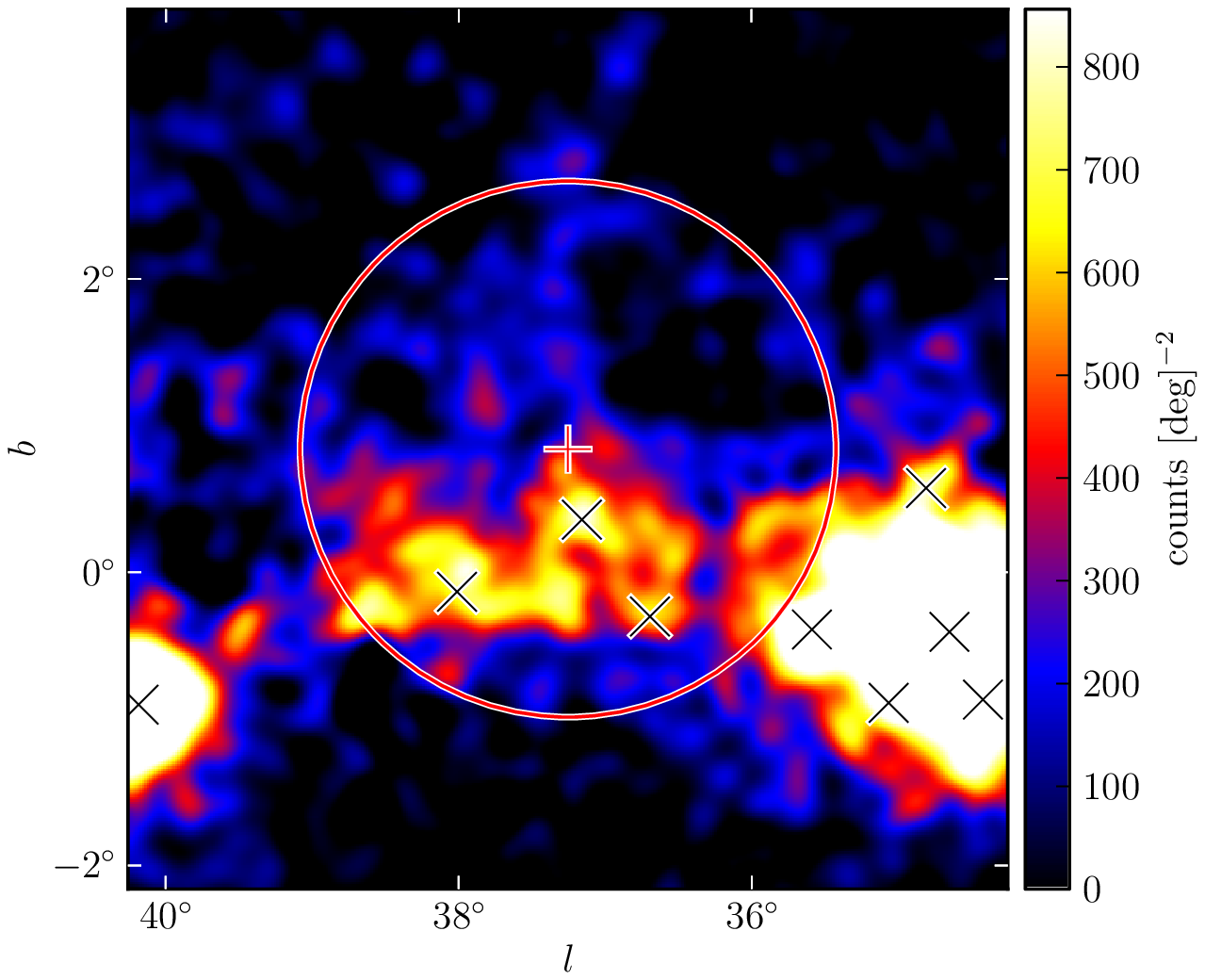}
    \else
    \plotone{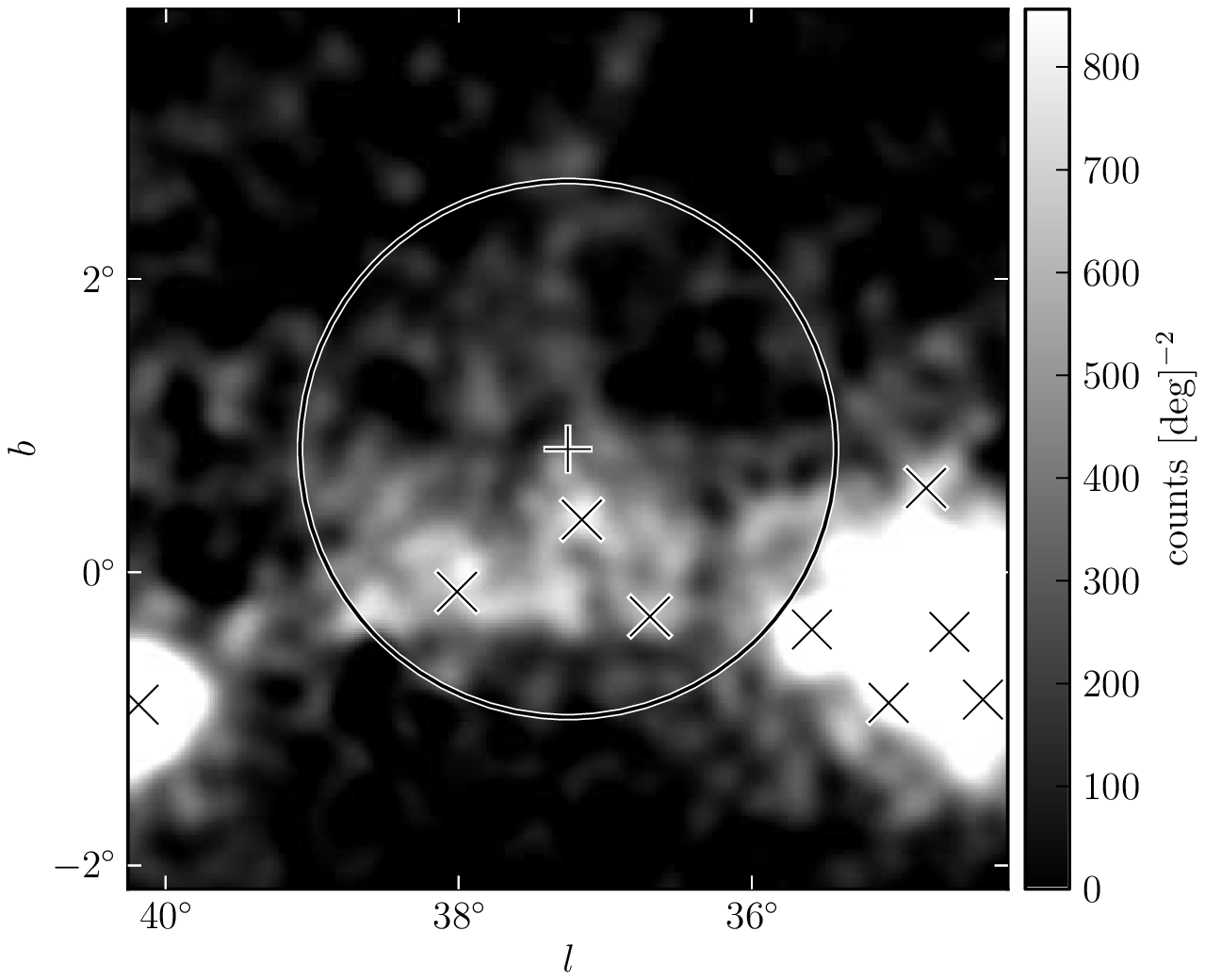}
    \fi
    \caption{
    A diffuse-emission-subtracted 1 \gev to 100 \gev counts map of the
    region around 2FGL\,J1856.2+0450c smoothed by a 0\fdg1 2D Gaussian
    kernel. The plus-shaped marker and circle (colored red in
    the online version) represent the center and size of the source
    fit with a
    radially-symmetric uniform disk spatial model.  The black
    crosses represent the positions of other 2FGL sources.  The extension
    is statistically significant, but the extension encompasses many 2FGL
    sources and the emission does not look to be uniform. Although
    the fit is statistically significant, it likely corresponds to
    residual features of inaccurately modeled diffuse emission picked
    up by the fit. 
    }
    \label{example_bad_fit}
\end{figure}

We found by visual inspection that in many
cases our results were strongly influenced by large-scale residuals in the
diffuse emission and hence the extension measure was unreliable.  This was
especially true in our analysis of sources in the 1 \gev to 100 \gev
energy range.  An example of such a case is 2FGL\,J1856.2+0450c analyzed
in the 1 \gev to 100 \gev energy range. Figure~\ref{example_bad_fit}
shows a diffuse-emission-subtracted smoothed counts map for
this source with the best
fit extension of the source overlaid. There appear to be large-scale
residuals in the diffuse emission in this region along the Galactic plane.
As a result, 2FGL\,J1856.2+0450c is fit to an extension of $\sim2\degree$
and the result is statistically significant with \tsext=45.4. However,
by looking at the residuals it is clear that this complicated region is
not well modeled. We manually discard sources like this.

We only selected extended source
candidates in regions that did not appear dominated by these issues and
where there was a multiwavelength
counterpart. Because of these systematic issues, this search can not be
expected to be complete and it is likely that there are other spatially
extended sources that this method missed.

For each candidate that was not biased by neighboring point-like
sources or by large-scale residuals in the diffuse emission model, we
improved the model of the region by deciding on a case by case basis which
background point-like sources should be kept.  We kept in our model the
sources that we
believed represented physically distinct sources and we removed 
sources that we believed were included in the 2FGL catalog to compensate
for residuals induced by not modeling the extension of the source.
Soft nearby point-like 2FGL sources that were not significant at higher energies
were frozen to the spectras predicted by 2FGL.
When deciding
which background sources to keep and which to remove, we used 
multiwavelength information about possibly extended source counterparts
to help guide our choice. For each extended source presented in 
Section~\ref{new_ext_srcs_section}, we describe any modifications from 2FGL
of the background model that were performed.
In Table~\ref{fake_2fgl_sources}, we summarize the sources in
the 2FGL catalog that we have concluded here correspond to residuals
induced by not modeling the extensions of nearby extended sources.

The best fit positions of nearby point-like sources can be influenced by
the extended source and vice versa.  Similarly,
the best fit positions of nearby point-like sources in the 2FGL catalog can be biased
by systematic issues at lower energies.
Therefore, after selecting the list of background sources, we iteratively
refit the positions and spectra of nearby background sources as well as
the positions and extensions of the analyzed spatially extended
sources until the overall fit converged globally.  For each extended
source, we will describe the positions of any relocalized background
sources.

After obtaining the overall best fit positions and extensions of all
of the sources in the region using \pointlike, we refit the spectral
parameters of the region using \gtlike.  With \gtlike, we obtained a
second measure of \tsext.  We only consider a source to be extended when
both \pointlike and \gtlike agree that $\tsext\ge16$. 
We further required that $\tsext\ge16$ using the 
alternative approach to modeling the diffuse emission
presented in Section~\ref{systematic_errors_on_extension}.
We then replaced the spatially extended source with two point-like
sources and refit the positions and spectra of the
two point-like sources to calculate \tsinc.
We only consider a source to be spatially extended, instead of being
the result of confusion of two point-like sources, if $\tsext>\tsinc$.
As was shown in Section~\ref{dual_localization_method}, this test is
fairly powerful at removing situations in which the emission actually
originates from two distinct point-like sources instead of one spatially
extended source.  On the other hand, it is still possible that longer
observations could resolve additional structure or new sources 
that the analysis cannot currently detect. 
Considering the very complicated morphologies
of extended sources observed at other wavelengths and the high density
of possible sources that are expected to emit at \gev energies, it is
likely that in some of these regions further observations will reveal
that the emission is significantly more complicated than the simple
radially-symmetric uniform disk model that we assume.

\section{New Extended Sources}
\label{new_ext_srcs_section}



Nine extended sources not included in the 2FGL catalog were found by our
extended source search. Two of these have been previously
studied in dedicated publications: RX\,J1713.7$-$3946 and Vela
Jr. \citep{rx_j1713_lat,vela_jr_lat}.
Two of these sources were
found when using photons with energies between 1 \gev and 100 \gev and seven
were found when using photons with energies between 10 \gev and 100 \gev.
For the sources found at energies above 10 \gev, we restrict our
analysis to higher energies because of the issues
of source confusion and diffuse emission modeling described in
Section~\ref{extended_source_search_method}.
The spectral and spatial properties of these nine sources are summarized
in Table~\ref{new_ext_srcs_table} and the results of our investigation of
systematic errors are presented in Table~\ref{alt_diff_model_results}.
Table~\ref{alt_diff_model_results}
also
compares the likelihood assuming the source is spatially
extended to the likelihood assuming 
that the emission originates from two independent point-like
sources. For these new extended sources, $\tsext>\tsinc$ 
so we conclude that 
the \gev emission does not originate from two physically
distinct point-like sources (see Section~\ref{dual_localization_method}).  
Table~\ref{alt_diff_model_results} also includes the
results of the extension fits using variations of the PSF and the Galactic
diffuse model described in Section~\ref{systematic_errors_on_extension}.
There is good agreement between \tsext and the fit size using the standard
analysis, the alternative approach to modeling the diffuse emission, and the alternative PSF.
This suggests that the sources are robust against mis-modeled features in the diffuse
emission model and uncertainties in the PSF.

\subsection{2FGL\,J0823.0$-$4246}
\label{section_2FGL_J0823.0-4246}

\begin{figure}
    \ifcolorfigure
      \plotone{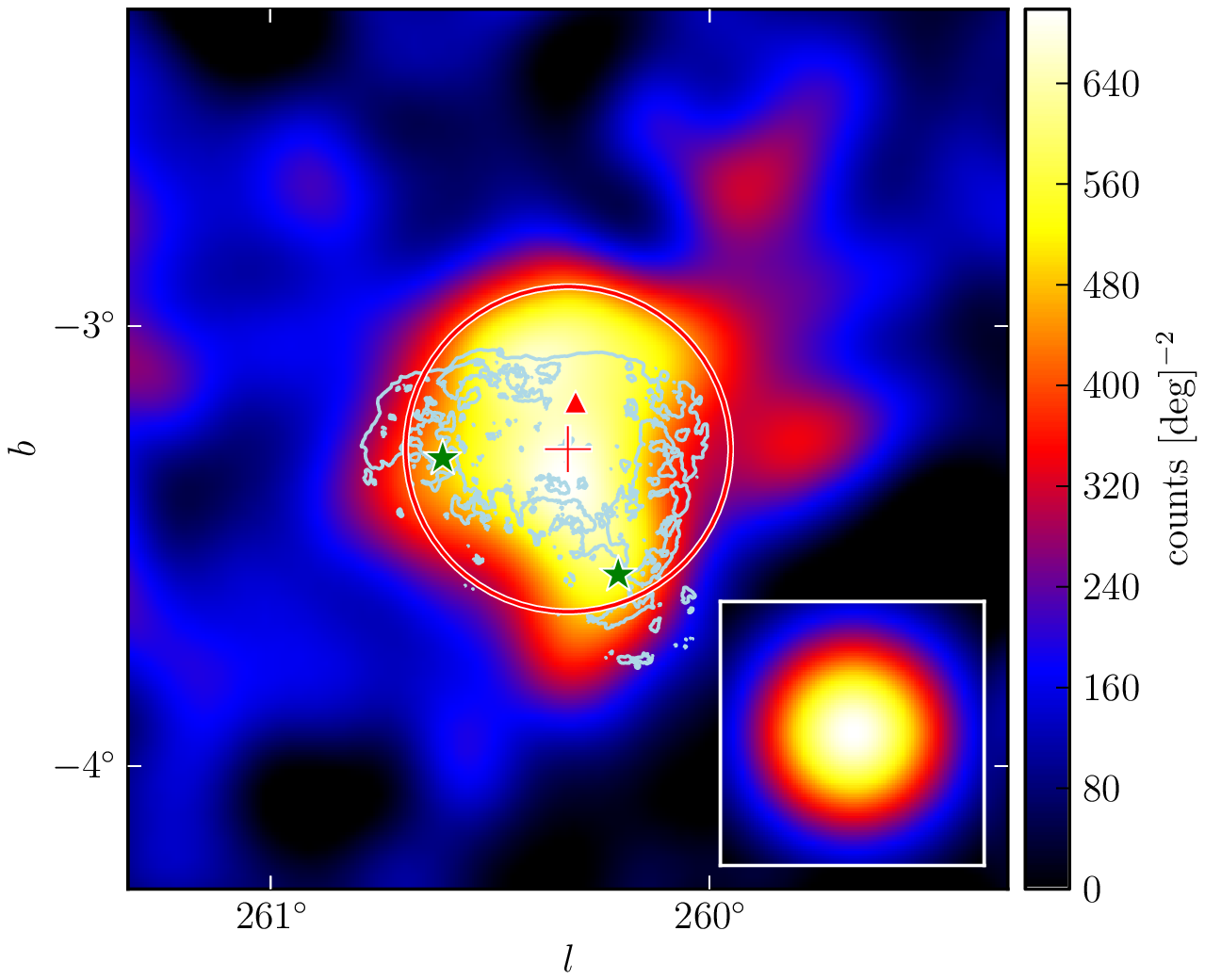}
    \else
      \plotone{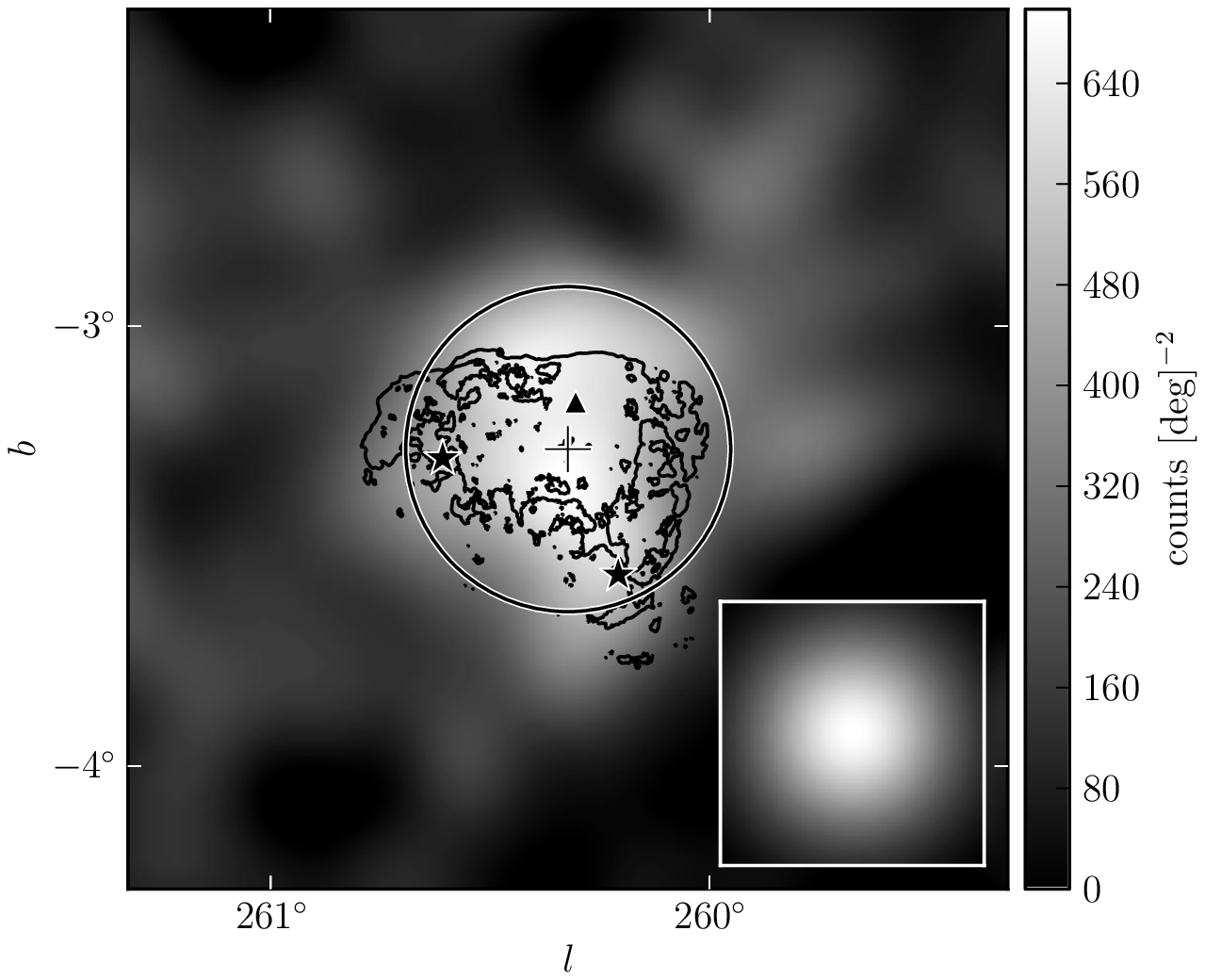}
    \fi
  \caption{A diffuse-emission-subtracted 1 \gev to 100 \gev counts map
  of 2FGL\,J0823.0$-$4246 smoothed by a 0\fdg1 2D
  Gaussian kernel.  The triangular marker 
  (colored red in the online version)
  represents the 2FGL position of this source.  The plus-shaped
  marker and the circle (colored
  red) represent 
  the best fit position and extension of this source assuming a
  radially-symmetric uniform disk model.  The two 
  star-shaped markers (colored
  green) represent 2FGL sources that were
  removed from the background model.
  From left to right, these sources are 2FGL\,J0823.4$-$4305 and 2FGL\,J0821.0$-$4254.
  The lower right inset is the model predicted emission from a point-like
  source with the same spectrum as 2FGL\,J0823.4$-$4305 smoothed by the
  same kernel.  This source is spatially coincident with the Puppis A
  SNR. The light blue contours correspond to the X-ray image of Puppis
  A observed by \rosat \citep{rosat_puppis_a}.
  }\label{1FGL_J0823.3-4248}
\end{figure}
\begin{figure}
    \ifcolorfigure
      \plotone{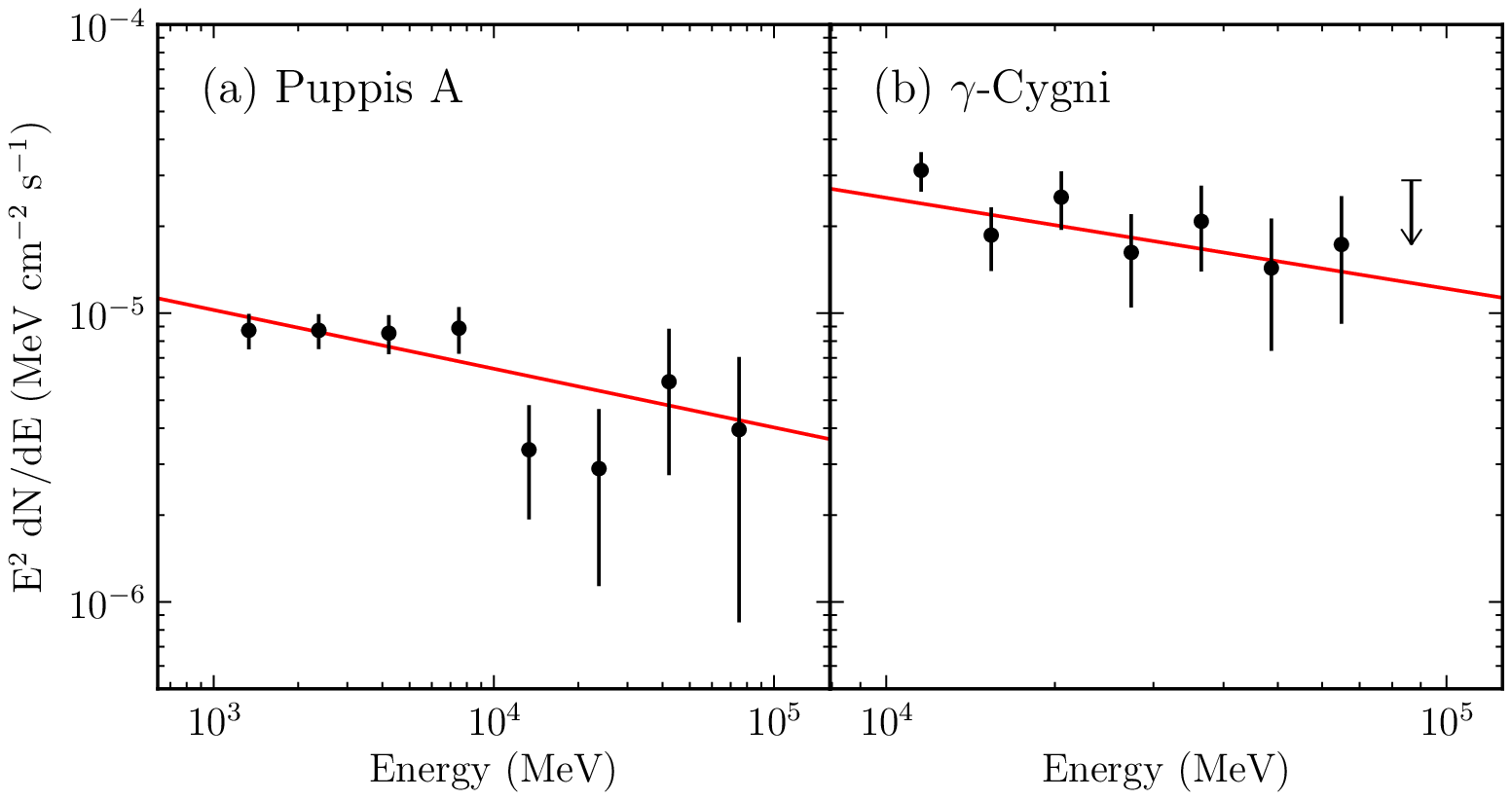}
    \else
      \plotone{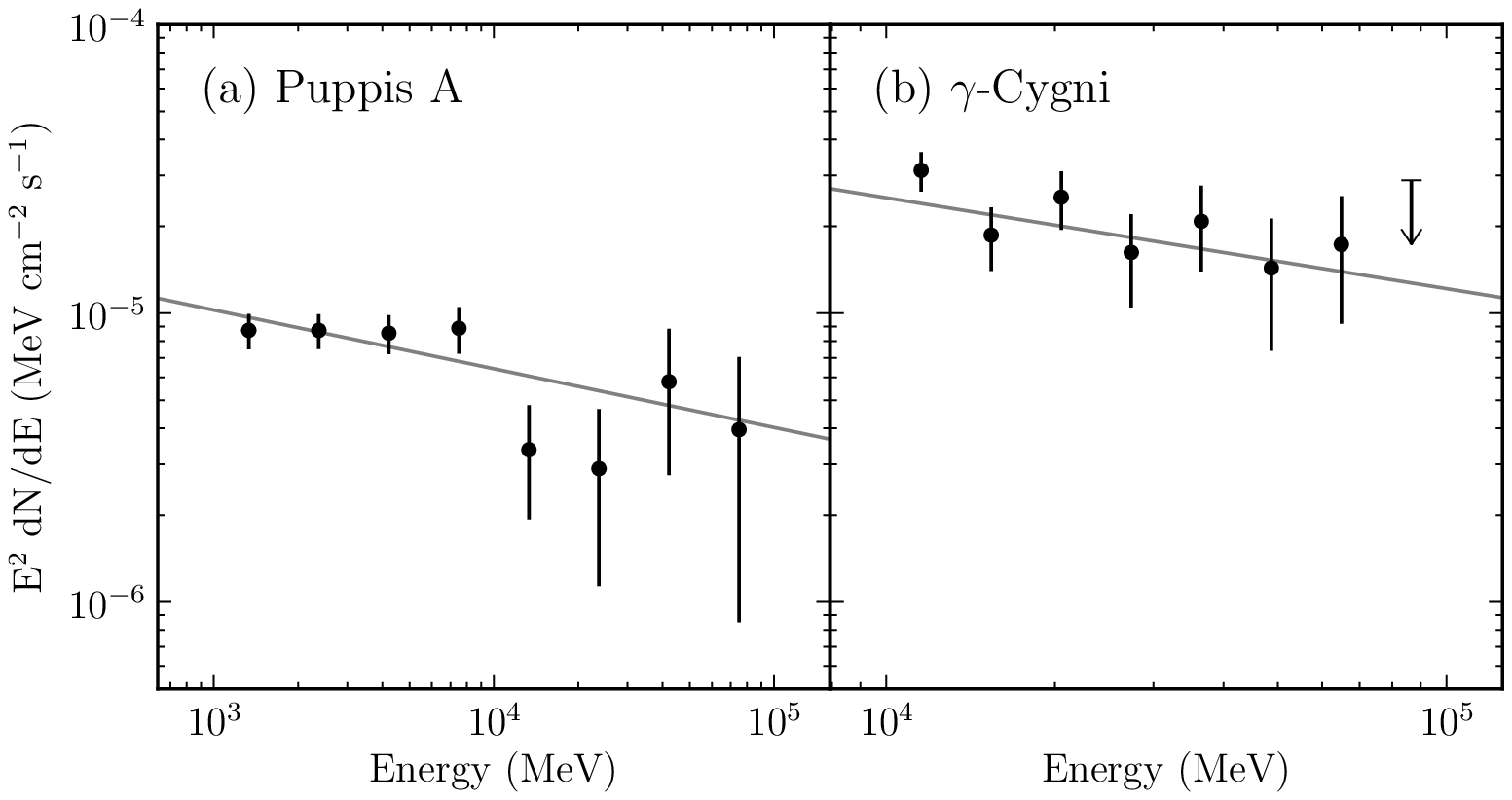}
    \fi
    \caption{
    The spectral energy distribution of the extended sources 
    Puppis A (2FGL\,J0823.0$-$4246) and $\gamma$-Cygni 
    (2FGL\,J2021.5+4026).
    The lines (colored red in the online version)
    are the best fit power-law spectral models of
    these sources. Puppis A has a spectral index of
    $2.21\pm0.09$ and $\gamma$-Cygni has an
    index of $2.42\pm0.19$.
    The spectral errors are statistical only.
    The upper limit is at the 95\% confidence level.
    }
    \label{snr_seds}
  \end{figure}

2FGL\,J0823.0$-$4246 was found by our search to be an extended source
candidate in the 1 \gev to 100 \gev energy range and is spatially
coincident with the SNR Puppis A.  Figure~\ref{1FGL_J0823.3-4248} shows
a counts map of this source. There are two nearby 2FGL sources 2FGL\,J0823.4$-$4305
and 2FGL\,J0821.0$-$4254 that are also coincident with the SNR but that
do not appear to represent physically distinct sources.
We conclude that
these nearby point-like sources
were included in the 2FGL catalog to compensate for residuals induced
by not modeling the extension of this source and
removed them from our model of the sky.
After removing these sources, 2FGL\,J0823.0$-$4246 was found to have an extension 
$\sigma=0\fdg37\pm0\fdg03_\stat\pm0\fdg02_\sys$ with 
$\tsext=48.0$.  Figure~\ref{snr_seds} shows the spectrum of
this source.

Puppis A has been studied in detail in radio \citep{puppis_a_vla}, 
and  X-ray \citep{rosat_puppis_a,suzaku_puppis_a}.
The fit extension of 2FGL\,J0823.0$-$4246
matches well the size of Puppis A in X-ray.  The distance of Puppis A was
estimated at 2.2 kpc \citep{reynoso_1995,reynoso_2003} and leads to a 1
\gev to 100 \gev luminosity of $\sim 3\times 10^{34}$ ergs$\,\s^{-1}$.
No molecular clouds have been observed directly adjacent to Puppis A
\citep{co_eastern_puppis_a}, similar to the LAT-detected Cygnus Loop SNR
\citep{cygnus_loop_lat}.  The luminosity of Puppis A is also smaller
than that of other SNRs believed to interact with molecular clouds
\citep{w51c,ic443,w44,w28,w49b_lat}.

\subsection{2FGL\,J0851.7$-$4635}
\label{section_2FGL_J0851.7-4635}

\begin{figure}
    \ifcolorfigure
      \plotone{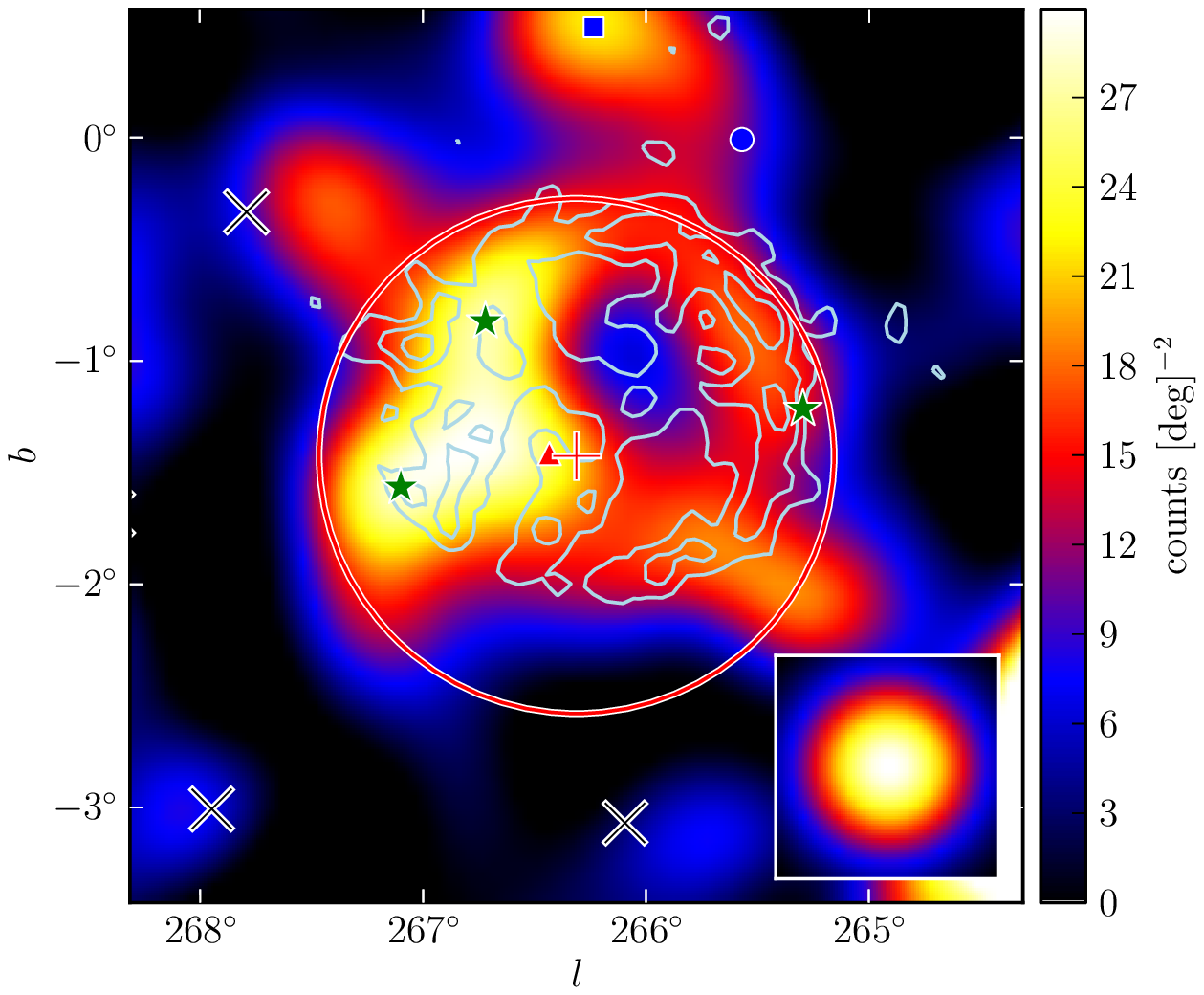}
    \else
      \plotone{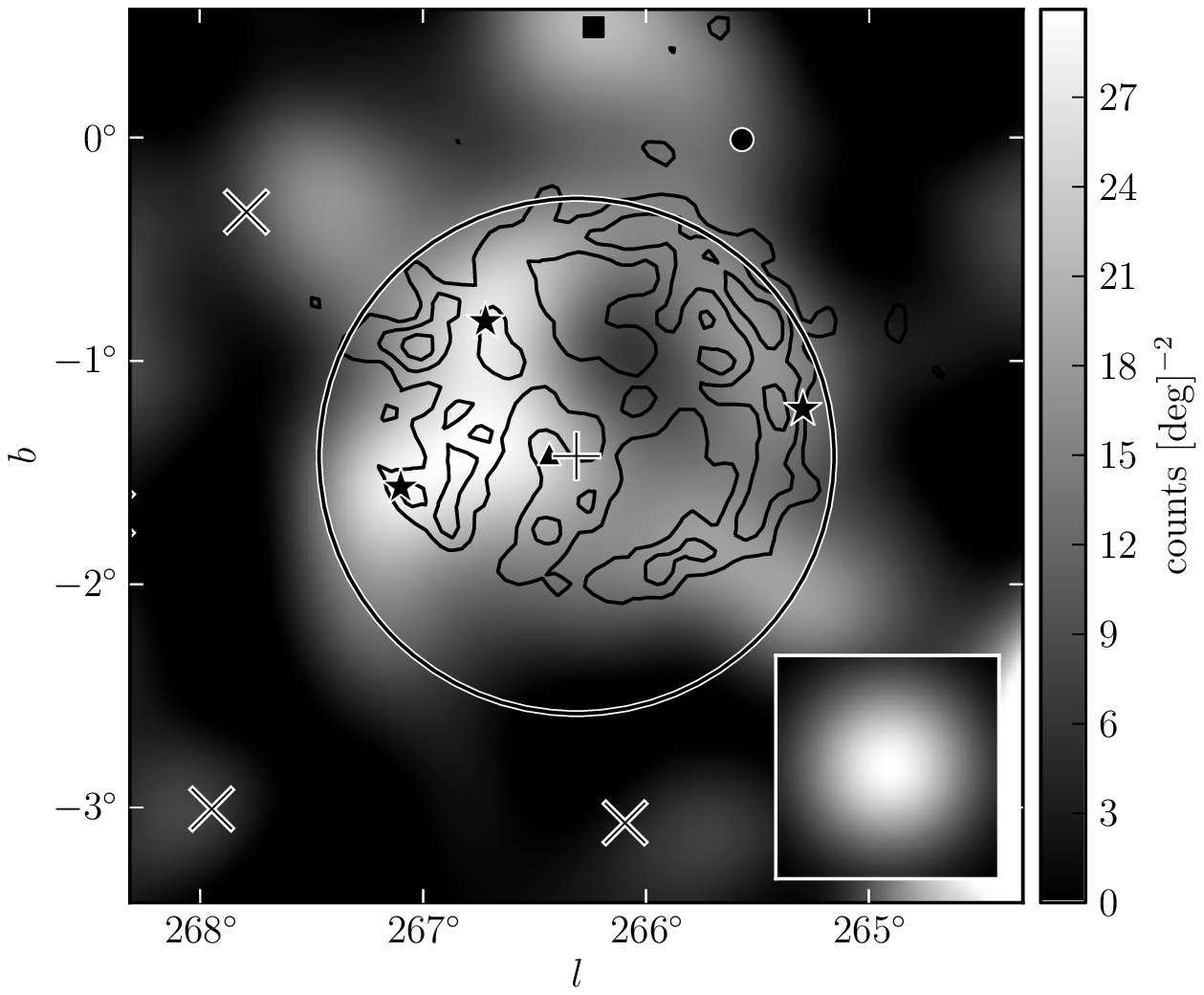}
    \fi
  \caption{A diffuse-emission-subtracted 10 \gev to 100 \gev counts map of
  2FGL\,J0851.7$-$4635 smoothed by a 0\fdg25 2D Gaussian
  kernel. The triangular marker (colored red in the electronic version)
  represents the 2FGL position of this source.  The plus-shaped marker
  and the circle (colored red) are the best fit position and extension of
  this source assuming a radially-symmetric uniform disk model.
  The three black crosses represent background 2FGL sources.
  The three star-shaped markers (colored green) represent other 2FGL sources
  that were removed from the background model.
  They are (from left to right) 2FGL\,J0853.5$-$4711, 2FGL\,J0855.4$-$4625, and 
  2FGL\,J0848.5$-$4535.
  The circular and square-shaped
  marker (colored blue) represents the 2FGL and relocalized position of another 2FGL source.  
  This extended source is spatially
  coincident with the Vela Jr. SNR.  The contours (colored light blue)
  correspond to the \tev image of Vela Jr.
  \citep{vela_jr_hess}.
  }\label{Vela_Jr}
\end{figure}

2FGL\,J0851.7$-$4635 was found by our search to be an extended source
candidate in the 10 \gev to 100 \gev energy range and is spatially
coincident with the SNR Vela Jr. This source was recently studied by the LAT
Collaboration in \cite{vela_jr_lat}.  Figure~\ref{Vela_Jr} shows a counts
map of the source.  Overlaid on Figure~\ref{Vela_Jr} are \tev
contours of Vela Jr. \citep{vela_jr_hess}.
There are three
point-like 2FGL sources 2FGL\,J0848.5$-$4535,
2FGL\,J0853.5$-$4711, and
2FGL\,J0855.4$-$4625
which correlate with the 
multiwavelength
emission of
this SNR but do not appear to be physically distinct sources.
They were most likely included in the 2FGL catalog to compensate for residuals
induced by not modeling the extension of Vela Jr. and were removed
from our model of the sky.  

With this model of the background, 2FGL\,J0851.7$-$4635 was found to
have an extension of $\sigma=1\fdg15\pm0\fdg08_\stat\pm0\fdg02_\sys$ with
$\tsext=86.8$.  The LAT size matches
well the \tev morphology of Vela Jr.  While fitting the extension
of 2FGL\,J0851.7$-$4635, we iteratively relocalized the position
of the nearby point-like 2FGL source 2FGL\,J0854.7$-$4501 to
$(l,b)=(266\fdg24,0\fdg49)$ to better fit its position at high energies.

\subsection{2FGL\,J1615.0$-$5051}
\label{section_2FGL_J1615.0-5051}

\begin{figure}
    \ifcolorfigure
      \plotone{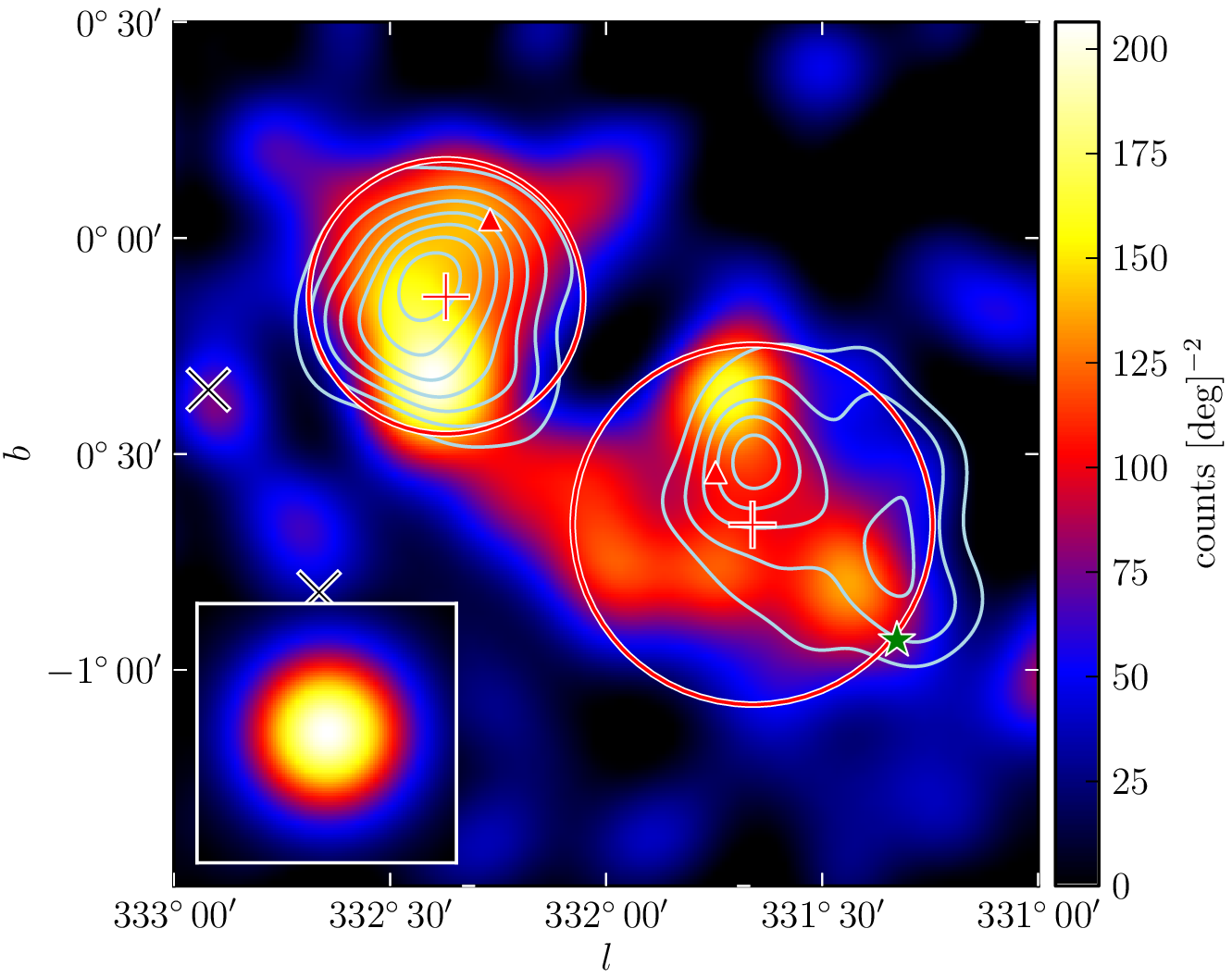}
    \else
      \plotone{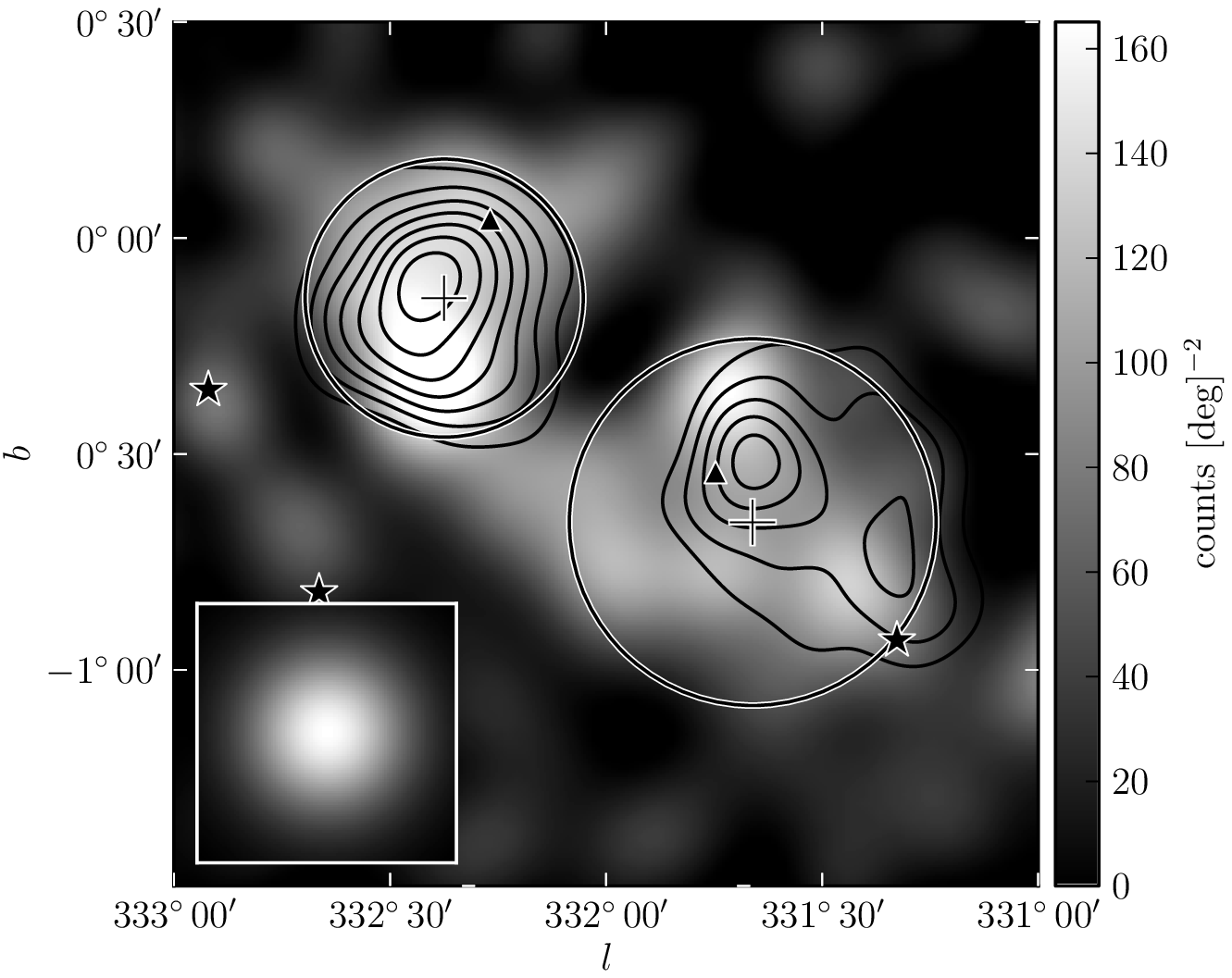}
    \fi
  \caption{
    A diffuse-emission-subtracted 10 \gev to 100 \gev counts
    map of 2FGL\,J1615.0$-$5051 (upper left)
    and 2FGL\,J1615.2$-$5138 (lower right) smoothed by a 0\fdg1
    2D Gaussian kernel.  The triangular markers (colored red in the
    electronic version) represent the 2FGL positions of these sources.
    The cross-shaped markers and the
    circles (colored red) represent the best fit
    positions and extensions of these sources assuming a radially
    symmetric uniform disk model.  
    The two black crosses represent  background 2FGL sources and
    the star-shaped
    marker (colored green) represents 2FGL J1614.9-5212, another 2FGL
    source that was removed from the background
    model. The contours (colored light blue) correspond to the \tev
    image of HESS\,J1616$-$508 (left) and HESS\,J1614$-$518 (right)
    \citep{hess_plane_survey}.   
    }\label{1FGL_J1613.6-5100c}
\end{figure}

2FGL\,J1615.0$-$5051 and 2FGL\,J1615.2$-$5138 were both found to be
extended source candidates in the 10 \gev to 100 \gev energy range. Because
they are less than $1\degree$ away from each other, they needed to be analyzed
simultaneously.  2FGL\,J1615.0$-$5051 is spatially coincident with
the extended \tev source HESS\,J1616$-$508 and 2FGL\,J1615.2$-$5138
is coincident with the extended \tev source HESS\,J1614$-$518.
Figure~\ref{1FGL_J1613.6-5100c} shows a counts map of these sources and
overlays the \tev contours of HESS\,J1616$-$508 and HESS\,J1614$-$518
\citep{hess_plane_survey}.  The figure shows that the 2FGL source
2FGL\,J1614.9$-$5212 is very close to 2FGL\,J1615.2$-$5138 and correlates
with the same extended \tev source as 2FGL\,J1615.2$-$5138.  We concluded
that this source was included in the 2FGL catalog to compensate for residuals induced
by not modeling the extension of 2FGL\,J1615.2$-$5138 and removed
it from our model of the sky.  

With this model of the sky, we iteratively fit the extensions of
2FGL\,J1615.0$-$5051 and 2FGL\,J1615.2$-$5138.
2FGL\,J1615.0$-$5051 was found to have an extension 
$\sigma=0\fdg32\pm0\fdg04_\stat\pm0\fdg01_\sys$ and
\tsext=16.7.

The \tev counterpart of 2FGL\,J1615.0$-$5051 was fit
with a radially-symmetric Gaussian surface brightness profile with
$\sigma=0\fdg136\pm0\fdg008$ \citep{hess_plane_survey}. This \tev size corresponds to a 68\%
containment radius of $\rsixeight=0\fdg21\pm0\fdg01$, comparable to the
LAT size $\rsixeight=0\fdg26\pm0\fdg03$.  Figure~\ref{hess_seds} shows
that the spectrum of 2FGL\,J1615.0$-$5051 at \gev energies connects to
the spectrum of HESS\,J1616$-$508 at \tev energies.

HESS\,J1616$-$508 is located in the region of two SNRs RCW103 (G332.4-04)
and Kes~32 (G332.4+0.1) but is not spatially coincident with either
of them \citep{hess_plane_survey}.  HESS\,J1616$-$508 is near three
pulsars PSR\,J1614$-$5048, PSR\,J1616$-$5109, and PSR\,J1617$-$5055.
\citep{discovery_of_PSR_J1617-5055,integral_HESS_J1616-508}.  Only
PSR\,J1617$-$5055 is energetically capable of powering the \tev emission
and \cite{hess_plane_survey} speculated that HESS\,J1616$-$508 could
be a PWN powered by this young pulsar.  Because HESS\,J1616$-$508 is
$9\arcmin$ away from PSR\,J1617$-$5055, this would require an asymmetric
X-ray PWNe to power the \tev emission.  \chandra ACIS observations
revealed an underluminous PWN of size $\sim1\arcmin$ around the
pulsar that was not oriented towards the \tev emission, rendering this
association uncertain \citep{discovery_of_pwn_for_PSR_J1617-5055}.
No other promising counterparts were observed at X-ray and soft
$\gamma$-ray energies by \suzaku \citep{suzakzu_HESS_J1616-508}, \swiftxrt,
IBIS/ISGRBI, BeppoSAX and \xmmnewton \citep{integral_HESS_J1616-508}.
\cite{discovery_of_pwn_for_PSR_J1617-5055} discovered additional
diffuse emission towards the center of HESS\,J1616$-$508 using archival
radio and infared observations. Deeper observations will likely be
necessary to understand this $\gamma$-ray source.

\subsection{2FGL\,J1615.2$-$5138}
\label{section_2FGL_J1615.2-5138}

2FGL\,J1615.2$-$5138 was found
 to have an extension $\sigma=0\fdg42\pm0\fdg04_\stat\pm0.02_\sys$
with $\tsext=46.5$.
To test for the possibility that
2FGL\,J1615.2$-$5138 is not spatially extended but instead composed
of two point-like sources (one of them represented in the 2FGL catalog by
2FGL\,J1614.9$-$5212), we refit 2FGL\,J1615.2$-$5138 as two point-like
sources.  Because $\tsinc=35.1$ is less than $\tsext=46.5$, we conclude that
this emission does not originate from two closely-spaced point-like sources.

2FGL\,J1615.2$-$5138 is spatially coincident with 
the extended \tev source HESS\,J1614$-$518.
H.E.S.S. measured a 2D Gaussian extension of $\sigma=0\fdg23\pm0\fdg02$
and $\sigma=0\fdg15\pm0\fdg02$ in the semi-major and
semi-minor axis. This corresponds to a 68\% containment size of
$\rsixeight=0\fdg35\pm0\fdg03$ and $0\fdg23\pm0\fdg03$,  consistent with
the LAT size $\rsixeight=0\fdg34\pm0\fdg03$.  Figure~\ref{hess_seds}
shows that the spectrum of 2FGL\,J1615.2$-$5138 at \gev energies connects
to the spectrum of HESS\,J1614$-$518 at \tev energies.  Further data
collected by H.E.S.S. in 2007 resolve a double peaked structure at
\tev energies but no spectral variation across this source, suggesting
that the emission is not the confusion of physically separate sources
\citep{closer_look_hess_j1614-518}.  This double peaked structure is
also hinted at in the LAT counts map in Figure~\ref{1FGL_J1613.6-5100c}
but is not very significant.  The \tev source was also detected by
CANGAROO-III \citep{cangaroo_j1614-518}.

There are five nearby pulsars, but none are luminous enough to
provide the energy output required to power the $\gamma$-ray
emission \citep{closer_look_hess_j1614-518}.  HESS\,J1614$-$518
is spatially coincident with a young open cluster Pismis 22
\citep{hess_1614_landi_atel,closer_look_hess_j1614-518}.  \suzaku detected
two promising X-ray candidates. Source A is an extended source consistent
with the peak of HESS\,J1614$-$518 and source B coincident with Pismis 22
and towards the center but in a relatively dim region of HESS\,J1614$-$518
\citep{suazku_hess_j1614_518}.  Three hypotheses have been presented to
explain this emission: either source A is an SNR powering the $\gamma$-ray
emission; source A is a PWN powered by an undiscovered pulsar in either
source A or B; and finally that the emission may arise from hadronic
acceleration in the stellar winds of Pismis 22 \citep{cangaroo_j1614-518}.

\subsection{2FGL\,J1627.0$-$2425c}
\label{section_2FGL_J1627.0-2425c}

\begin{figure}
    \ifcolorfigure
      \plotone{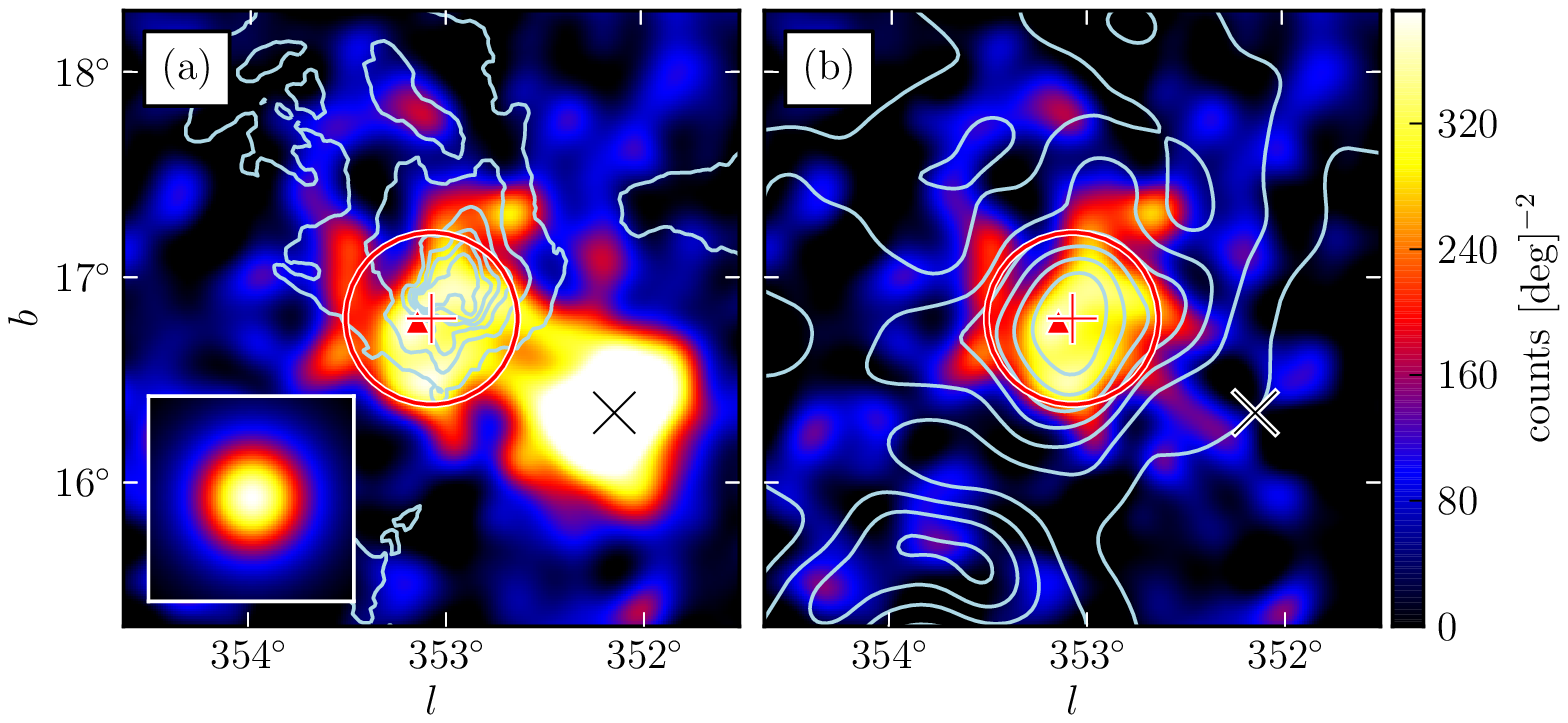}
    \else
      \plotone{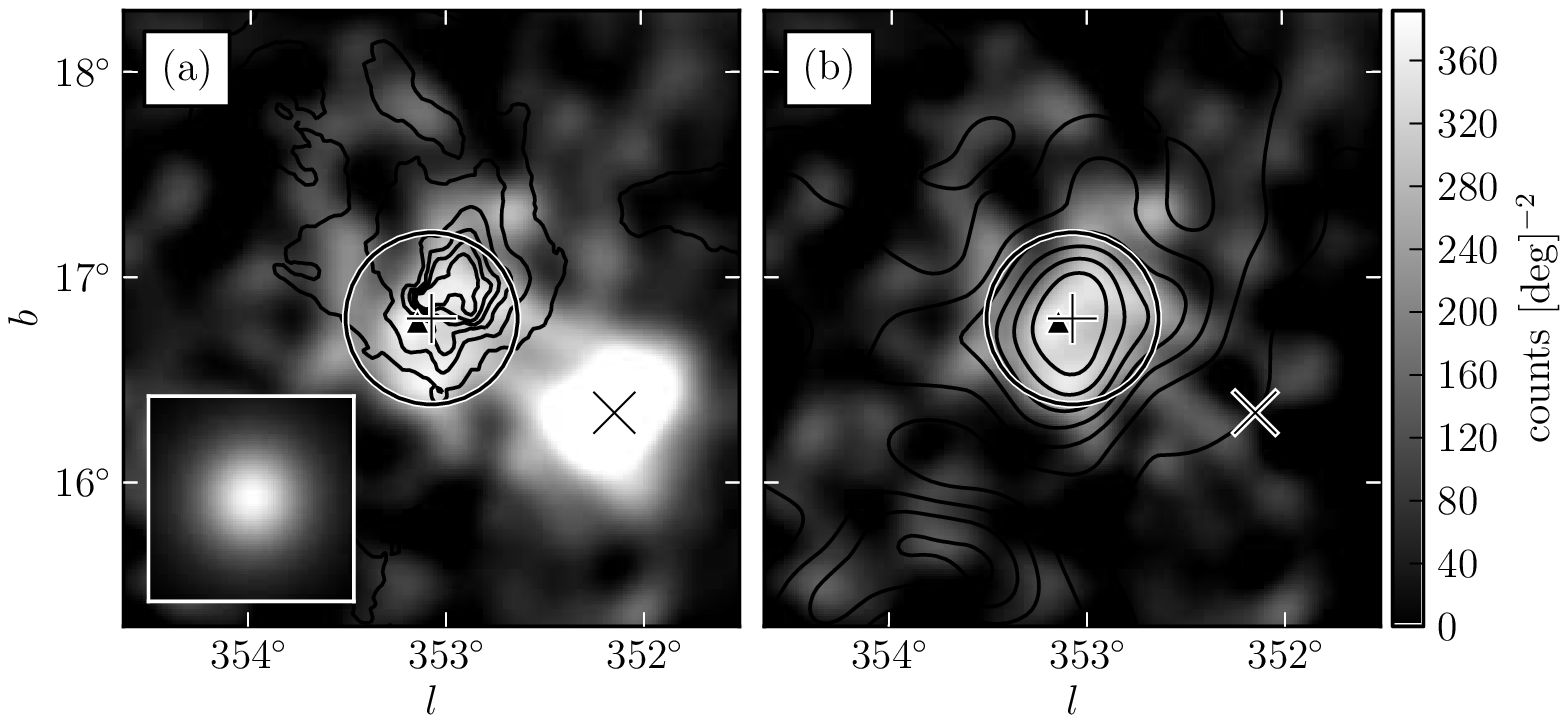}
    \fi
  \caption{
  A diffuse-emission-subtracted 1 \gev to 100 \gev counts map of (a) the region
  around 2FGL\,J1627.0$-$2425 smoothed by a 0\fdg1 2D Gaussian kernel and (b)
  with the emission from 2FGL\,J1625.7$-$2526
  subtracted.  The triangular marker 
  (colored red in the
  online version) represents the 2FGL position of this source.
  The plus-shaped marker and the circle (colored red) 
  represent the best fit position and extension of this
  source assuming a radially-symmetric uniform disk model
  and the black cross represents a background 2FGL source. 
  The
  contours in (a) correspond to the 100 $\mu$m image observed by
  IRAS \citep{iras_rho_ophiuci}.  The contours in (b) correspond to
  CO ($J=1\rightarrow 0$) emission integrated from $-$8 $\km\,\s^{-1}$
  to 20 $\km\,\s^{-1}$.  They are from \cite{co_rho_ophiuci}, were cleaned using
  the moment-masking technique \citep{masking_moment_2011}, and have
  been smoothed by a 0\fdg25 2D Gaussian kernel.
  }\label{1FGL_J1628.6-2419c}
\end{figure}

2FGL\,J1627.0$-$2425c was found by our search to
have an extension $\sigma=0\fdg42\pm0\fdg05_\stat\pm0\fdg16_\sys$ with
$\tsext=32.4$
using photons with energies between 1 \gev and 100 \gev.  
Figure \ref{1FGL_J1628.6-2419c} shows a counts map of this source.

This source is in a region of remarkably complicated diffuse emission.
Even though it is $16\degree$ from the Galactic plane, this source is on
top of the core of the Ophiuchus molecular cloud which contains massive
star-forming regions that are bright in infrared.  The region also has
abundant molecular and atomic gas traced by CO and H~I and significant 
dark gas found only by its association with dust emission
\citep{isabelle_dark_gass}. Embedded star-forming regions make it even
more challenging to measure the column density of dust.  Infared and 
CO ($J=1\rightarrow 0$)
contours are overlaid on Figure~\ref{1FGL_J1628.6-2419c} and show good
spatial correlation with the \gev emission \citep{iras_rho_ophiuci,co_rho_ophiuci}.
This source might 
represent $\gamma$-ray emission from the interactions of cosmic rays with
interstellar gas which has not been accounted for in the LAT diffuse
emission model.

\subsection{2FGL\,J1632.4$-$4753c}
\label{section_2FGL_J1632.4-4753c}

\begin{figure}
    \ifcolorfigure
      \plotone{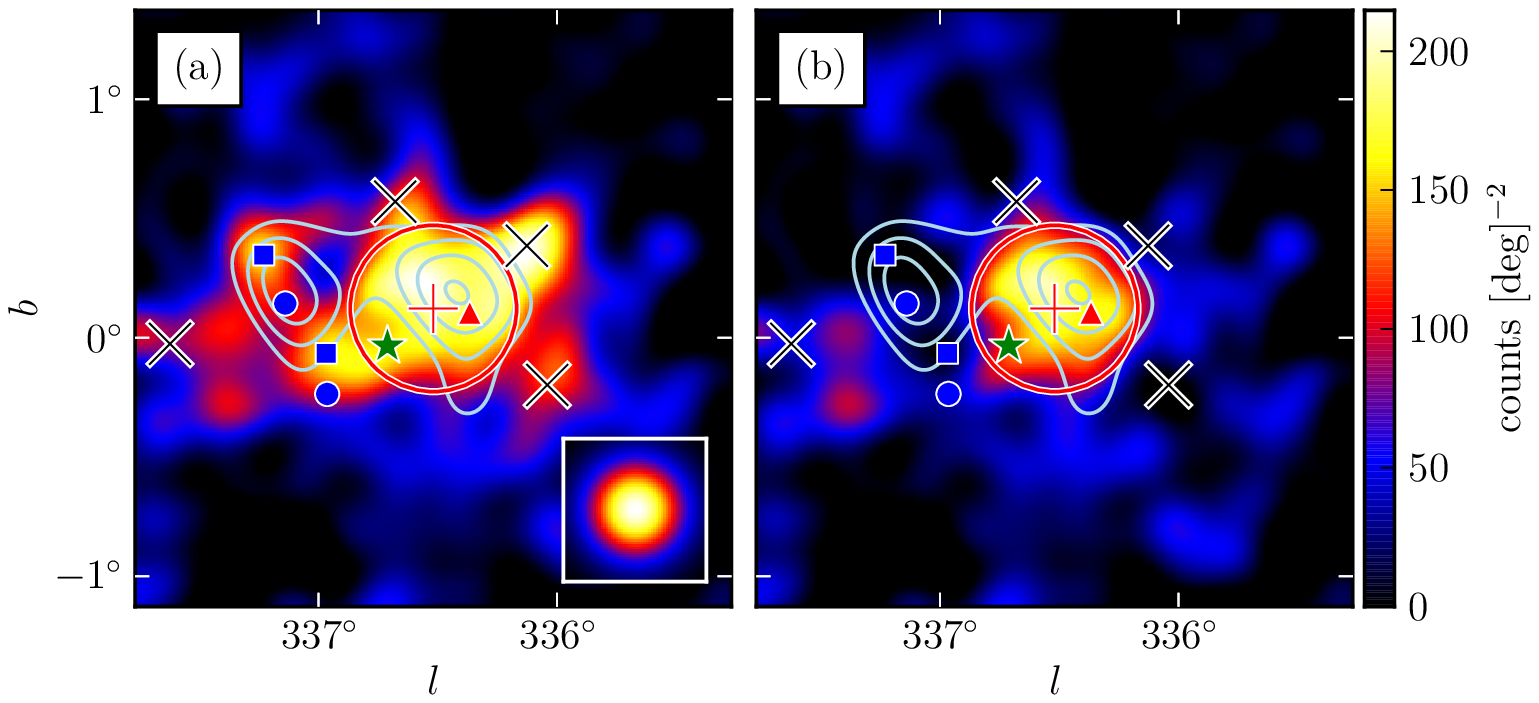}
    \else
      \plotone{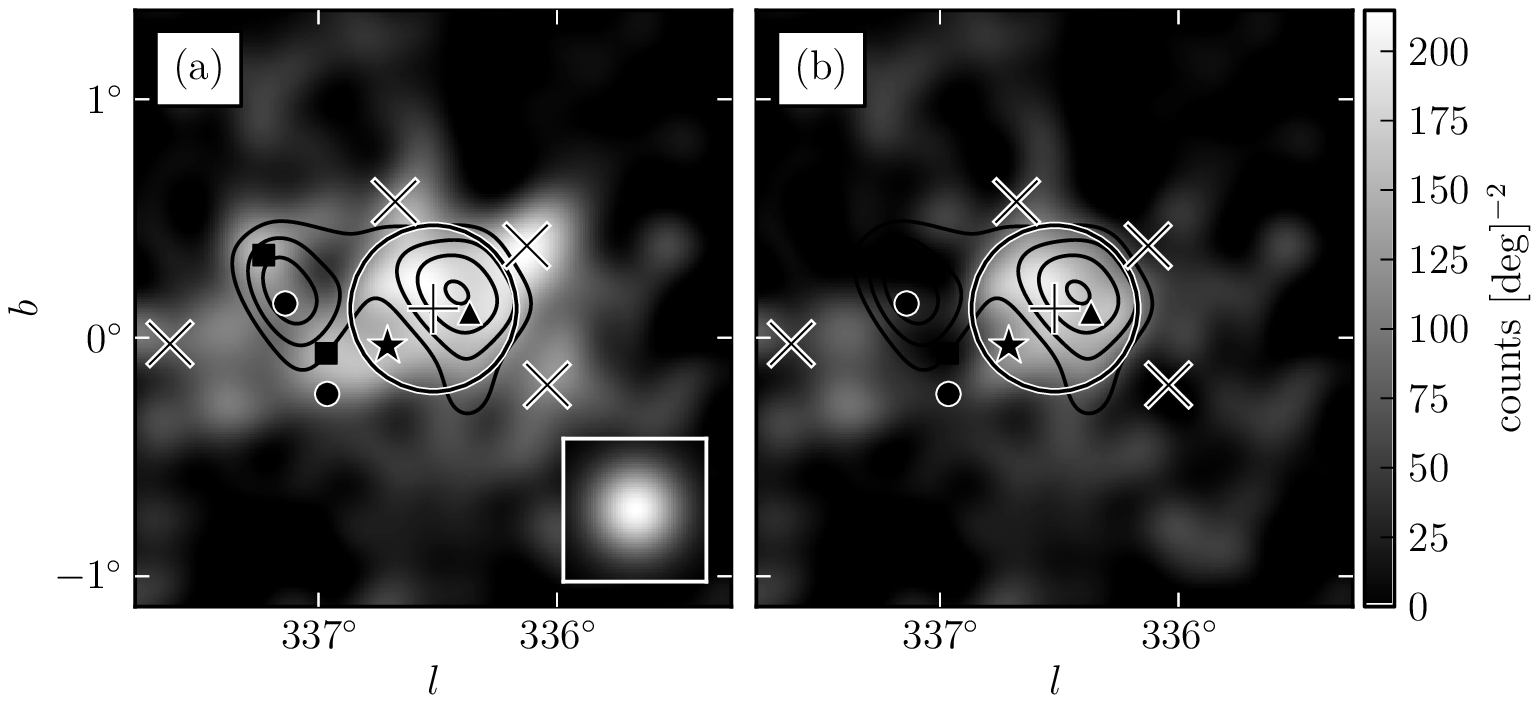}
    \fi
  \caption{
  A diffuse-emission-subtracted 10 \gev to 100 \gev counts map of
  2FGL\,J1632.4$-$4753c (a) smoothed by a 0\fdg1 2D Gaussian
  kernel and (b) with the emission from the background sources subtracted.  
  The triangular marker (colored red in the electronic version)
  represents the 2FGL position of this source.  The plus-shaped marker
  and the circle (colored red) are the best fit position and extension of
  2FGL\,J1632.4$-$4753c assuming a radially-symmetric uniform disk model.  
  The four black crosses represent background 2FGL
  sources subtracted in (b).  The 
  circular and square-shaped markers (colored blue) represent 
  the 2FGL and relocalized positions respectively of two
  additional background 2FGL sources subtracted in (b).
  The star-shaped marker (colored green) represents 
  2FGL\,J1634.4$-$4743c, another 2FGL source that was removed from the background model.
  The contours (colored light blue) correspond to the \tev image of
  HESS\,J1632$-$478 \citep{hess_plane_survey}.
  }\label{1FGL_J1632.9-4802c}
\end{figure}

2FGL\,J1632.4$-$4753c was found by our search to be an extended source
candidate in the 10 \gev to 100 \gev energy range but is in a crowded
region of the sky.  It is spatially coincident with the \tev source
HESS\,J1632$-$478.  Figure~\ref{1FGL_J1632.9-4802c}a shows a counts
map of this source and overlays \tev contours of HESS\,J1632$-$478
\citep{hess_plane_survey}.  There are six nearby point-like 2FGL
sources that appear to represent physically distinct sources
and were included in our background model: 2FGL\,J1630.2$-$4752, 2FGL\,J1631.7$-$4720c, 2FGL\,J1632.4$-$4820c, 2FGL\,J1635.4$-$4717c,
2FGL\,J1636.3$-$4740c, and
2FGL\,J1638.0$-$4703c. On the
other hand, one point-like 2FGL source 2FGL\,J1634.4$-$4743c  correlates
with the extended \tev source and at \gev energies does
not appear physically separate. 
It is very close to the position of 2FGL\,J1632.4$-$4753c
and does not show spatially separated emission in the observed photon distribution.
We therefore removed this source from our model of the
background.  Figure~\ref{1FGL_J1632.9-4802c}b shows the same region with
the background sources subtracted.  With this model, 2FGL\,J1632.4$-$4753c
was found to have an extension $\sigma=0\fdg35\pm0\fdg04_\stat\pm0\fdg02_\sys$
with $\tsext=26.9$.  While fitting the
extension of 2FGL\,J1632.4$-$4753c, we iteratively relocalized
2FGL\,J1635.4$-$4717c to $(l,b)=(337\fdg23,0\fdg35)$ and
2FGL\,J1636.3$-$4740c to $(l,b)=(336\fdg97,-0\fdg07)$.

H.E.S.S measured an extension of $\sigma=0\fdg21\pm0\fdg05$ and
$0\fdg06\pm0\fdg04$ along the semi-major and semi-minor axes when
fitting HESS\,J1632$-$478 with an elliptical 2D Gaussian surface
brightness profile.  This corresponds to a 68\% containment size
$\rsixeight=0\fdg31\pm0\fdg08$ and $0\fdg09\pm0\fdg06$ along
the semi-major and semi-minor axis, consistent with the LAT size
$\rsixeight=0\fdg29\pm0\fdg04$.  Figure~\ref{hess_seds} shows that
the spectrum of 2FGL\,J1632.4$-$4753c at \gev energies connects to the
spectrum of HESS\,J1632$-$478 at \tev energies.

\cite{hess_plane_survey} argued that HESS\,J1632$-$478
is positionally coincident with the hard X-ray source
IGR\,J1632$-$4751 observed by \asca, INTEGRAL, and \xmmnewton
\citep{asca_plane_survey,Igr_J16320-4751_circ,xmm_newton_IGR_J16320-4751},
but this source is suspected to be a Galactic X-Ray Binary so the
$\gamma$-ray extension disfavors the association.  Further observations
by \xmmnewton discovered point-like emission coincident with the peak
of the H.E.S.S. source surrounded by extended emission of size
$\sim32\arcsec\times15\arcsec$ \citep{hess_j1632_478_xmm_newton}.
They found in archival MGPS-2 data a spatially coincident extended
radio source \citep{most_survey_galactic_plane} and argued for a single
synchrotron and inverse Compton process producing the radio, X-ray, and \tev emission,
likely due to a PWN.  The increased size at \tev energies compared
to X-ray energies has previously been observed in several aging PWNe
including HESS\,J1825$-$137 \citep{hess_j1825_xmm_newton,hess_j1825_hess},
HESS\,J1640$-$465 \citep{hess_plane_survey,xmm_newton_hess_j_1640-466},
and Vela X \citep{vela_x_rosat,vela_x_hess} and can be explained by
different synchrotron cooling times for the electrons that produce X-rays
and $\gamma$-rays.

\subsection{2FGL\,J1712.4$-$3941}
\label{section_2FGL_J1712.4-3941}

\begin{figure}
    \ifcolorfigure
      \plotone{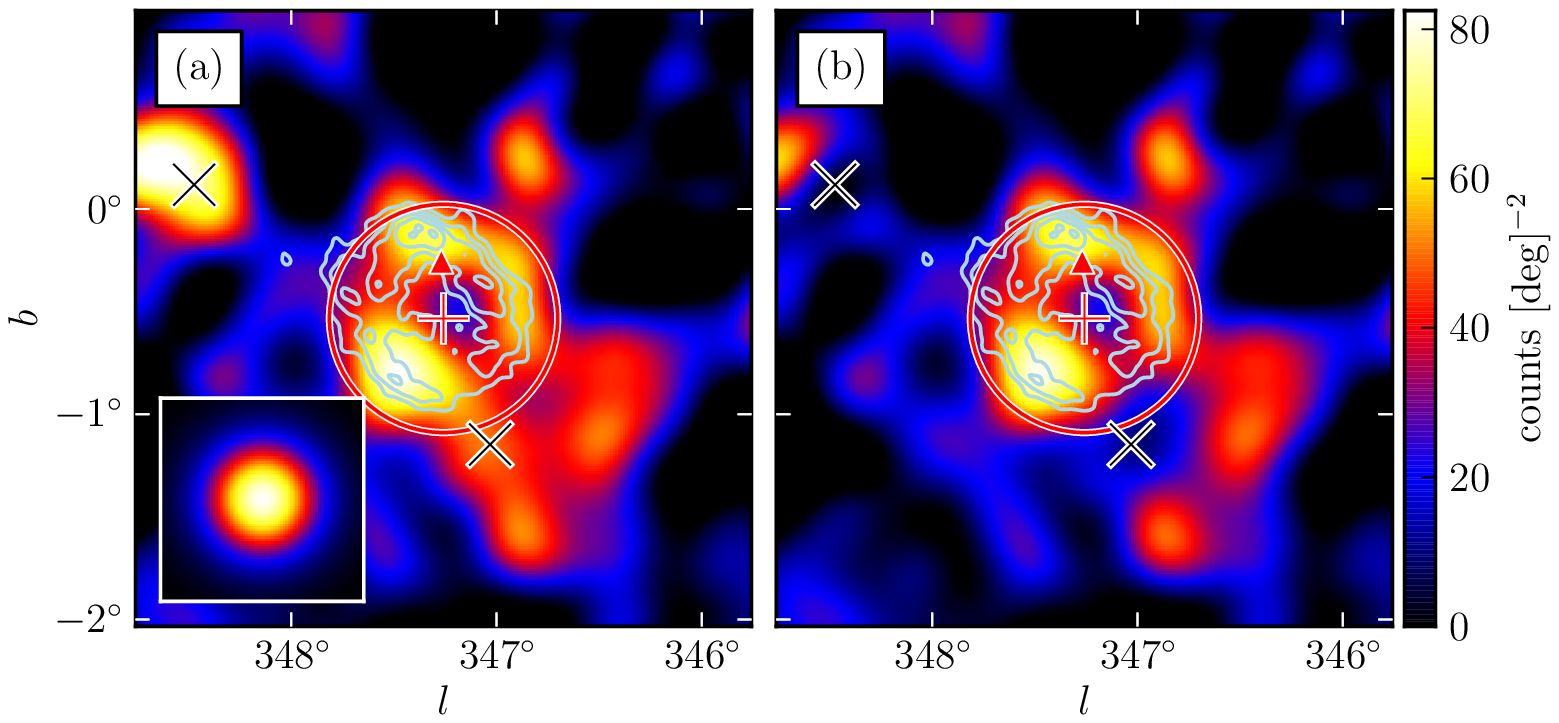}
    \else
      \plotone{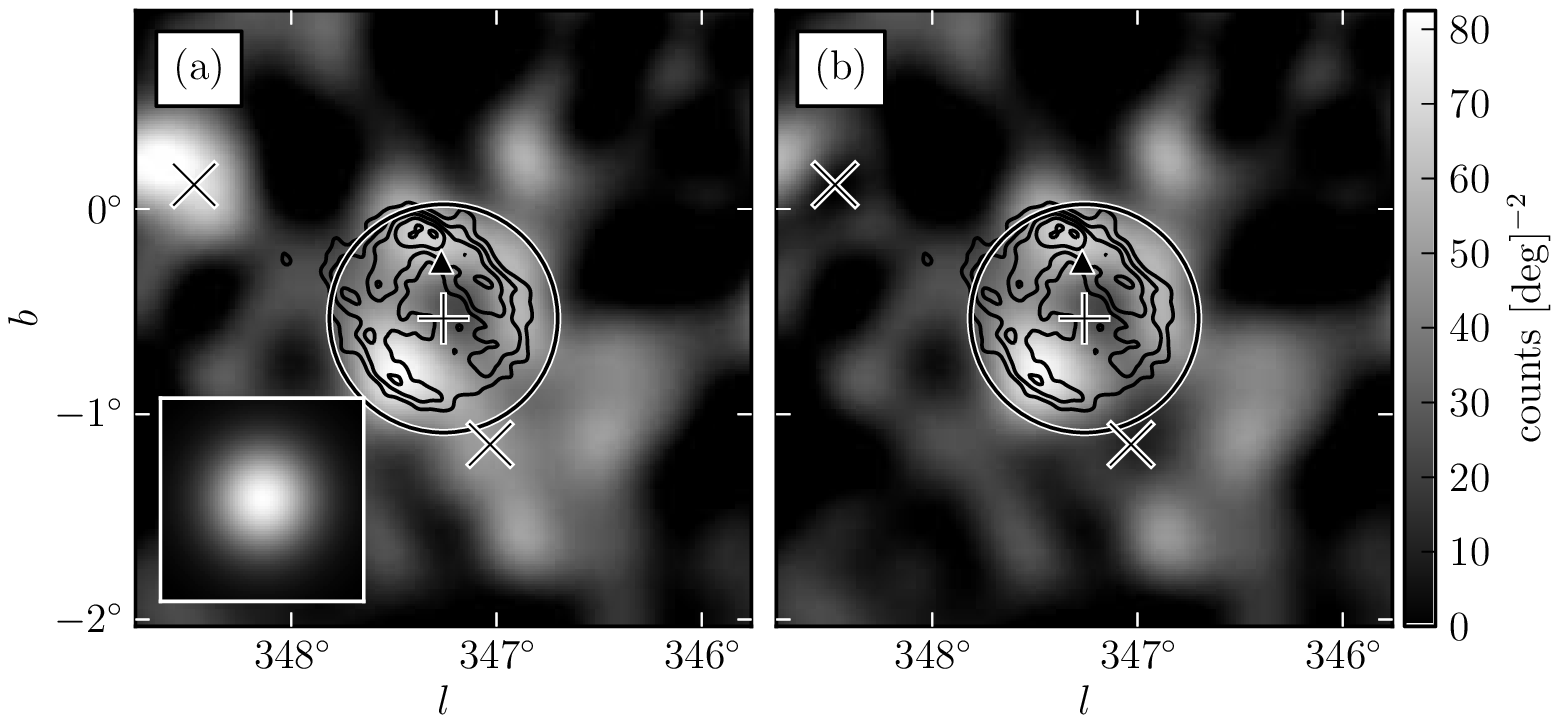}
    \fi
  \caption{
  A diffuse-emission-subtracted 10 \gev to 100 \gev counts map
  of 2FGL\,J1712.4$-$3941 (a) smoothed by a 0\fdg15 2D
  Gaussian kernel and (b) with the emission from the background sources
  subtracted.  This source is spatially coincident with RX\,J1713.7$-$3946
  and was recently studied in \cite{rx_j1713_lat}.  The triangular marker
  (colored red in the online version) represents the 2FGL position of
  this source.  The plus-shaped marker and the circle (colored red) are
  the best fit position and extension of this source assuming a radially
  symmetric uniform disk model.  
  The two black crosses represent background 2FGL sources subtracted in (b).
  The contours (colored light blue)
  correspond to the \tev image \citep{rx_j1713_hess}.  
  }\label{2FGL_J1712.4-3941}
\end{figure}

2FGL\,J1712.4$-$3941 was found by our search to be spatially extended
using photons with energies between 1 \gev and 100 \gev.  This source
is spatially coincident with the SNR RX\,J1713.7$-$3946 and was
recently studied by the LAT Collaboration in \cite{rx_j1713_lat}.
To avoid issues related to uncertainties in the nearby Galactic
diffuse emission at lower energy, we restricted our analysis only
to energies above 10 \gev.  Figure~\ref{2FGL_J1712.4-3941} shows a
smoothed counts map of the source. Above 10 \gev, the \gev emission
nicely correlates with the \tev contours of RX\,J1713.7$-$3946
\citep{rx_j1713_hess} and 2FGL\,J1712.4$-$3941 fit to an extension 
$\sigma=0\fdg56\pm0\fdg04_\stat\pm0\fdg02_\sys$ with $\tsext=38.5$.

\subsection{2FGL\,J1837.3$-$0700c}
\label{section_2FGL_J1837.3-0700c}

\begin{figure}
    \ifcolorfigure
      \plotone{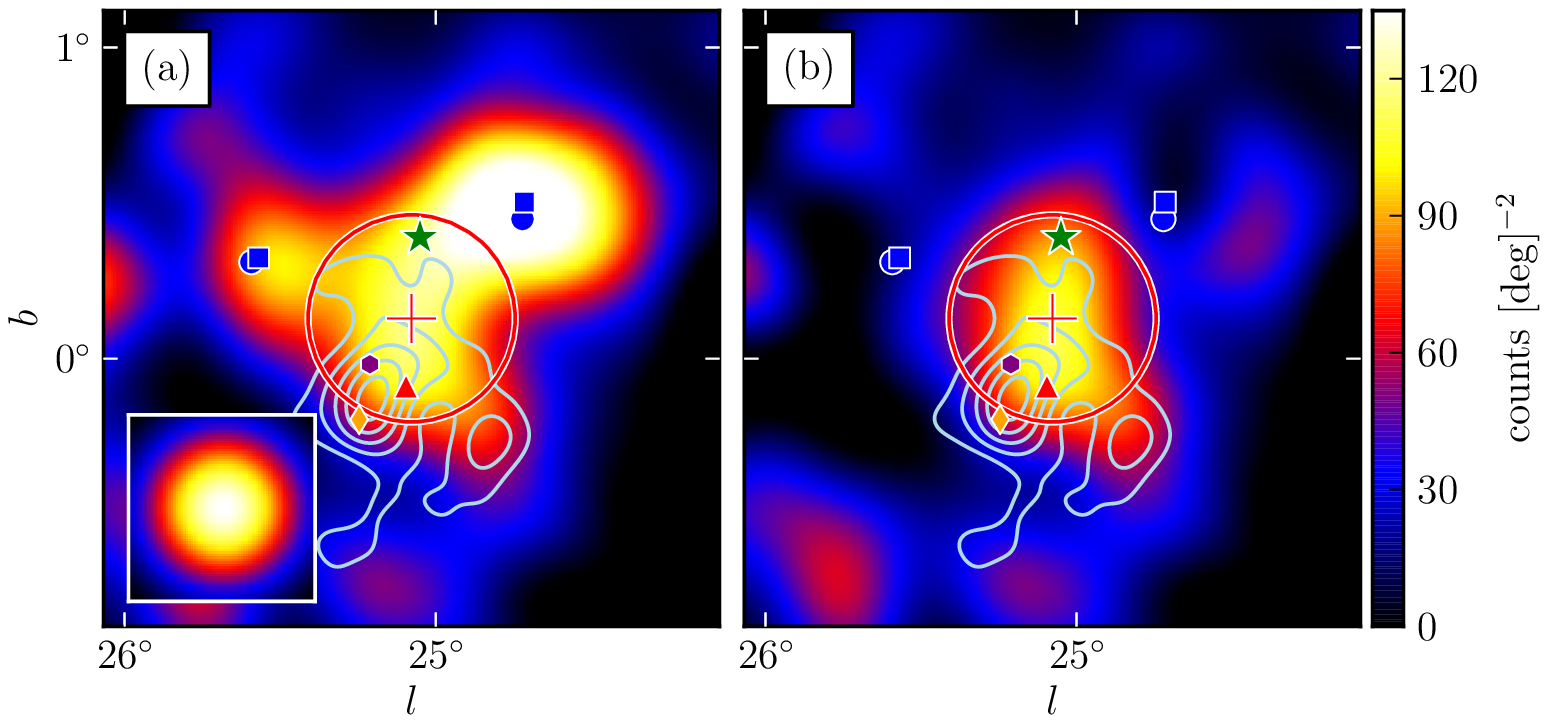}
    \else
      \plotone{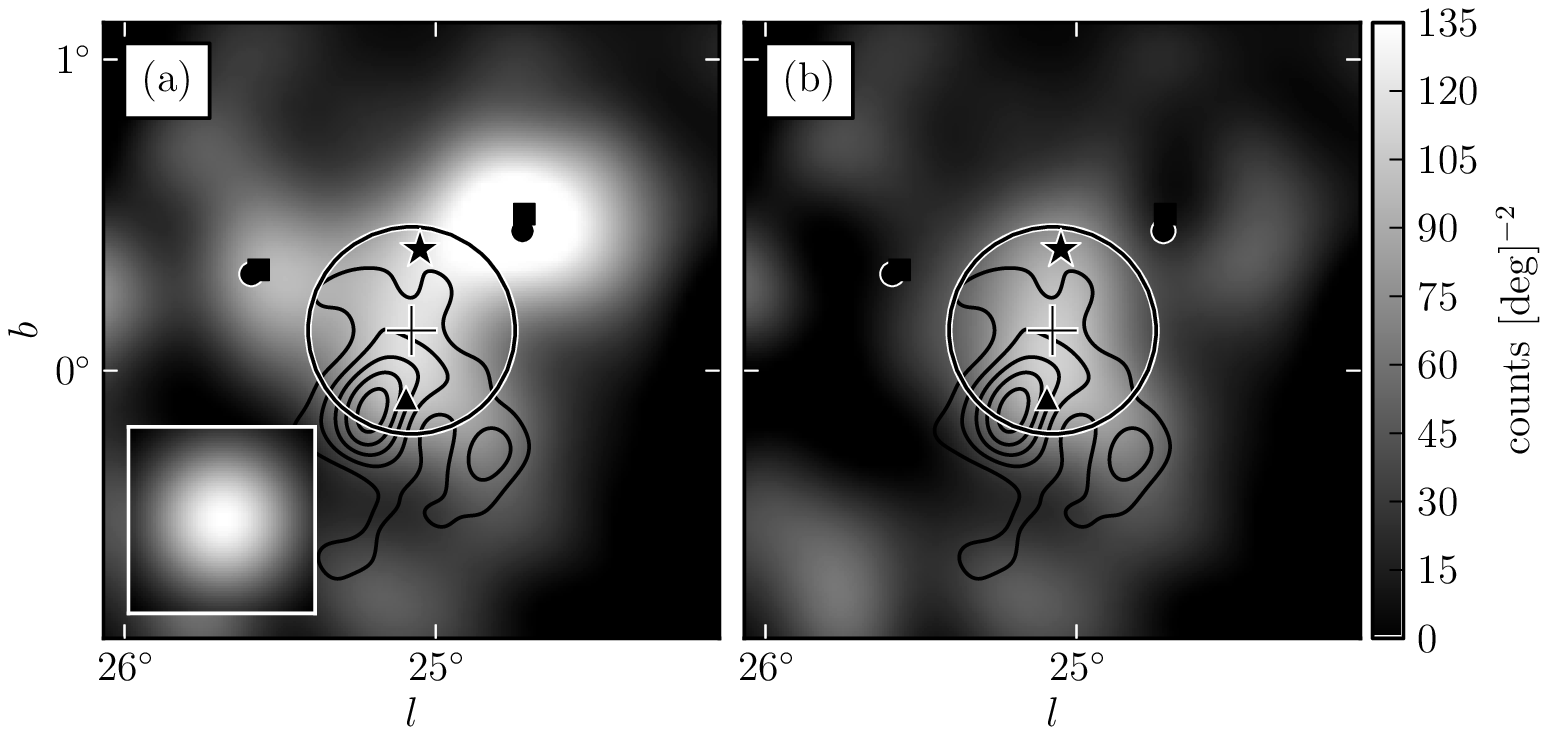}
    \fi
  \caption{
  A diffuse-emission-subtracted 10 \gev to 100 \gev counts map of the
  region around 2FGL\,J1837.3$-$0700c (a) smoothed by a 0\fdg15 2D Gaussian
  kernel and (b) with the emission from the background sources subtracted.
  The triangular marker (colored red in the online version) represents
  the 2FGL
  position of this source. 
  The plus-shaped marker and 
  the circle (colored red) represent the best fit position and extension
  of 2FGL\,J1837.3$-$0700c assuming a radially-symmetric uniform disk model. 
  The circular and square-shaped markers (colored
  blue) represent the 2FGL and the relocalized positions respectively of
  two background 2FGL sources subtracted in (b).  The star-shaped marker
  (colored green) represents 2FGL\,J1835.5$-$0649, another 2FGL source that was removed from the
  background model.  The contours (colored light blue) correspond to
  the \tev image of HESS\,J1837$-$069
  \citep{hess_plane_survey}.
  The diamond-shaped marker (colored orange) represents the position of PSR\,J1838$-$0655
  and the hexagonal-shaped marker (colored purple) represents the position AX\,J1837.3$-$0652
  \citep{pulsations_HESS_J1837-069}.
  }\label{1FGL_J1837.5-0659c}
\end{figure}

2FGL\,J1837.3$-$0700c was found by our search to be an extended source
candidate in the 10 \gev to 100 \gev energy range and is spatially
coincident with the \tev source HESS\,J1837$-$069.  This source is
in a complicated region.  Figure~\ref{1FGL_J1837.5-0659c}a shows a
smoothed counts map of the region and overlays the \tev contours of
HESS\,J1837$-$069 \citep{hess_plane_survey}.  There are two very nearby
point-like 2FGL sources, 2FGL\,J1836.8$-$0623c and 2FGL\,J1839.3$-$0558c,
that clearly represent distinct sources.  On the other hand, there is
another source 2FGL\,J1835.5$-$0649 located between the three sources that
appears to correlate with the \tev morphology of HESS\,J1837$-$069 but
at \gev energies
does not appear to represent a physically distinct source.  We concluded
that this source was included in the 2FGL catalog to compensate for residuals induced by
not modeling the extension of this source and removed it from our model
of the sky.  Figure~\ref{1FGL_J1837.5-0659c}b shows a
counts map of this region after subtracting these background sources.
After removing 2FGL\,J1835.5$-$0649,
we tested for
source confusion by fitting 
2FGL\,J1837.3$-$0700c
instead as two point-like sources.
Because $\tsinc=10.8$ is less than $\tsext=18.5$, we conclude that this emission
does not originate from two nearby point-like sources.

With this model, 2FGL\,J1837.3$-$0700c was found to have an
extension $\sigma=0\fdg33\pm0\fdg07_\stat\pm0\fdg05_\sys$.
While fitting the extension of 2FGL\,J1837.3$-$0700c,
we iteratively relocalized the two closest
background sources along with the extension of 2FGL\,J1837.3$-$0700c but
their positions did not significantly change.  2FGL\,J1834.7$-$0705c
moved to $(l,b)=(24\fdg77,0\fdg50)$, 2FGL\,J1836.8$-$0623c moved
to $(l,b)=(25\fdg57,0\fdg32)$. 

H.E.S.S. measured an extension of
$\sigma=0\fdg12\pm0\fdg02$ and $0\fdg05\pm0\fdg02$ 
of the coincident \tev source HESS\,J1837$-$069 
along the semi-major and semi-minor axis when fitting this source
with an elliptical 2D Gaussian surface brightness profile.  This corresponds
to a 68\% containment radius of $\rsixeight=0\fdg18\pm0\fdg03$ and
$0\fdg08\pm0\fdg03$ along the semi-major and semi-minor axis. The
size is not significantly different from the LAT 68\% containment
radius of $\rsixeight=0\fdg27\pm0\fdg07$ (less than $2\sigma$).
Figure~\ref{hess_seds} shows that the spectrum of 2FGL\,J1837.3$-$0700c
at \gev energies connects to the spectrum of HESS\,J1837$-$069 at \tev
energies.

HESS\,J1837$-$069 is coincident with the hard and steady X-ray source
AX\,J1838.0$-$0655 \citep{hard_x-ray_asca}.  This source was discovered
by RXTE to be a pulsar (PSR J1838-0655) 
sufficiently luminous to power the \tev emission
and was resolved by \chandra to be a bright point-like source surrounded
by a $\sim2\arcmin$ nebula \citep{pulsations_HESS_J1837-069}. The
$\gamma$-ray emission may be powered by this pulsar.  The hard spectral
index and spatial extension of 2FGL\,J1837.3$-$0700c disfavor a pulsar
origin of the LAT emission and suggest instead that the \gev and \tev
emission both originate from the pulsar's wind.  There is another
X-ray point-like source AX\,J1837.3$-$0652 near HESS\,J1837$-$069
\citep{hard_x-ray_asca} that was also resolved into a point-like
and diffuse component \citep{pulsations_HESS_J1837-069}.  Although no
pulsations have been detected from it, it could also be a pulsar powering
some of the $\gamma$-ray emission.

\subsection{2FGL\,J2021.5+4026}
\label{section_2FGL J2021.5+4026}


The source 2FGL\,J2021.5+4026 is associated with the $\gamma$-Cygni SNR 
and has been speculated
to originate from the interaction of accelerated particles in the SNR
with dense molecular clouds \citep{pollock_1985,gaisser_1998}. This
association was disfavored when the \gev emission from this source
was detected to be pulsed \citep[PSR\,J2021+4026,][]{first_lat_pulsar_cat}.
This pulsar was also observed by AGILE \citep{gamma_cygni_agile}.

\begin{figure}
    \ifcolorfigure
      \plotone{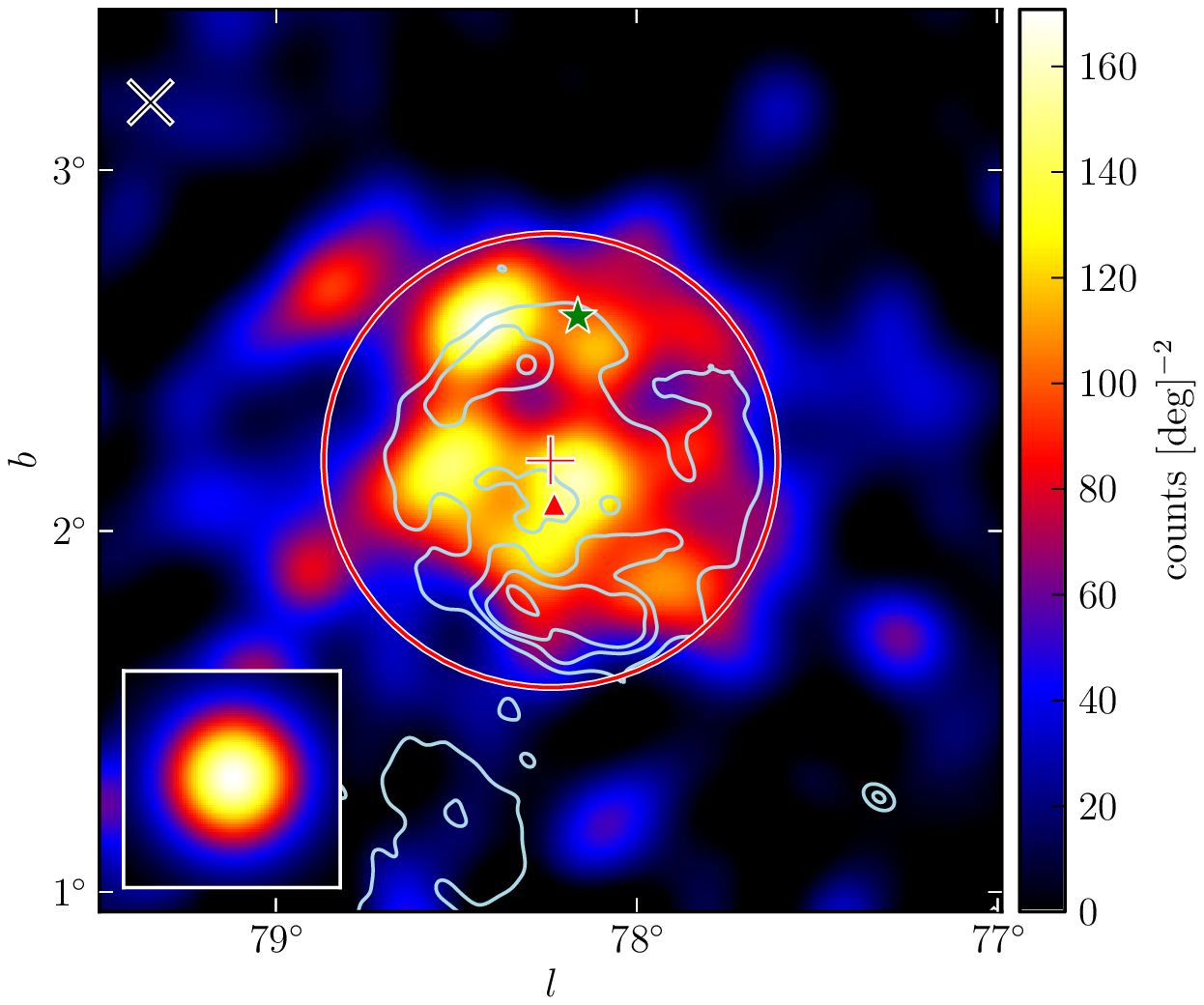}
    \else
      \plotone{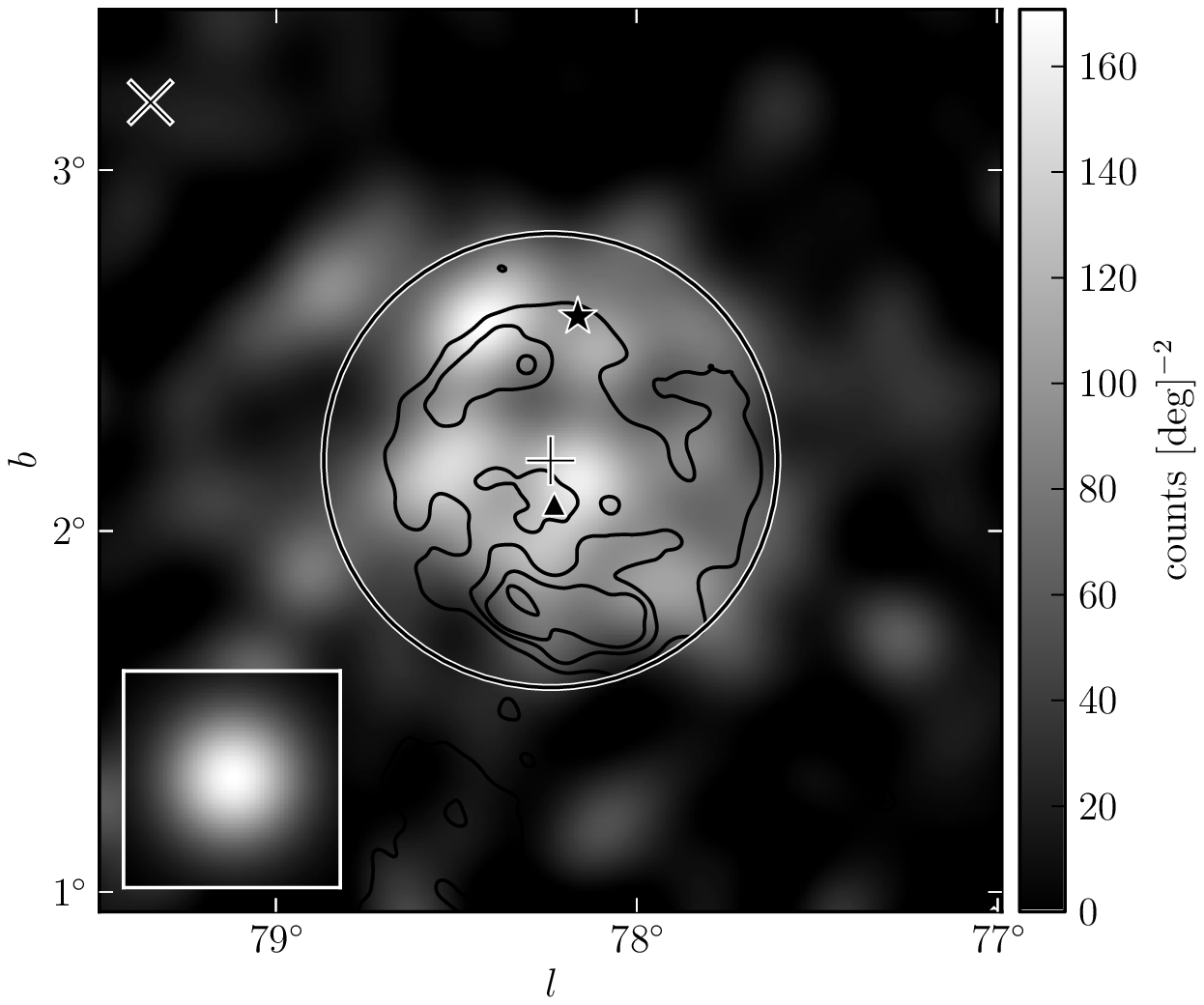}
    \fi
  \caption{A diffuse-emission-subtracted 
  10 \gev to 100 \gev counts map of the
  region around 2FGL\,J2021.5+4026 smoothed by a 0\fdg1 2D Gaussian
  kernel. The triangular marker (colored red in the online version)
  represents the 2FGL position of this source.  The plus-shaped
  marker and the circle (colored red) represent the best fit position
  and extension of 2FGL\,J2021.5+4026 assuming a radially-symmetric
  uniform disk model.  
  The star-shaped marker (colored green)
  represents 2FGL\,J2019.1+4040,
  a 2FGL source that was removed from the background model.
  2FGL\,J2021.5+4026
  is spatially coincident with the $\gamma$-Cygni SNR.  The contours
  (colored light blue) correspond to the 408MHz image of $\gamma$-Cygni
  observed by the Canadian Galactic Plane Survey \citep{canadian_galactic_plane_survey}.
  }\label{1FGL_J2020.0+4049}
\end{figure}

Looking at the same region at energies above 10 \gev, the pulsar is
no longer significant but we instead found in our search an extended
source candidate.  Figure~\ref{1FGL_J2020.0+4049} shows a counts map
of this source and overlays radio contours of $\gamma$-Cygni from the
Canadian Galactic Plane Survey \citep{canadian_galactic_plane_survey}.
There is good spatial overlap between the SNR and the \gev emission.

There is a nearby source 2FGL\,J2019.1+4040 that correlates with the radio
emission of $\gamma$-Cygni and at \gev energies 
does not appear to represent a physically
distinct source.  We concluded that it was included in the 2FGL catalog to compensate
for residuals induced by not modeling the extension of $\gamma$-Cygni and
removed it from our model of the sky.  With this model, 2FGL\,J2021.5+4026
was found to have an extension $\sigma=0\fdg63\pm0\fdg05_\stat\pm0\fdg04_\sys$
with $\tsext=128.9$.  Figure~\ref{snr_seds}
shows its spectrum.  The inferred size of this source at \gev energies
well matches the radio size of $\gamma$-Cygni.  Milagro detected
a $4.2\sigma$ excess at energies $\sim 30$ \tev from this location
\citep{lat_bsl,milagro_bright_source_list}.  VERITAS also detected an
extended source VER\,J2019+407 coincident with the SNR above 200 \gev
and suggested that the \tev emission could be a shock-cloud interaction
in $\gamma$-Cygni \citep{veritas_gamma_cygni}.




\section{Discussion}

\begin{figure}
      \ifcolorfigure
      \plotone{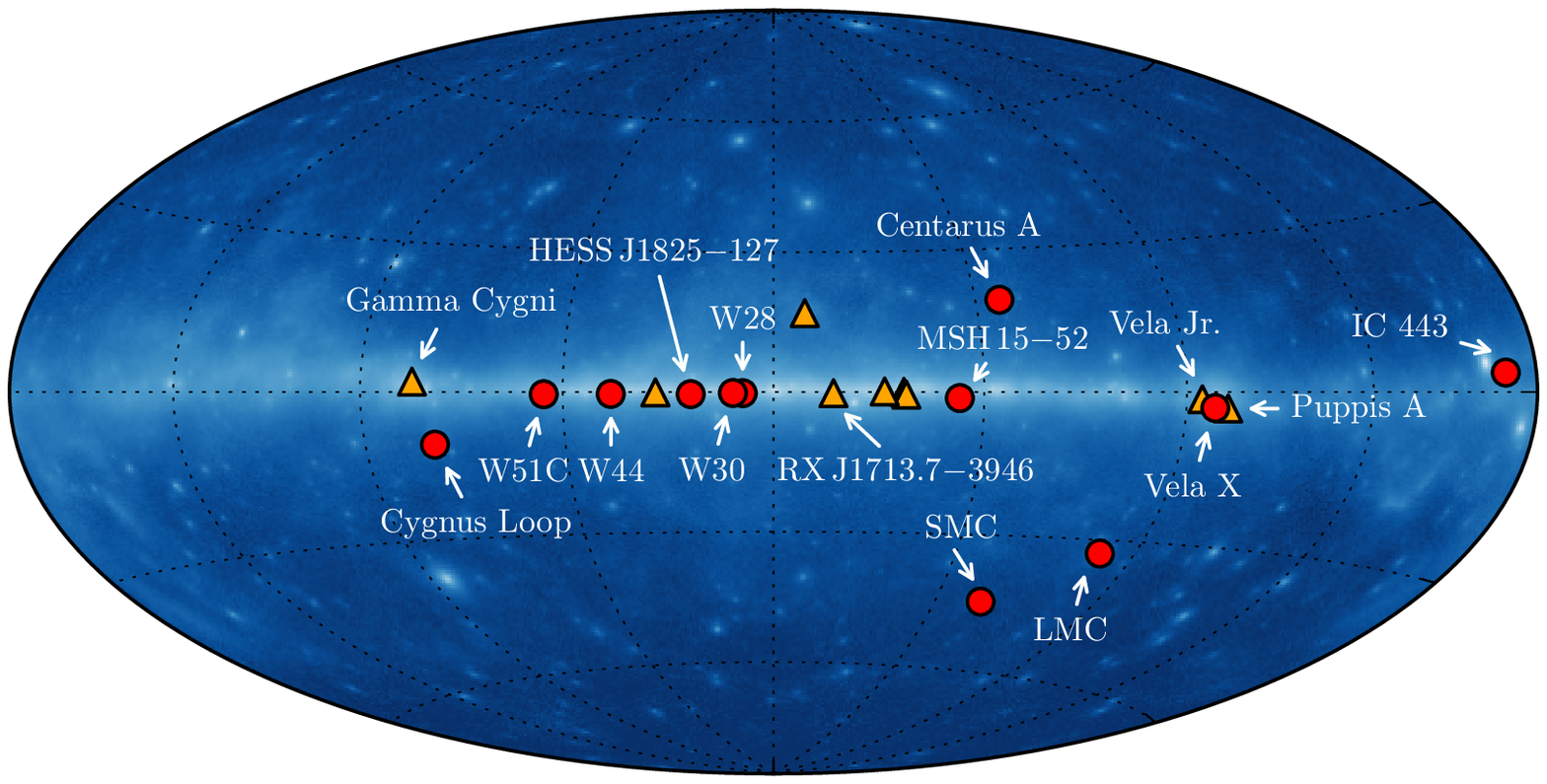}
      \else
      \plotone{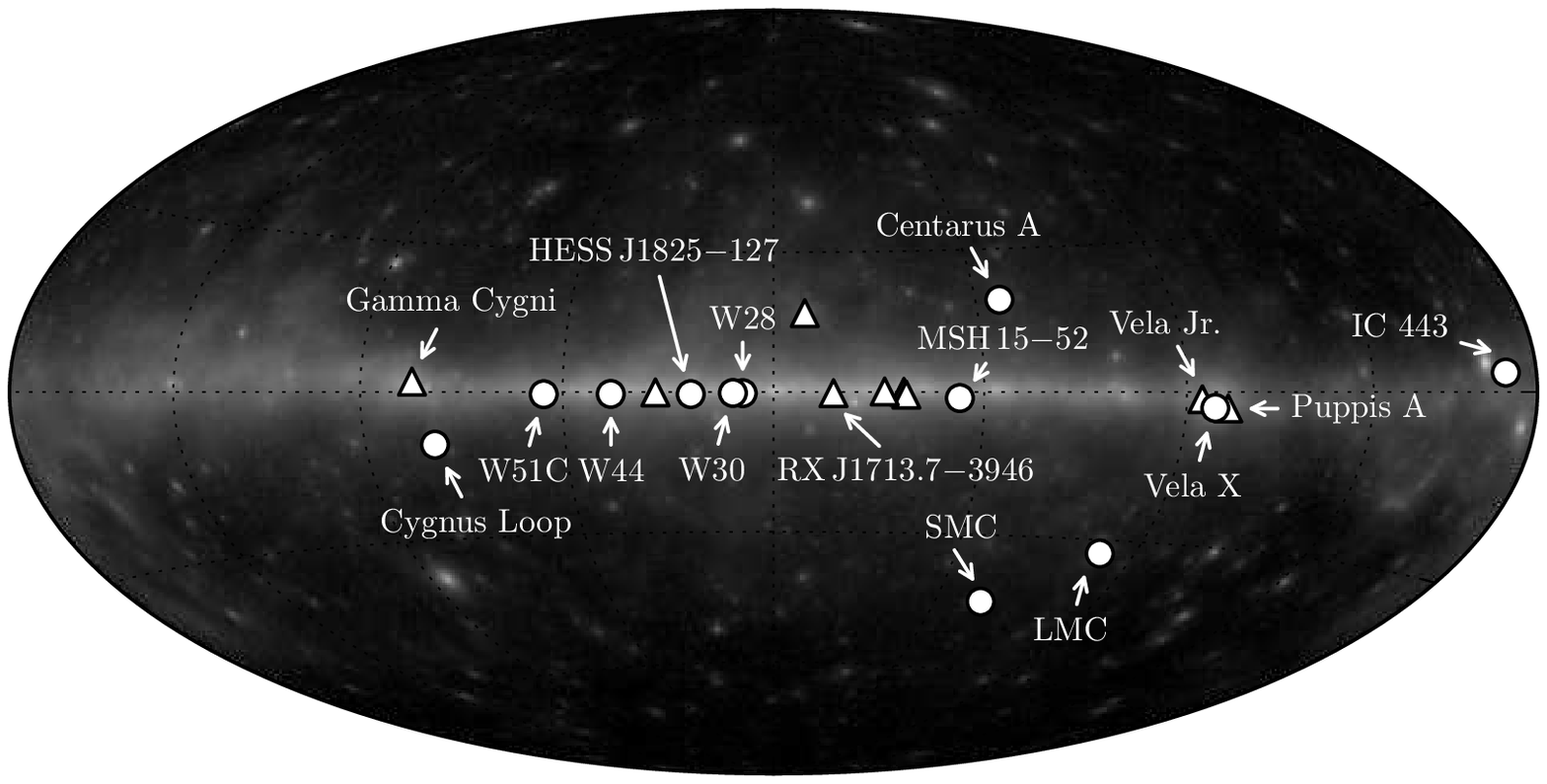}
      \fi
      \caption{The 21
      spatially extended sources detected by the LAT
      at \gev energies 
      with 2 years of data.  The twelve extended sources included in
      2FGL are represented by the circular markers (colored red in the online
      version).  The nine new extended sources are represented by
      the triangular markers (colored orange).
      The source positions are overlaid on a 100 \mev to 100 \gev 
      Aitoff projection sky map of the LAT data in Galactic coordinates.
}
\label{allsky_extended_sources}
  \end{figure}

Twelve extended sources were included in the 2FGL catalog and two additional extended
sources were studied in dedicated publications.  Using 2 years of
LAT data and a new analysis method, we presented the detection of seven
additional extended sources.  We also reanalyzed the spatial extents of the
twelve extended sources in the 2FGL catalog and the two additional sources.  The 21
extended LAT sources are located primarily along the Galactic plane
and their locations are shown in Figure~\ref{allsky_extended_sources}.
Most of the LAT-detected extended sources are expected to be of Galactic
origin as the distances of extragalactic sources (with the exception of
the local group Galaxies) are typically too large to be able to resolve
them at $\gamma$-ray energies.

\begin{figure}
    \ifcolorfigure
      \plotone{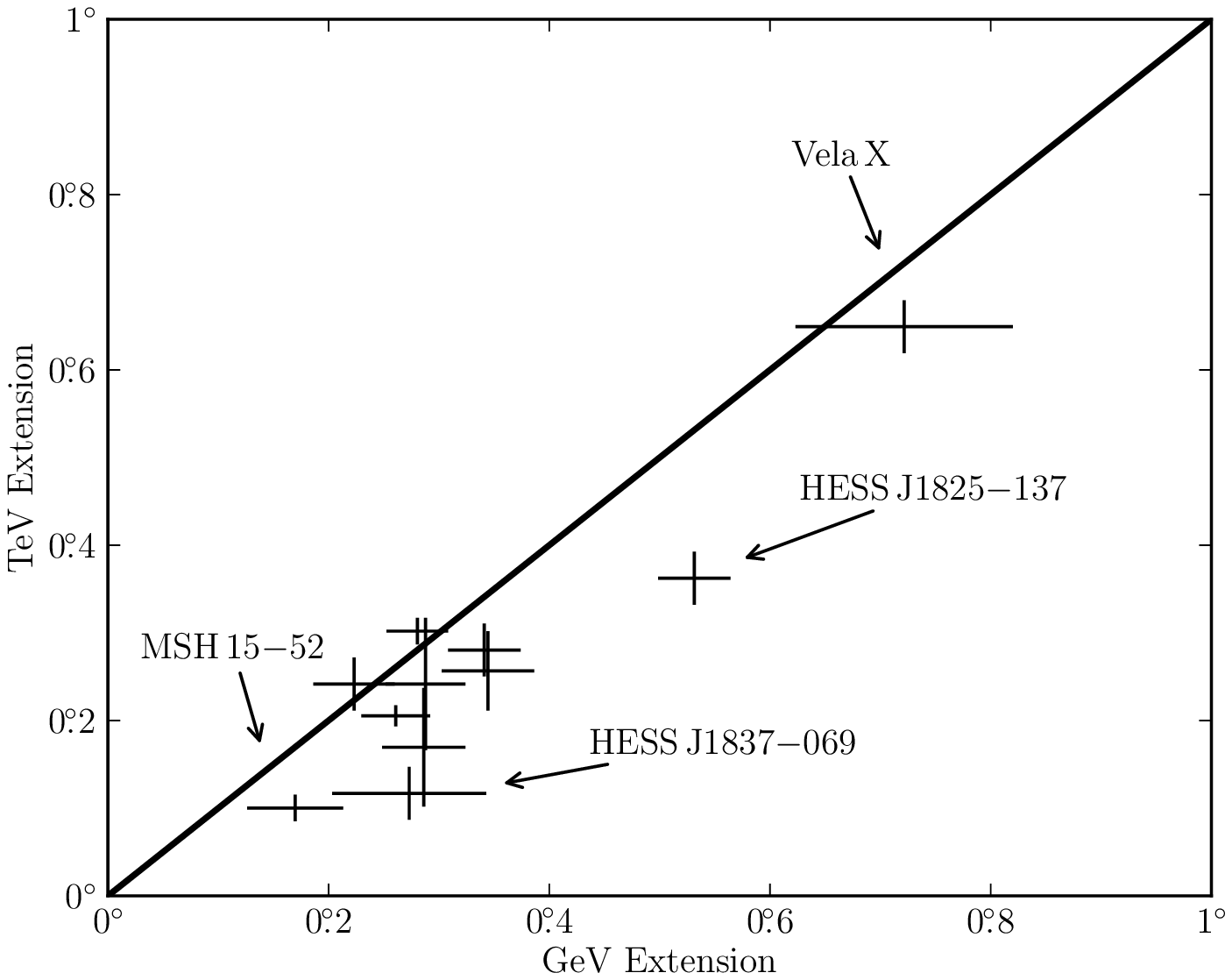}
    \else
      \plotone{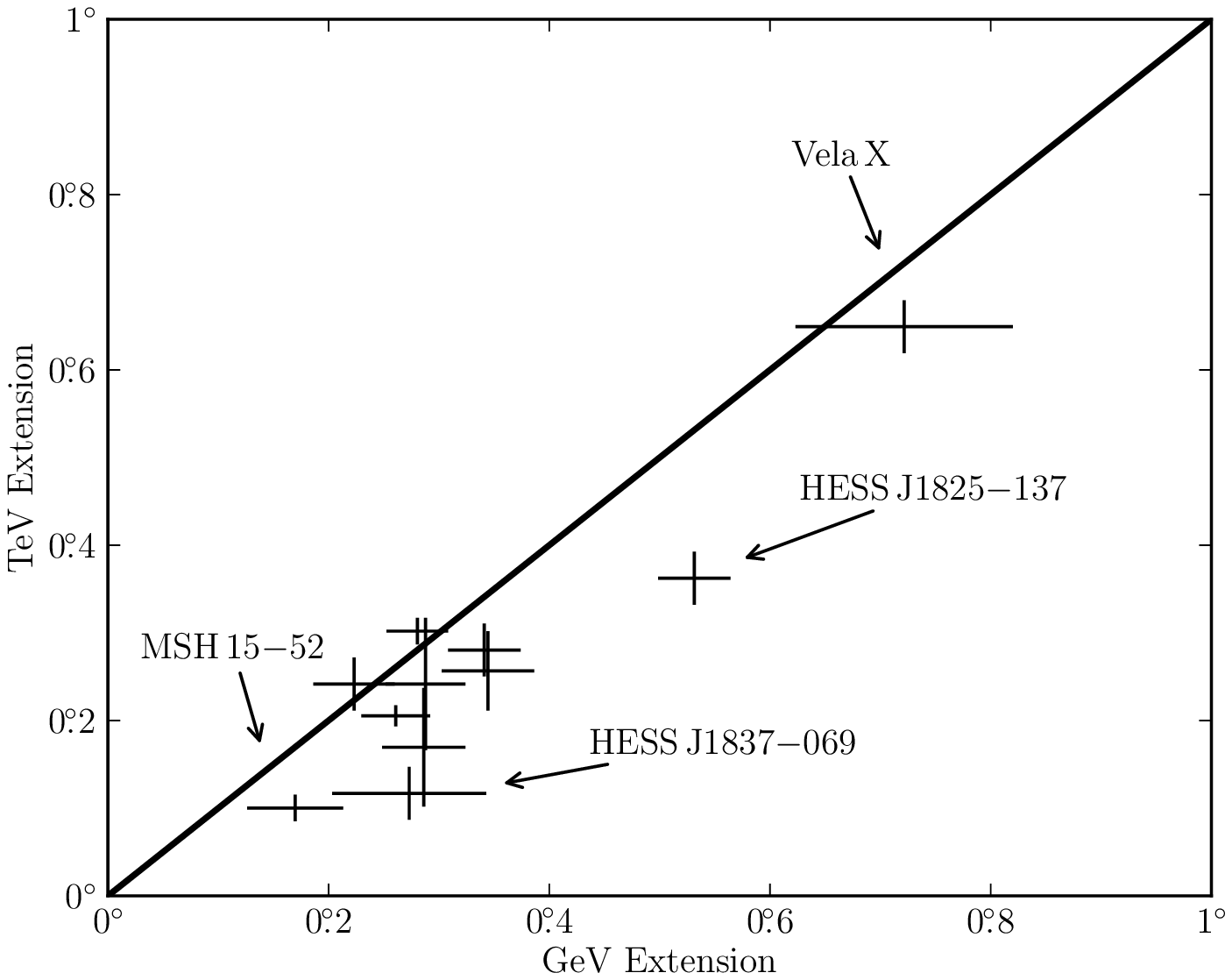}
      \fi
    \caption{
    A comparison of the sizes of extended sources detected at
    both \gev and \tev energies.  The \tev sizes of W30,
    2FGL\,J1837.3$-$0700c, 2FGL\,J1632.4$-$4753c, 2FGL\,J1615.0$-$5051,
    and 2FGL\,J1615.2$-$5138 are from \cite{hess_plane_survey}.
    The \tev sizes of MSH\,15$-$52, HESS\,J1825$-$137,
    Vela X, Vela Jr., RX\,J1713.7$-$3946 and W28 are from
    \cite{msh_15_52_hess,hess_j1825_hess,vela_x_hess,vela_jr_hess,rx_j1713_hess,w28_hess}.
    The \tev size of IC~443 is from \cite{ic443_veritas} and
    W51C is from \cite{w51c_with_magic_at_fermi_symposium}.  The \tev
    sizes of MSH\,15$-$52, HESS\,J1614$-$518, HESS\,J1632$-$478, and
    HESS\,J1837$-$069 have only been reported with an elliptical 2D
    Gaussian fit and so the plotted sizes are the geometric 
    mean of the semi-major and semi-minor axis.
    The LAT extension of
    Vela X is from \cite{velax}. 
    The \tev sources were fit assuming a 2D Gaussian surface brightness profile
    so the plotted \gev and \tev extensions were first converted to
    \rsixeight (see Section~\ref{compare_source_size}).  
    Because of
    their large sizes, the shape of RX\,J1713.7$-$3946 and Vela Jr.
    were not directly fit at \tev energies and so are not included
    in this comparison. On the other hand, dedicated publications by the 
    LAT collaboration on
    these sources showed that their morphologies are consistent
    \citep{rx_j1713_lat,vela_jr_lat}.
    The LAT
    extension errors are the statistical and systematic errors added
    in quadrature. 
}\label{gev_vs_tev_plot}
  \end{figure}

For the LAT extended sources also seen at \tev energies,
Figure~\ref{gev_vs_tev_plot} shows that there is a good correlation
between the sizes of the sources at \gev and \tev energies. Even so,
the sizes of PWNe are expected to vary across the \gev and \tev energy
range and the size of HESS\,J1825$-$137 is significantly larger at
\gev than \tev energies \citep{fermi_hess_j1825}.  It is interesting
to compare the sizes of other PWN candidates at \gev and \tev energies,
but definitively measuring a difference in size would require a more in-depth analysis
of the LAT data using the same elliptical Gaussian spatial model.

\begin{figure}
    \ifcolorfigure
      \plotone{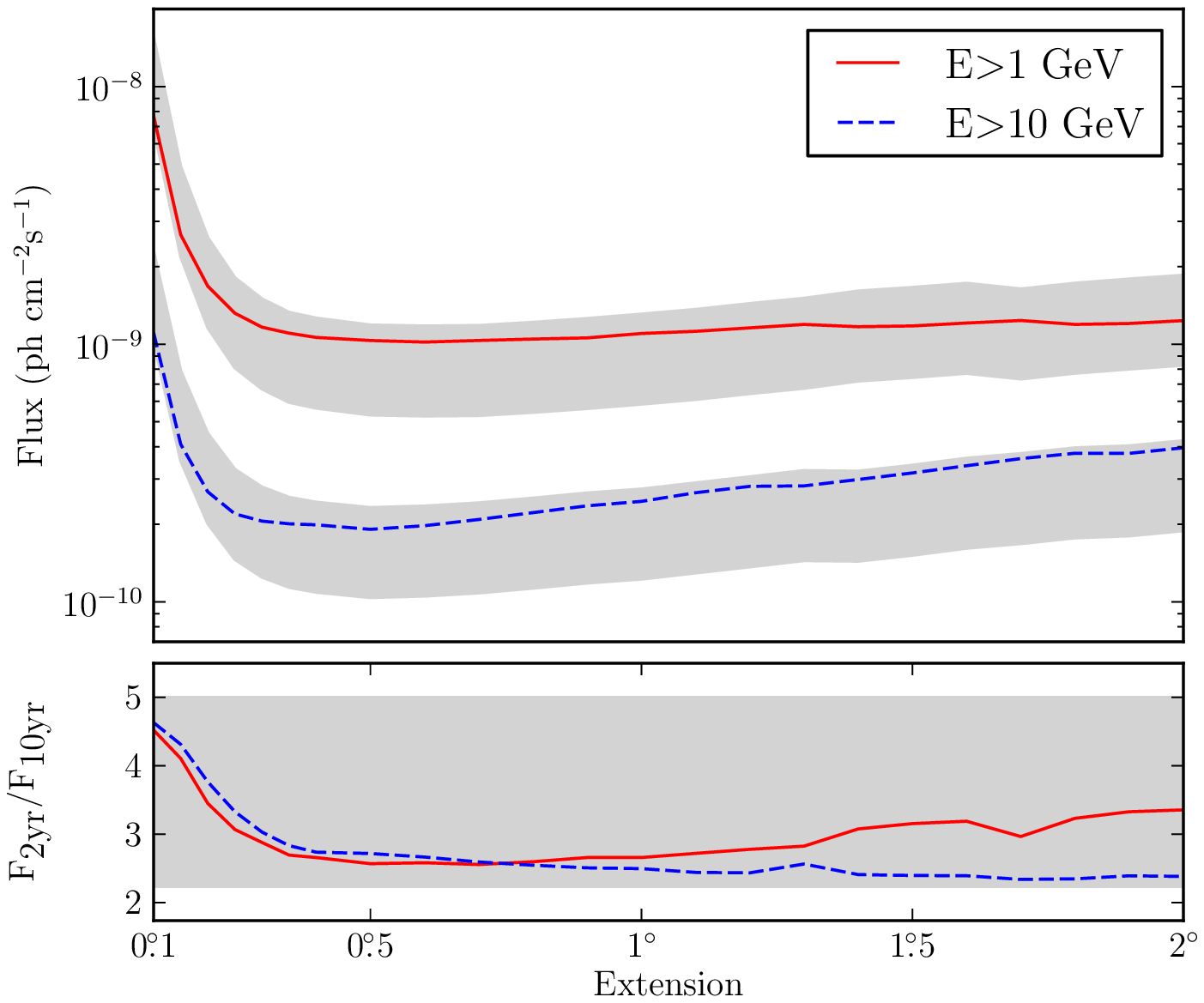}
    \else
      \plotone{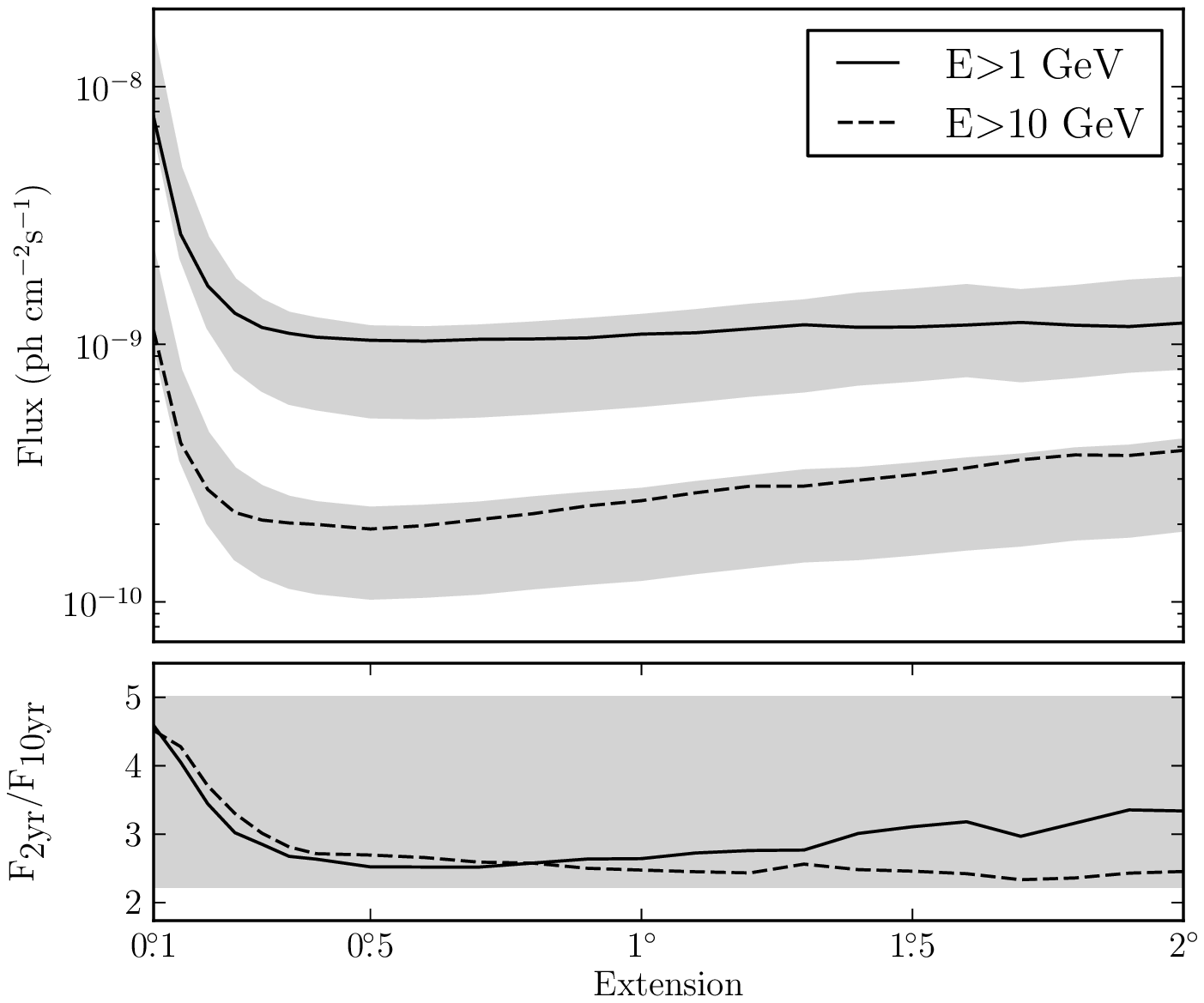}
    \fi
    \caption{
    The projected detection threshold of the LAT to extension after 10 years
    for a power-law
    source of spectral index 2 against 10 times the isotropic background
    in the energy range from 1 \gev to 100 \gev (solid line colored red in the electronic
    version) and 10 \gev to 100 \gev
    (dashed line colored blue). The shaded gray regions represent the
    detection threshold assuming the sensitivity improves from 2 to 10
    years by the square root of the exposure (top edge) and linearly with
    exposure (bottom edge).  The lower plot shows the factor increase in
    sensitivity.  For small extended sources, the detection threshold of the LAT
    to the extension of a source will improve by a factor larger than the square root of 
    the exposure.
    }\label{time_sensitivity}
  \end{figure}

\begin{figure}
    \ifcolorfigure
      \plotone{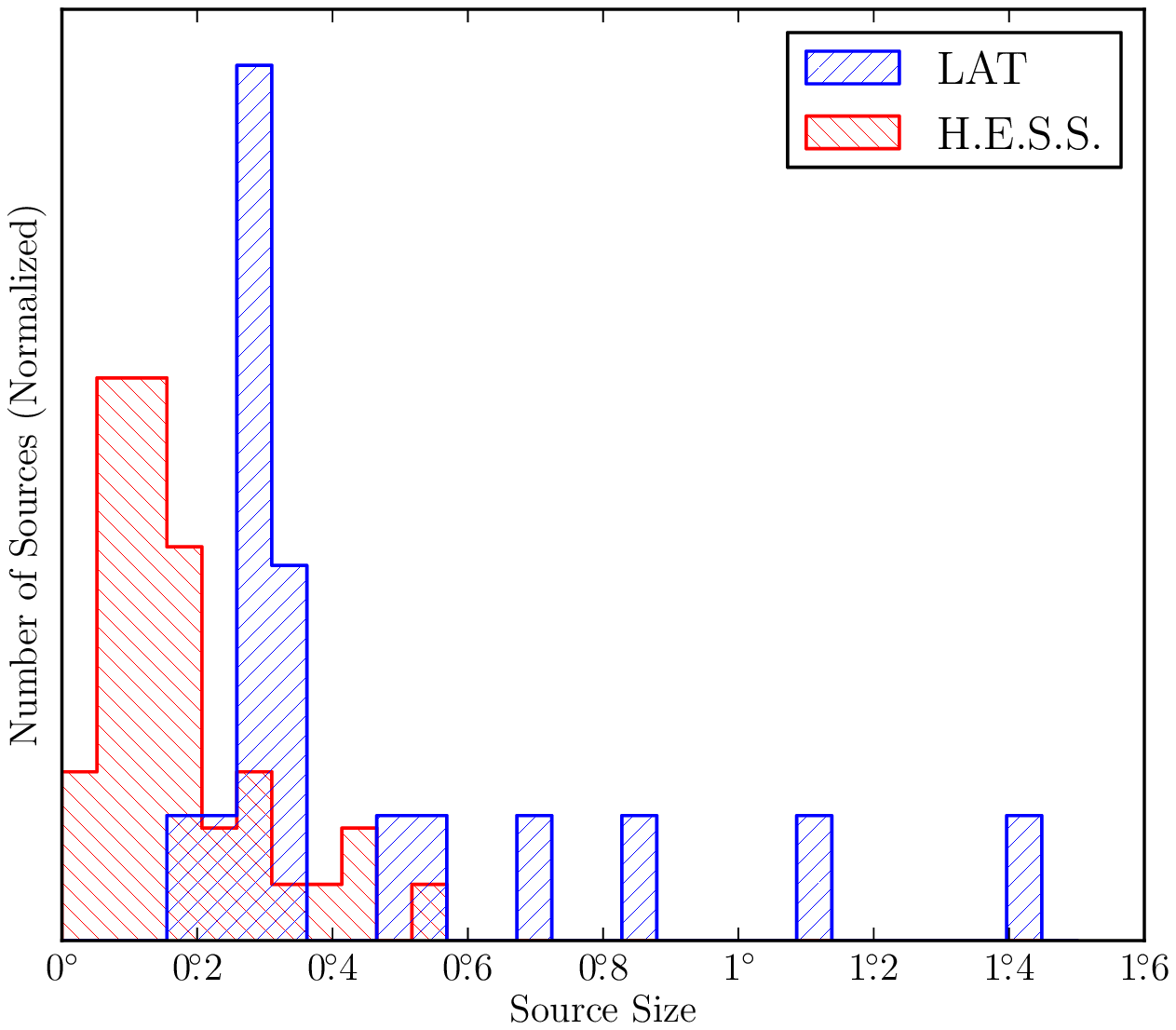}
    \else
      \plotone{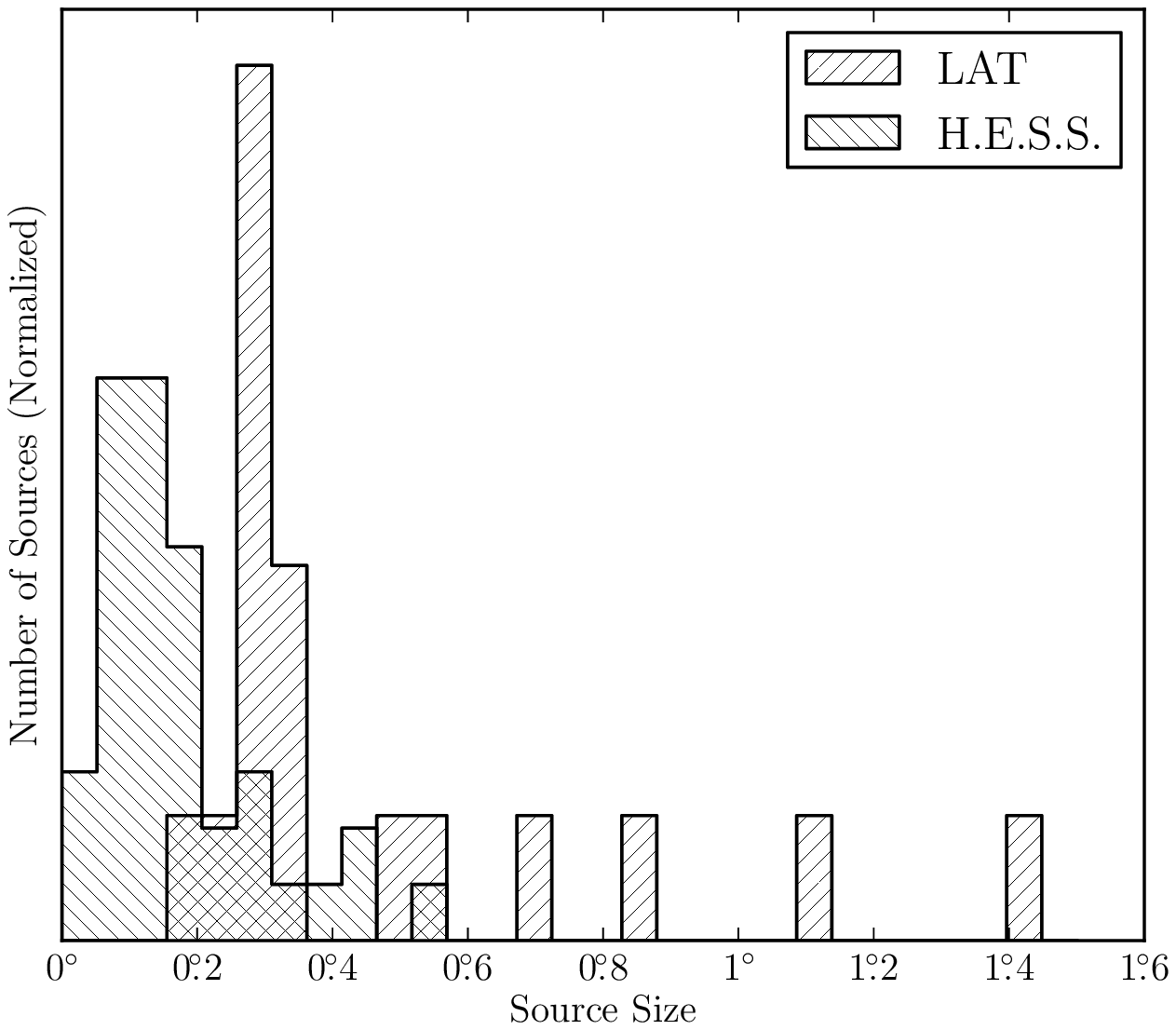}
    \fi
    \caption{
    The distributions of the sizes of 18 extended LAT sources
    at \gev energies
    (colored blue in the electronic version) and the sizes of the
    40 extended H.E.S.S. sources at \tev energies
    (colored red).  
    The H.E.S.S. sources were fit with a 2D Gaussian surface
    brightness profile so the LAT and H.E.S.S. sizes were first converted
    to \rsixeight. 
    The \gev size of Vela X is taken from \cite{velax}.  
    Because of their large sizes, the shape of RX\,J1713.7$-$3946 and
    Vela Jr. were not directly fit at \tev energies
    and are not included in this comparison.
    Centaurus A is not included because of its large size.
    }\label{gev_vs_tev_histogram}
  \end{figure}

Figure~\ref{gev_vs_tev_histogram} compares the sizes of the 21 extended
LAT sources to the 42 extended H.E.S.S. sources.\footnote{The 
\tev extension of
the 42 extended H.E.S.S. sources comes from the H.E.S.S. Source
Catalog \url{http://www.mpi-hd.mpg.de/hfm/HESS/pages/home/sources/}.}
Because of the large
field of view and all-sky coverage, the LAT can more easily measure
larger sources.  On the other hand, the 
better
angular resolution of air Cherenkov detectors allows them to measure a
population of extended sources below the resolution limit of the LAT (currently 
about $\sim0\fdg2$).  \fermi has a 5 year nominal mission lifetime with
a goal of 10 years of operation.  As Figure~\ref{time_sensitivity} shows,
the low background of the LAT at high energies allows its sensitivity 
to
these smaller sources to improve by a factor greater than the square root
of the relative exposures.  With increasing exposure, the LAT will likely begin to
detect and resolve some of these smaller \tev sources.

\begin{figure}
    \ifcolorfigure
      \plotone{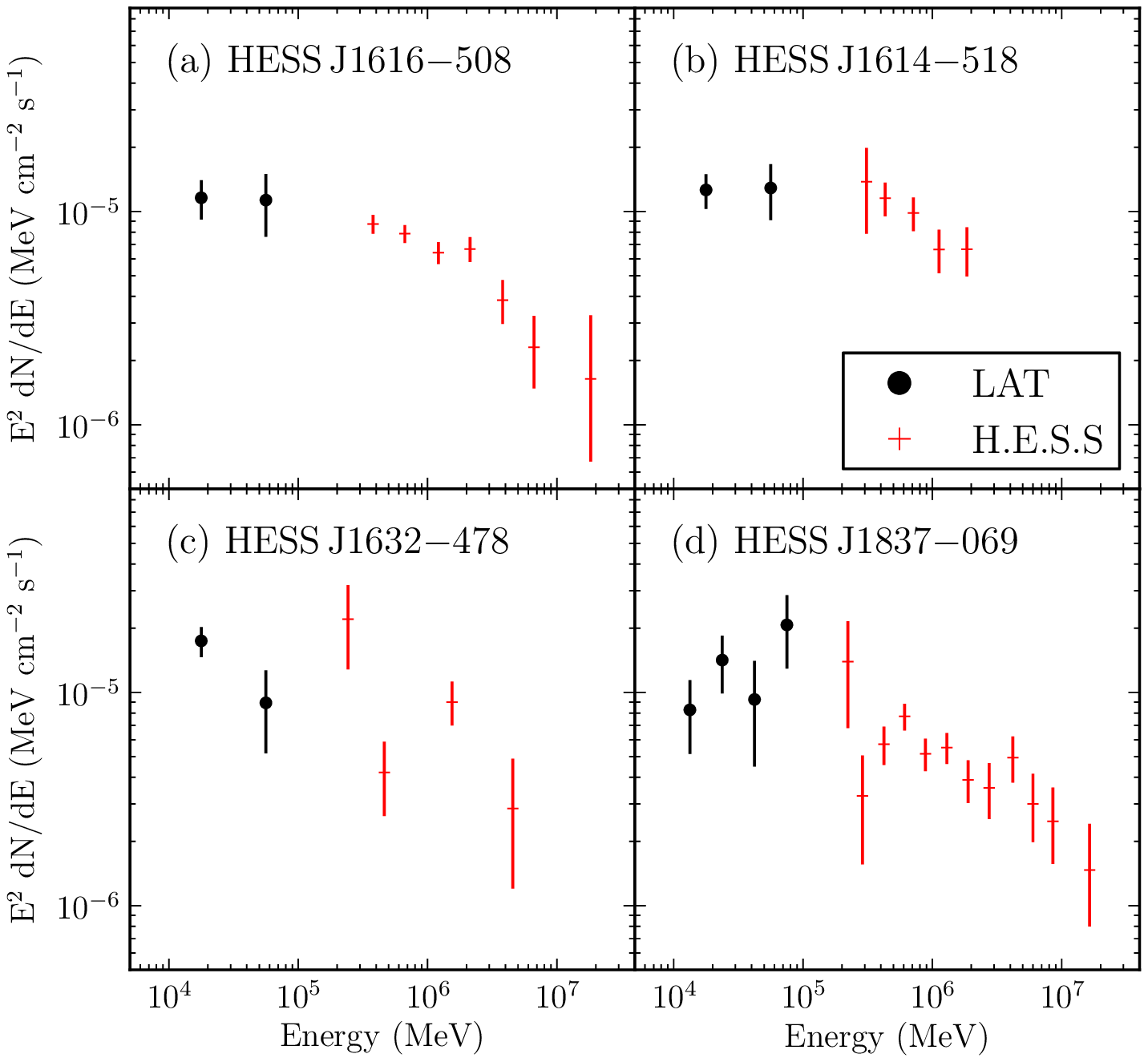}
    \else
      \plotone{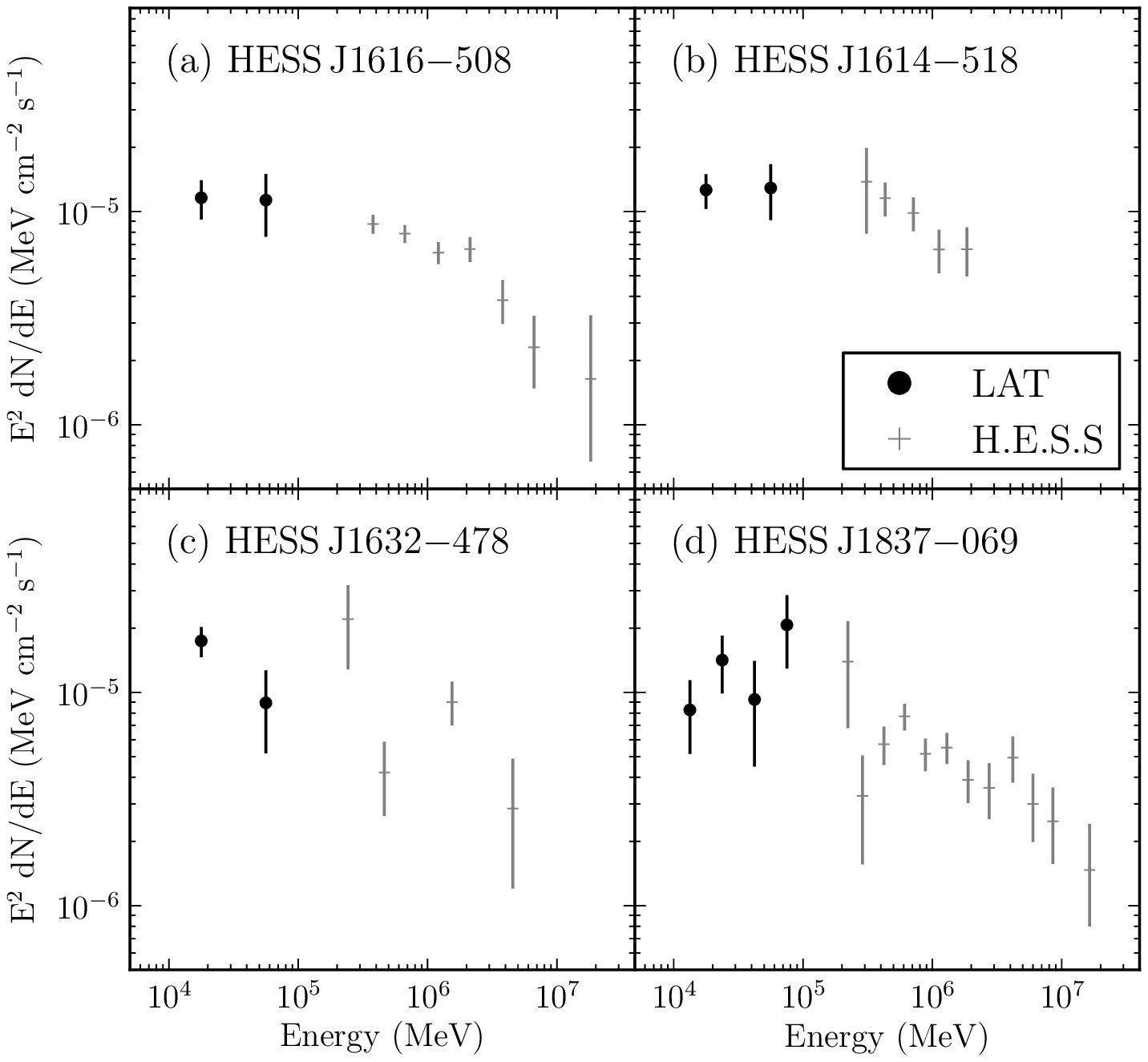}
    \fi
    \caption{
    The spectral energy distribution of four extended
    sources associated with unidentified
    extended \tev sources.  The black points
    with circular markers are obtained by the LAT. The points with
    plus-shaped markers (colored red in the electronic version) are
    for the associated H.E.S.S sources.  (a) the
    LAT SED of 2FGL\,J1615.0$-$5051 together with the H.E.S.S. SED
    of HESS\,J1616$-$508. (b) 2FGL\,J1615.2$-$5138
    and HESS\,J1614$-$518. (c) 2FGL\,J1632.4$-$4753c
    and HESS\,J1632$-$478. (d) 2FGL\,J1837.3$-$0700c
    and HESS\,J1837$-$069. The H.E.S.S. data points are from
    \citep{hess_plane_survey}. Both LAT and H.E.S.S. spectral errors are
    statistical only.}
    \label{hess_seds}
  \end{figure}

\begin{figure}
    \ifcolorfigure
      \plotone{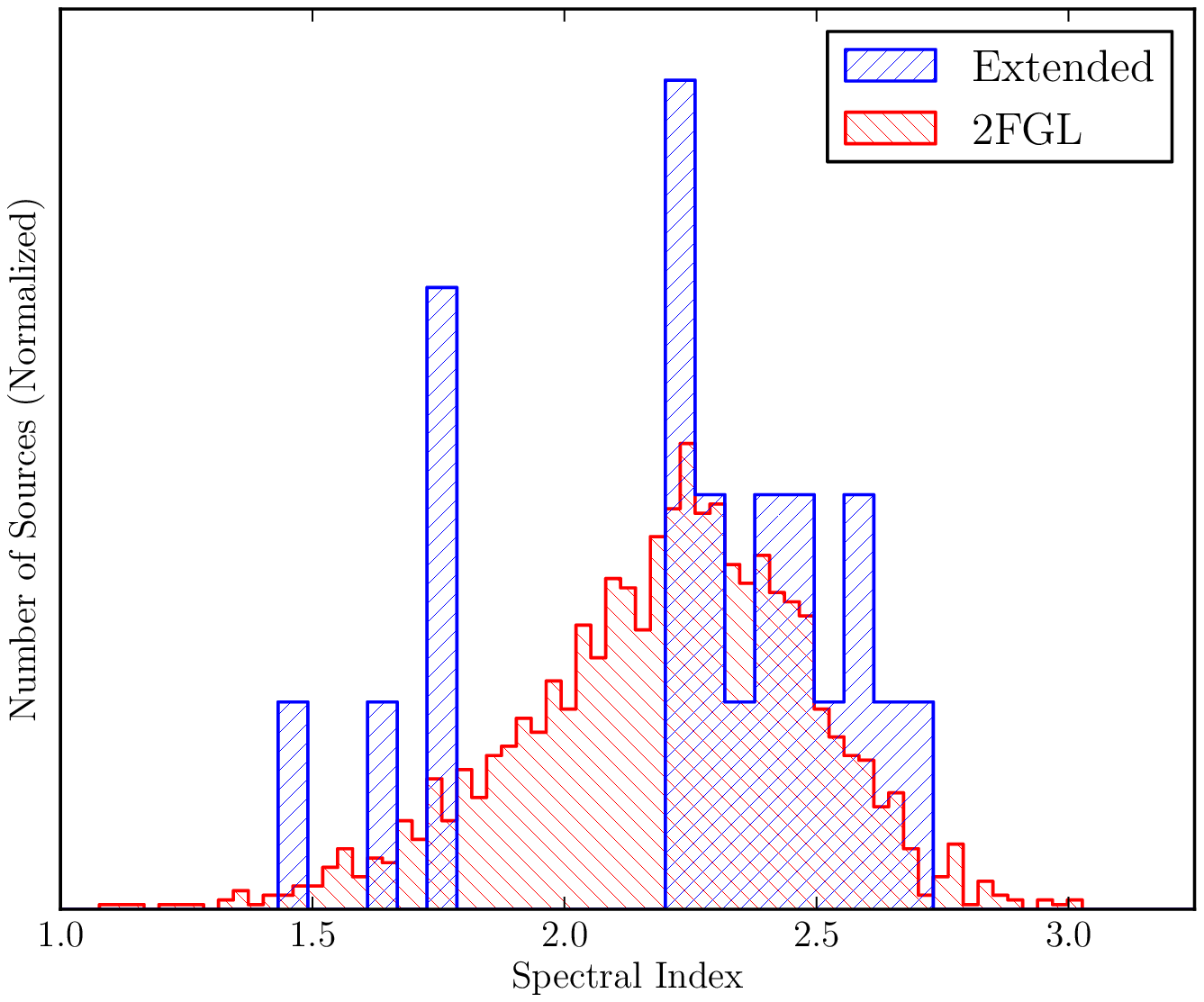}
    \else
      \plotone{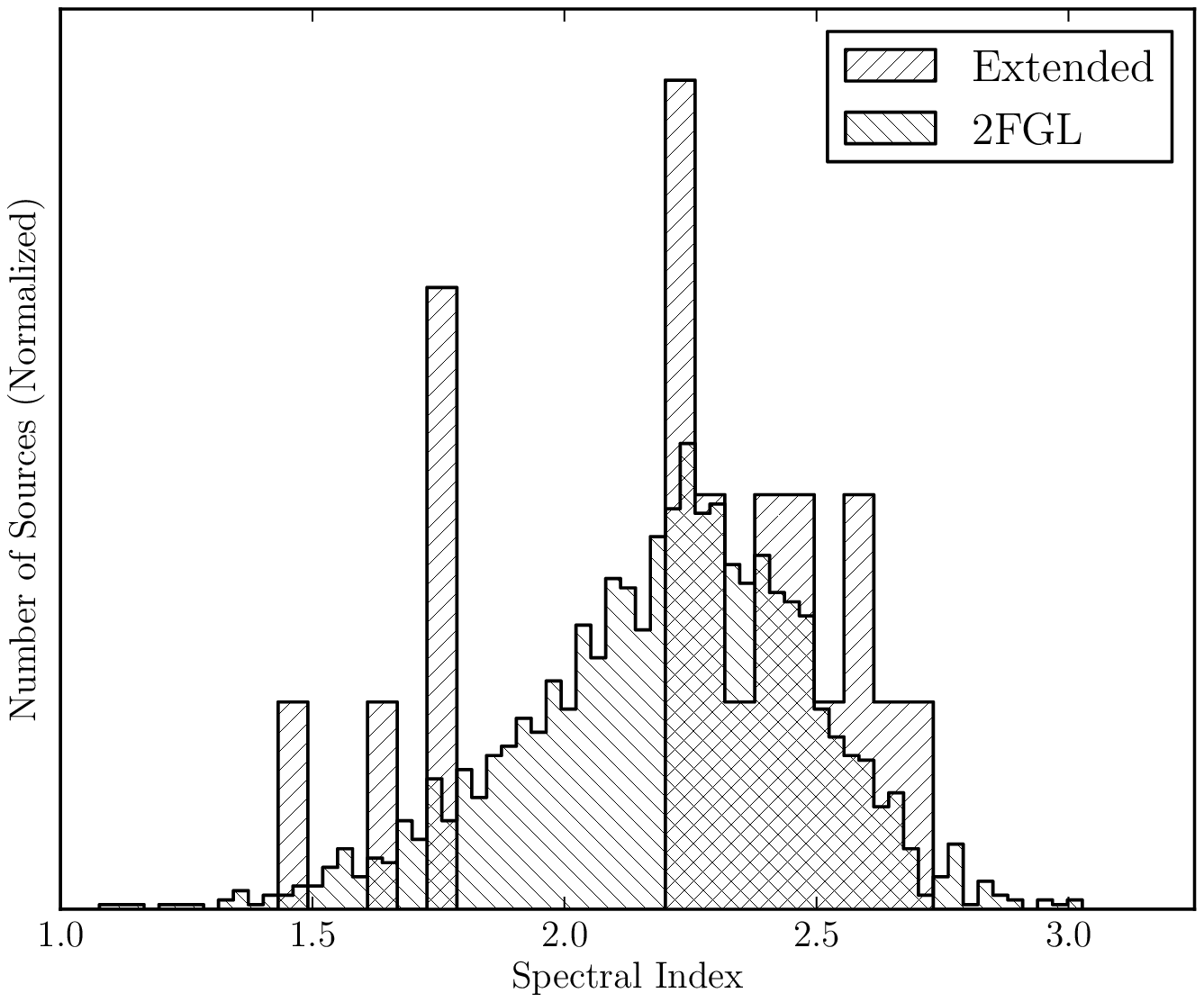}
    \fi
    \caption{
    The distribution of spectral indices of the 1873 2FGL sources
    (colored red in the electronic version) and the 21 spatially extended
    sources (colored blue).  The index of Centaurus A is taken from
    \cite{second_cat} and the index of Vela X is taken from \cite{velax}.
    }\label{compare_index_2FGL}
  \end{figure}

Figure~\ref{compare_index_2FGL} compares the spectral indices of LAT
detected extended sources and of all sources in the 2FGL catalog. This, and
Tables~\ref{known_extended_sources} and~\ref{new_ext_srcs_table},
show that the LAT observes a population of hard extended sources
at energies above 10 \gev.  Figure~\ref{hess_seds} shows that
the spectra of four of these sources (2FGL\,J1615.0$-$5051,
2FGL\,J1615.2$-$5138, 2FGL\,J1632.4$-$4753c, and 2FGL\,J1837.3$-$0700c)
at \gev energies connects to the spectra of their H.E.S.S. counterparts
at \tev energies. This is also true of Vela Jr., HESS\,J1825$-$137
\citep{fermi_hess_j1825}, and RX\,J1713.7$-$3946 \citep{rx_j1713_lat}.
It is likely that the \gev and \tev emission from these sources originates
from the same population of high-energy particles.

Many of the \tev-detected extended sources now seen at \gev energies
are currently unidentified and further multiwavelength follow-up
observations will be necessary to understand these particle accelerators.
Extending the spectra of these \tev sources towards lower energies
with LAT observations may help to determine the origin and nature of
the high-energy emission.

The \textit{Fermi} LAT Collaboration acknowledges generous ongoing support
from a number of agencies and institutes that have supported both the
development and the operation of the LAT as well as scientific data analysis.
These include the National Aeronautics and Space Administration and the
Department of Energy in the United States, the Commissariat \`a l'Energie Atomique
and the Centre National de la Recherche Scientifique / Institut National de Physique
Nucl\'eaire et de Physique des Particules in France, the Agenzia Spaziale Italiana
and the Istituto Nazionale di Fisica Nucleare in Italy, the Ministry of Education,
Culture, Sports, Science and Technology (MEXT), High Energy Accelerator Research
Organization (KEK) and Japan Aerospace Exploration Agency (JAXA) in Japan, and
the K.~A.~Wallenberg Foundation, the Swedish Research Council and the
Swedish National Space Board in Sweden.

Additional support for science analysis during the operations phase is gratefully
acknowledged from the Istituto Nazionale di Astrofisica in Italy and the Centre National d'\'Etudes Spatiales in France.

This research has made use of
pywcsgrid2, an open-source plotting package for
Python\footnote{\url{http://leejjoon.github.com/pywcsgrid2/}}. The authors acknowledge the
use of HEALPix\footnote{\url{http://healpix.jpl.nasa.gov/}} \citep{healpix}.

\bibliography{extended_search}

\clearpage

\clearpage

\clearpage
\begin{deluxetable}{rrrrrr}
\tabletypesize{\scriptsize}
\tablecaption{Monte Carlo Spectral Parameters
\label{ts_ext_num_sims}
}
\tablecolumns{6}
\tablewidth{0pt}
\tablehead{
\colhead{Spectral Index}&
\colhead{Flux\tablenotemark{(a)}}&
\colhead{$N_{1-100\gev}$}&
\colhead{$\langle\ts\rangle_{1-100\gev}$}&
\colhead{$N_{10-100\gev}$}&
\colhead{$\langle\ts\rangle_{10-100\gev}$}\\
\colhead{}&
\colhead{(\phflux)}&
\colhead{}&
\colhead{}&
\colhead{}&
\colhead{}
}

\startdata
\multicolumn{6}{c}{Isotropic Background} \\
\hline
      1.5 &  $3\times 10^{-7}$ &           18938 &  22233    &     18938   &    8084     \\
          &          $10^{-7}$ &           19079 &   5827    &     19079   &    2258     \\
          &  $3\times 10^{-8}$ &           19303 &   1276    &     19303   &     541     \\
          &          $10^{-8}$ &           19385 &    303    &     19381   &     142     \\
          &  $3\times 10^{-9}$ &           18694 &     62    &     12442   &      43     \\
      \hline
        2 &          $10^{-6}$ &           18760 &  22101    &     18760   &    3033     \\
          &  $3\times 10^{-7}$ &           18775 &   4913    &     18775   &     730     \\
          &          $10^{-7}$ &           18804 &   1170    &     18803   &     192     \\
          &  $3\times 10^{-8}$ &           18836 &    224    &     15256   &      50     \\
          &          $10^{-8}$ &           17060 &     50    &   \nodata   & \nodata     \\
      \hline
      2.5 &  $3\times 10^{-6}$ &           18597 &  19036    &     18597   &     786    \\
          &          $10^{-6}$ &           18609 &   4738    &     18608   &     208    \\
          &  $3\times 10^{-7}$ &           18613 &    954    &     15958   &      53    \\
          &          $10^{-7}$ &           18658 &    203    &   \nodata   & \nodata    \\
          &  $3\times 10^{-8}$ &           14072 &     41    &   \nodata   & \nodata    \\
      \hline                                                
        3 &          $10^{-5}$ &           18354 &  19466    &     18354   &     215    \\
          &  $3\times 10^{-6}$ &           18381 &   4205    &     15973   &      54    \\
          &          $10^{-6}$ &           18449 &    966    &   \nodata   & \nodata    \\
          &  $3\times 10^{-7}$ &           18517 &    174    &   \nodata   & \nodata    \\
          &          $10^{-7}$ &           13714 &     41    &   \nodata   & \nodata    \\
\cutinhead{Galactic Diffuse and Isotropic Background\tablenotemark{(b)}}
      1.5 &  $2.3\times 10^{-8}$ &           90741 &     63    &   \nodata   & \nodata     \\
        2 &  $1.2\times 10^{-7}$ &           92161 &     60    &   \nodata   & \nodata     \\
      2.5 &  $4.5\times 10^{-7}$ &           86226 &     47    &   \nodata   & \nodata    \\
        3 &  $2.0\times 10^{-6}$ &           94412 &     61    &   \nodata   & \nodata    \\
\enddata

\tablenotetext{(a)}{Integral 100 \mev to 100 \gev flux.}
\tablenotetext{(b)}{
For the Galactic simulations, the quoted fluxes are
the fluxes for sources placed 
in the Galactic center. The actual fluxes are scaled by Equation~\ref{scale_flux_by_background}.
}

\tablecomments{
    A list of the spectral models of the simulated point-like sources
    which were tested for extension.  For each model, the number of
    statistically independent simulations and the average value of \ts is
    also tabulated.  
    The top rows are the simulations on top of an isotropic background and
    the bottom rows are the simulations on top of the Galactic diffuse and isotropic
    background.
}
\end{deluxetable}

\clearpage

\clearpage

\clearpage

\clearpage

\clearpage

\clearpage

\clearpage

  \clearpage
  \thispagestyle{empty}
\begin{deluxetable}{rr|rrrrrrrrrrrrrrrrrrrrr}
\tablecolumns{22}
\tabletypesize{\scriptsize}
\rotate
\tablewidth{0pt}
\tablecaption{Extension Detection Threshold
\label{all_sensitivity_table}
}
\tablehead{
\colhead{$\gamma$}&       
\colhead{BG}&
\colhead{$0.1$}&
\colhead{$0.2$}&
\colhead{$0.3$}&
\colhead{$0.4$}&
\colhead{$0.5$}&
\colhead{$0.6$}&
\colhead{$0.7$}&
\colhead{$0.8$}&
\colhead{$0.9$}&
\colhead{$1.0$}&
\colhead{$1.1$}&
\colhead{$1.2$}&
\colhead{$1.3$}&
\colhead{$1.4$}&
\colhead{$1.5$}&
\colhead{$1.6$}&
\colhead{$1.7$}&
\colhead{$1.8$}&
\colhead{$1.9$}&
\colhead{$2.0$}
}
\startdata
\multicolumn{22}{c}{E$>$1 \gev} \\
\hline
     1.5 &      $1\times$ &      148.1 &       23.3 &       11.3 &        8.0 &        7.2 &        6.9 &        6.7 &        6.8 &        7.1 &        7.4 &        7.6 &        7.9 &        8.1 &        8.5 &        9.2 &        9.9 &        9.1 &        9.2 &        9.0 &       10.3 \\
         &     $10\times$ &      148.4 &       29.0 &       18.7 &       15.2 &       15.4 &       15.0 &       16.1 &       16.0 &       16.8 &       17.7 &       18.2 &       19.3 &       20.9 &       22.5 &       23.8 &       24.8 &       21.3 &       22.8 &       23.4 &       23.7 \\
         &    $100\times$ &      186.8 &       55.0 &       43.4 &       40.7 &       41.0 &       41.8 &       40.9 &       40.9 &       42.7 &       43.6 &       38.4 &       39.9 &       40.6 &       38.4 &       36.9 &       36.3 &       37.1 &       38.8 &       37.2 &       37.6 \\
       2 &      $1\times$ &      328.4 &       43.4 &       18.9 &       13.4 &       11.2 &       10.4 &       10.2 &       10.2 &       10.2 &       10.4 &       10.7 &       10.9 &       11.2 &       11.5 &       12.4 &       12.6 &       13.0 &       13.4 &       14.0 &       14.4 \\
         &     $10\times$ &      341.0 &       55.9 &       32.3 &       27.6 &       26.5 &       25.4 &       25.6 &       25.9 &       27.4 &       26.8 &       27.8 &       28.7 &       29.8 &       30.1 &       31.0 &       31.5 &       31.7 &       34.0 &       34.3 &       35.9 \\
         &    $100\times$ &      420.5 &      128.3 &       90.2 &       77.3 &       73.3 &       70.8 &       67.5 &       64.3 &       64.2 &       64.1 &       62.8 &       63.6 &       61.7 &       61.9 &       58.4 &       59.0 &       61.4 &       63.3 &       60.1 &       58.1 \\
     2.5 &      $1\times$ &      627.1 &       75.6 &       29.8 &       19.3 &       15.5 &       13.5 &       12.8 &       12.6 &       12.5 &       12.5 &       12.6 &       12.9 &       12.9 &       13.1 &       13.5 &       13.7 &       14.3 &       14.8 &       15.2 &       15.8 \\
         &     $10\times$ &      638.9 &       99.1 &       52.1 &       39.1 &       34.6 &       33.0 &       32.5 &       32.5 &       32.8 &       33.2 &       34.1 &       34.3 &       34.5 &       35.1 &       36.6 &       36.9 &       35.5 &       36.0 &       36.5 &       37.3 \\
         &    $100\times$ &      795.0 &      262.1 &      140.9 &      104.3 &       90.4 &       81.2 &       77.2 &       75.1 &       69.7 &       70.9 &       66.5 &       65.6 &       64.9 &       64.0 &       58.9 &       58.1 &       60.2 &       58.4 &       57.5 &       55.8 \\
       3 &      $1\times$ &      841.5 &      110.6 &       43.2 &       25.5 &       18.7 &       16.1 &       14.4 &       13.6 &       13.3 &       13.2 &       13.1 &       13.1 &       13.4 &       13.6 &       13.5 &       13.8 &       14.2 &       14.4 &       14.8 &       15.4 \\
         &     $10\times$ &      921.6 &      151.3 &       69.1 &       47.8 &       40.7 &       37.1 &       35.5 &       34.5 &       35.1 &       35.5 &       35.3 &       35.3 &       35.4 &       35.5 &       36.8 &       37.6 &       35.3 &       35.4 &       36.3 &       36.6 \\
         &    $100\times$ &     1124.1 &      282.9 &      181.1 &      119.8 &      100.7 &       91.1 &       84.3 &       77.9 &       73.3 &       71.8 &       67.6 &       66.4 &       65.5 &       63.9 &       59.0 &       58.6 &       58.8 &       57.5 &       55.4 &       54.4 \\
\cutinhead{E$>$10 \gev}
     1.5 &      $1\times$ &       44.6 &        8.0 &        4.3 &        3.2 &        2.7 &        2.6 &        2.5 &        2.5 &        2.4 &        2.5 &        2.5 &        2.6 &        2.7 &        2.8 &        2.9 &        2.9 &        3.1 &        3.2 &        3.3 &        3.4 \\
         &     $10\times$ &       45.2 &        9.2 &        5.8 &        5.0 &        4.9 &        4.9 &        5.0 &        5.2 &        5.3 &        5.7 &        5.9 &        6.3 &        6.6 &        6.5 &        6.8 &        7.6 &        7.8 &        8.2 &        8.5 &        8.7 \\
         &    $100\times$ &       47.3 &       13.4 &       11.6 &       10.6 &       10.8 &       10.8 &       12.0 &       12.7 &       13.2 &       13.7 &       15.3 &       16.1 &       17.2 &       18.2 &       18.9 &       19.5 &       20.4 &       21.0 &       21.7 &       22.9 \\
       2 &      $1\times$ &       49.7 &        8.4 &        4.4 &        3.3 &        2.8 &        2.6 &        2.6 &        2.6 &        2.6 &        2.6 &        2.7 &        2.7 &        2.8 &        2.9 &        3.0 &        3.2 &        3.2 &        3.4 &        3.5 &        3.5 \\
         &     $10\times$ &       48.6 &        9.5 &        6.0 &        5.2 &        5.0 &        5.2 &        5.2 &        5.3 &        5.4 &        5.8 &        6.4 &        6.6 &        7.0 &        7.1 &        7.5 &        8.0 &        8.3 &        8.6 &        9.0 &        9.2 \\
         &    $100\times$ &       51.8 &       14.7 &       11.8 &       11.5 &       11.5 &       11.9 &       13.2 &       14.0 &       14.3 &       15.3 &       16.2 &       16.9 &       18.4 &       19.2 &       19.8 &       21.0 &       22.0 &       22.8 &       23.2 &       24.3 \\
     2.5 &      $1\times$ &       53.1 &        9.1 &        4.5 &        3.3 &        2.8 &        2.7 &        2.6 &        2.5 &        2.5 &        2.6 &        2.7 &        2.7 &        2.8 &        2.8 &        2.9 &        3.1 &        3.2 &        3.3 &        3.5 &        3.6 \\
         &     $10\times$ &       53.7 &       10.5 &        6.3 &        5.4 &        5.1 &        5.1 &        5.3 &        5.4 &        5.7 &        6.0 &        6.3 &        6.6 &        6.8 &        6.9 &        7.5 &        8.1 &        8.3 &        8.6 &        8.9 &        9.2 \\
         &    $100\times$ &       57.0 &       15.6 &       12.7 &       11.9 &       11.8 &       12.2 &       13.1 &       14.3 &       14.6 &       15.2 &       16.3 &       17.0 &       18.8 &       19.2 &       19.9 &       21.0 &       21.9 &       22.3 &       23.3 &       23.7 \\
       3 &      $1\times$ &       55.5 &        9.4 &        4.8 &        3.4 &        2.9 &        2.7 &        2.6 &        2.5 &        2.5 &        2.5 &        2.6 &        2.7 &        2.7 &        2.8 &        2.9 &        3.0 &        3.1 &        3.2 &        3.4 &        3.4 \\
         &     $10\times$ &       56.0 &       10.5 &        6.2 &        5.3 &        5.1 &        5.1 &        5.1 &        5.3 &        5.5 &        5.7 &        5.9 &        6.4 &        6.4 &        6.6 &        7.0 &        7.8 &        8.0 &        8.3 &        8.6 &        8.9 \\
         &    $100\times$ &       60.3 &       16.2 &       12.7 &       11.7 &       11.8 &       12.2 &       12.6 &       13.8 &       14.2 &       14.6 &       15.8 &       16.5 &       17.6 &       18.5 &       19.4 &       19.8 &       20.7 &       21.0 &       21.8 &       22.5 \\
\enddata
\tablecomments{
      The detection threshold to resolve spatially extended
      sources with a uniform disk spatial model for a two-year exposure
      The threshold is calculated for
      sources of varying energy
      ranges, spectral indices, and background levels.  The sensitivity
      was calculated against
      a Sreekumar-like isotropic background
      and the second column is the factor that the simulated background
      was scaled by. The remaining columns are varying sizes of the source. 
      The table quotes
      integral fluxes in the analyzed energy range (1 \gev to 100 \gev or 10 \gev to 100
      \gev) in units of $10^{-10}$ \phflux.
      }
\end{deluxetable}

\clearpage

\clearpage

\clearpage

\begin{deluxetable}{lrrrrrrrr}
\tablecolumns{9}
\rotate
\tabletypesize{\footnotesize}
\tablewidth{0pt}
\tablecaption{Analysis of the twelve extended sources included in the 2FGL catalog
\label{known_extended_sources}
}
\tablehead{
\colhead{Name}&
\colhead{\glon}&
\colhead{\glat}&
\colhead{$\sigma$}&
\colhead{\ts}&
\colhead{\tsext}&
\colhead{Pos Err}&
\colhead{Flux\tablenotemark{(a)}}&
\colhead{Index}\\
\colhead{}&
\colhead{(deg.)}&
\colhead{(deg.)}&
\colhead{(deg.)}&
\colhead{}&
\colhead{}&
\colhead{(deg.)}&
\colhead{}&
\colhead{}
}

\startdata
\multicolumn{9}{c}{E$>$1 \gev} \\
\hline
SMC                                  &     302.59 &   $-$44.42 & $  1.32 \pm   0.15 \pm   0.31 $ &       95.0 &       52.9 &   0.14 & $    2.7 \pm     0.3$ & $   2.48 \pm    0.19$  \\
LMC                                  &     279.26 &   $-$32.31 & $  1.37 \pm   0.04 \pm   0.11 $ &     1127.9 &      909.9 &   0.04 & $   13.6 \pm     0.6$ & $   2.43 \pm    0.06$  \\
IC~443                               &     189.05 &       3.04 & $  0.35 \pm   0.01 \pm   0.04 $ &    10692.9 &      554.4 &   0.01 & $   62.4 \pm     1.1$ & $   2.22 \pm    0.02$  \\
Vela X                               &     263.34 &    $-$3.11 & $                         0.88$ &            &            &        &                       &                        \\
Centaurus A                          &     309.52 &      19.42 &                        $\sim10$ &            &            &        &                       &                        \\
W28                                  &       6.50 &    $-$0.27 & $  0.42 \pm   0.02 \pm   0.05 $ &     1330.8 &      163.8 &   0.01 & $   56.5 \pm     1.8$ & $   2.60 \pm    0.03$  \\
W30                                  &       8.61 &    $-$0.20 & $  0.34 \pm   0.02 \pm   0.02 $ &      464.8 &       76.0 &   0.02 & $   29.1 \pm     1.5$ & $   2.56 \pm    0.05$  \\
W44                                  &      34.69 &    $-$0.39 & $  0.35 \pm   0.02 \pm   0.02 $ &     1917.0 &      224.8 &   0.01 & $   71.2 \pm     0.5$ & $   2.66 \pm    0.00$  \\
W51C                                 &      49.12 &    $-$0.45 & $  0.27 \pm   0.02 \pm   0.04 $ &     1823.4 &      118.9 &   0.01 & $   37.2 \pm     1.3$ & $   2.34 \pm    0.03$  \\
Cygnus Loop                          &      74.21 &    $-$8.48 & $  1.71 \pm   0.05 \pm   0.06 $ &      357.9 &      246.0 &   0.06 & $   11.4 \pm     0.7$ & $   2.50 \pm    0.10$  \\
\cutinhead{E$>$10 \gev}
MSH\,15$-$52\tablenotemark{(b)}      &     320.39 &    $-$1.22 & $  0.21 \pm   0.04 \pm   0.04 $ &       76.3 &        6.6 &   0.03 & $    0.6 \pm     0.1$ & $   2.20 \pm    0.22$  \\
HESS\,J1825$-$137\tablenotemark{(b)} &      17.56 &    $-$0.47 & $  0.65 \pm   0.04 \pm   0.02 $ &       59.7 &       33.8 &   0.05 & $    1.6 \pm     0.2$ & $   1.63 \pm    0.22$  \\
\enddata

\tablenotetext{(a)}{
Integral Flux in units of $10^{-9}$ \phflux and integrated in the fit
energy range (either 1 \gev to 100 \gev or 10 \gev to 100 \gev).
}
\tablenotetext{(b)}{
The discrepancy in the best fit spectra of MSH\,15$-$52 and HESS\,J1825$-$137
compared to \cite{msh1552} and
\cite{fermi_hess_j1825} is due to fitting over a different energy range.
}

\tablecomments{
All sources were fit using a radially-symmetric uniform disk spatial model.
\glon and \glat are Galactic longitude
and latitude of the best fit extended source respectively.  The first
error on $\sigma$ is statistical and the second is systematic (see
Section~\ref{systematic_errors_on_extension}).  
The errors on the integral fluxes and the spectral indices
are statistical only.
Pos Err is the error on
the position of the source.  Vela X and the Centaurus A Lobes were
not fit in our analysis but are included for completeness.
}
\end{deluxetable}

\clearpage

\clearpage
\thispagestyle{empty}
\begin{deluxetable}{lrrrrrrrrl}
\tablecolumns{10}
\tabletypesize{\small}
\rotate
\tablewidth{0pt}
\tablecaption{Extension fit for the nine additional extended sources
\label{new_ext_srcs_table}
}
\tablehead{
\colhead{Name}&
\colhead{\glon}&
\colhead{\glat}&
\colhead{$\sigma$}&
\colhead{\ts}&
\colhead{\tsext}&
\colhead{Pos Err}&
\colhead{Flux\tablenotemark{(a)}}&
\colhead{Index}&
\colhead{Counterpart}\\
\colhead{}&
\colhead{(deg.)}&
\colhead{(deg.)}&
\colhead{(deg.)}&
\colhead{}&
\colhead{}&
\colhead{(deg.)}&
\colhead{}&
\colhead{}&
\colhead{}
}

\startdata
\multicolumn{10}{c}{E$>$1 \gev} \\
\hline
2FGL\,J0823.0$-$4246                       &     260.32 &    $-$3.28 & $  0.37 \pm   0.03 \pm   0.02 $ &      322.2 &       48.0 &   0.02 & $    8.4 \pm     0.6$ & $   2.21 \pm    0.09$ &                  Puppis A \\
2FGL\,J1627.0$-$2425c                      &     353.07 &      16.80 & $  0.42 \pm   0.05 \pm   0.16 $ &      139.9 &       32.4 &   0.04 & $    6.3 \pm     0.6$ & $   2.50 \pm    0.14$ &                 Ophiuchus \\
\cutinhead{E$>$10 \gev}
2FGL\,J0851.7$-$4635                       &     266.31 &    $-$1.43 & $  1.15 \pm   0.08 \pm   0.02 $ &      116.6 &       86.8 &   0.07 & $    1.3 \pm     0.2$ & $   1.74 \pm    0.21$ &                  Vela Jr. \\
2FGL\,J1615.0$-$5051                       &     332.37 &    $-$0.13 & $  0.32 \pm   0.04 \pm   0.01 $ &       50.4 &       16.7 &   0.04 & $    1.0 \pm     0.2$ & $   2.19 \pm    0.28$ &         HESS\,J1616$-$508 \\
2FGL\,J1615.2$-$5138                       &     331.66 &    $-$0.66 & $  0.42 \pm   0.04 \pm   0.02 $ &       76.1 &       46.5 &   0.04 & $    1.1 \pm     0.2$ & $   1.79 \pm    0.26$ &         HESS\,J1614$-$518 \\
2FGL\,J1632.4$-$4753c                      &     336.52 &       0.12 & $  0.35 \pm   0.04 \pm   0.02 $ &       64.4 &       26.9 &   0.04 & $    1.4 \pm     0.2$ & $   2.66 \pm    0.30$ &         HESS\,J1632$-$478 \\
2FGL\,J1712.4$-$3941\tablenotemark{(b)}    &     347.26 &    $-$0.53 & $  0.56 \pm   0.04 \pm   0.02 $ &       59.4 &       38.5 &   0.05 & $    1.2 \pm     0.2$ & $   1.87 \pm    0.22$ &        RX\,J1713.7$-$3946 \\
2FGL\,J1837.3$-$0700c                      &      25.08 &       0.13 & $  0.33 \pm   0.07 \pm   0.05 $ &       47.0 &       18.5 &   0.07 & $    1.0 \pm     0.2$ & $   1.65 \pm    0.29$ &         HESS\,J1837$-$069 \\
2FGL\,J2021.5+4026                         &      78.24 &       2.20 & $  0.63 \pm   0.05 \pm   0.04 $ &      237.2 &      128.9 &   0.05 & $    2.0 \pm     0.2$ & $   2.42 \pm    0.19$ &            $\gamma$-Cygni \\
\enddata

\tablenotetext{(a)}{
Integral Flux in units of $10^{-9}$ \phflux and integrated in the fit
energy range (either 1 \gev to 100 \gev or 10 \gev to 100 \gev).
}

\tablenotetext{(b)}{
The discrepancy in the best fit spectra of 2FGL\,J1712.4$-$3941 compared
to \cite{rx_j1713_lat} is due to fitting over a different energy range.
}

\tablecomments{
    The columns in this table have the same meaning as those in
    Table~\ref{known_extended_sources}.
    RX\,J1713.7$-$3946 and Vela Jr. were previously studied in dedicated publications \citep{rx_j1713_lat,vela_jr_lat}.
}
\end{deluxetable}

\clearpage
\thispagestyle{empty}
\begin{deluxetable}{lrrrrrrrrrr}
  \tabletypesize{\small}
  \tablecolumns{11}
  \rotate
  \tablewidth{0pt}
  \tablecaption{Dual localization, alternative PSF, and alternative approach to modeling the diffuse emission
  \label{alt_diff_model_results}
  }
  \tablehead{
  \colhead{Name}&
  \colhead{$\ts_\pointlike$}&
  \colhead{$\ts_\gtlike$}&
  \colhead{$\ts_\altdiff$}&
  \colhead{$\tsext_\pointlike$}&
  \colhead{$\tsext_\gtlike$}&
  \colhead{$\tsext_\altdiff$}&
  \colhead{$\sigma$}&
  \colhead{$\sigma_\altdiff$}&
  \colhead{$\sigma_\altpsf$}&
  \colhead{$\tsinc$}\\
  \colhead{}&
  \colhead{}&
  \colhead{}&
  \colhead{}&
  \colhead{}&
  \colhead{}&
  \colhead{}&
  \colhead{(deg.)}&
  \colhead{(deg.)}&
  \colhead{(deg.)}&
  \colhead{}
  }
  \startdata
\multicolumn{11}{c}{E$>$1 \gev} \\
\hline
2FGL\,J0823.0$-$4246      &                331.9 &                322.2 &                356.0 &                 60.0 &                 48.0 &                 56.0 &                 0.37 &                 0.39 &                 0.39 &                 23.0 \\
2FGL\,J1627.0$-$2425c     &                154.8 &                139.9 &                105.7 &                 39.4 &                 32.4 &                 24.8 &                 0.42 &                
 0.40 &                 0.58 &                 24.5 \\
\cutinhead{E$>$10 \gev}
2FGL\,J0851.7$-$4635      &                115.2 &                116.6 &                123.1 &                 83.9 &                 86.8 &                 89.8 &                 1.15 &                 1.16 &                 1.17 &                 15.5 \\
2FGL\,J1615.0$-$5051\tablenotemark{(a)} &                 48.2 &                 50.4 &                 56.6 &                 15.2 &                 16.7 &                 17.8 &                 0.32 &                 0.33 &                 0.32 &                 13.1 \\
2FGL\,J1615.2$-$5138      &                 75.0 &                 76.1 &                 83.8 &                 42.9 &                 46.5 &                 54.1 &                 0.42 &                 0.43 &                 0.43 &                 35.1 \\
2FGL\,J1632.4$-$4753c     &                 64.5 &                 64.4 &                 66.8 &                 23.0 &                 26.9 &                 25.5 &                 0.35 &                 0.36 &                 0.37 &                 10.9 \\
2FGL\,J1712.4$-$3941      &                 59.8 &                 59.4 &                 39.9 &                 38.4 &                 38.5 &                 30.7 &                 0.56 &                 0.55 &                 0.53 &                  2.7 \\
2FGL\,J1837.3$-$0700c     &                 44.5 &                 47.0 &                 39.2 &                 17.6 &                 18.5 &                 16.1 &                 0.33 &                 0.32 &                 0.38 &                 10.8 \\
2FGL\,J2021.5+4026        &                239.1 &                237.2 &                255.8 &                139.1 &                128.9 &                138.0 &                 0.63 &                 0.65 &                 0.59 &                 37.3 \\
  \enddata

\tablenotetext{(a)}{
Using \pointlike, \tsext for 2FGL\,J1615.0$-$5051 was sligthly below
16 when the source was fit in the 10 \gev to 100 \gev energy range. To
confirm the extension measure, the extension was refit in \pointlike
using a slightly lower energy.  In the 5.6 \gev to 100 \gev energy range,
we obtained a consistent extension and \tsext=28.0.  In the rest of
this paper, we quote the $E>10 \gev$ results for consistency with the
other sources.
}

\tablecomments{
$\ts_\pointlike$, $\ts_\gtlike$, and $\ts_\altdiff$ are the test
statistic values from \pointlike, \gtlike, and \gtlike with the alternative
approach to modeling the diffuse emission respectively.  $\tsext_\pointlike$, $\tsext_\gtlike$,
and $\tsext_\altdiff$ are the TS values from \pointlike, \gtlike, and \gtlike with the alternative 
approach to modeling the diffuse emission respectively.  $\sigma$, $\sigma_\altdiff$, and $\sigma_\altpsf$
are the fit sizes assuming
a radially-symmetric uniform disk model
with the standard analysis, the alternative 
approach to modeling the diffuse emission, and the alternative PSF respectively.  
}
\end{deluxetable}

\clearpage
\begin{deluxetable}{lr}
\tablecolumns{2}
\tablewidth{0pt}
\tablecaption{Nearby Residual-induced Sources
\label{fake_2fgl_sources}
}
\tablehead{
\colhead{Extended Source}&
\colhead{Residual-induced Sources}
}
\startdata
\hline
2FGL\,J0823.0$-$4246    & 2FGL\,J0821.0$-$4254, 2FGL\,J0823.4$-$4305 \\
2FGL\,J1627.0$-$2425c   & \nodata \\
2FGL\,J0851.7$-$4635    & 2FGL\,J0848.5$-$4535, 2FGL\,J0853.5$-$4711, 2FGL\,J0855.4$-$4625 \\
2FGL\,J1615.0$-$5051    & \nodata \\
2FGL\,J1615.2$-$5138    & 2FGL\,J1614.9$-$5212  \\
2FGL\,J1632.4$-$4753c   & 2FGL\,J1634.4$-$4743c \\
2FGL\,J1712.4$-$3941    & \nodata \\
2FGL\,J1837.3$-$0700c   & 2FGL\,J1835.5$-$0649 \\
2FGL\,J2021.5+4026      & 2FGL\,J2019.1+4040 \\
\enddata

\tablecomments{
For each new extended source, we list nearby 2FGL
soruces that we have concluded here correspond to residuals induced by not
modeling the extensions of nearby extended sources.
}
\end{deluxetable}

\clearpage

\clearpage

\clearpage

\clearpage

\clearpage

\clearpage

\end{document}